\documentclass[twocolumn,prd,noshowpacs,nofootinbib,amsmath,amssymb,superscriptaddress]{revtex4}
\usepackage{graphicx}
\usepackage{amssymb}
\usepackage{amsmath}
\usepackage{amsfonts}
\usepackage{latexsym}
\usepackage{hyperref}
\usepackage[all]{xy}
\usepackage{slashed}
\usepackage{physics}
\usepackage{accents}
\usepackage{afterpage}
\usepackage{booktabs,siunitx,dcolumn}

\newcommand{\be}{\begin{eqnarray}}
\newcommand{\ee}{\end{eqnarray}}

\newcommand{\meV}{~\mathrm{meV}}
\newcommand{\eV}{~\mathrm{eV}}
\newcommand{\keV}{~\mathrm{keV}}
\newcommand{\MeV}{~\mathrm{MeV}}
\newcommand{\GeV}{~\mathrm{GeV}}

\newcommand{\cm}{~\mathrm{cm}}
 
\newcommand{\fm}{~\mathrm{fm}}
\newcommand{\um}{~\mu\mathrm{m}}
\newcommand{\nm}{~\mathrm{nm}}
\newcommand{\kms}{~\mathrm{km/s}}
\newcommand{\K}{~\mathrm{K}}

\newcommand{\AAA}{~\mathrm{\AA}}

\newcommand{\alphaM}{\alpha_{\rm _M}}
\newcommand{\alphaMsense}{\hat{\alpha}_{\rm _M}}
\newcommand{\alphaDM}{\alpha_{\rm _{DM}}}
\newcommand{\gM}{g_{\rm _{M}}}
\newcommand{\gDM}{g_{\rm _{DM}}}

\newcommand{\mmed}{m}
\newcommand{\mDM}{M}
\newcommand{\mDMesc}{\mDM^{\mathrm{esc}}}
\newcommand{\mN}{M_{_N}}

\newcommand{\pin}{{\bf k}}
\newcommand{\pout}{{\bf k}'}
\newcommand{\pmid}{{\bf p}}
\newcommand{\pell}{\boldsymbol{\ell}}

\newcommand{\lambdamed}{\lambdabar_{\mmed}}
\newcommand{\lambdaDM}{\lambdabar_{\rm _{DM}}}
\newcommand{\rhoDM}{\rho_{\rm _{DM}}}
\newcommand{\vDM}{\bar{v}_{\rm _{DM}}}
\newcommand{\vE}{v_{\odot}}
\newcommand{\vEsc}{v_{\mathrm{esc}}}
\newcommand{\Znorm}{Z}

\newcommand{\TDM}{T_{\rm{DM}}}
\newcommand{\DeltaX}{\Delta x}
\newcommand{\gammaDM}{\gamma_{\rm _{DM}}}
\newcommand{\gammaOther}{\gamma_{\mathrm{other}}}

\newcommand{\vis}{\gamma_{\mathrm{vis}}}
\newcommand{\DeltaNum}{\Delta B}
\newcommand{\barNum}{B_0}
\newcommand{\Num}{B}

\newcommand{\mrn}{\mathrm{mrn}}
\newcommand{\eve}{\mathrm{eve}}

\newcommand{\alphafac}{\alpha}
\newcommand{\betafac}{\beta}
\newcommand{\sigmafac}{\sigma}

\newcommand{\Cndx}{C_{\bf k \DeltaX}}
\newcommand{\Sndx}{S_{\bf k \DeltaX}}
\newcommand{\Cchi}{C_{\bf \vE \DeltaX}}
\newcommand{\Schi}{S_{\bf \vE \DeltaX}}
\newcommand{\Cnq}{C_{\bf k q}}
\newcommand{\Snq}{S_{\bf k q}}

\newcommand{\Ycoeff}{Y}
\newcommand{\chistat}{\chi}
\newcommand{\Gammacount}{\Gamma}
\newcommand{\Trun}{T_{\mathrm{run}}}

\newcommand{\sDM}{s_{_{\mathrm{DM}}}}
\newcommand{\phiDM}{\phi_{_{\mathrm{DM}}}}
\newcommand{\DphiDM}{{\accentset{\sim}{\phi}}_{_{\mathrm{DM}}}}
\newcommand{\Dsdm}{{\accentset{\sim}{s}}_{_{\mathrm{DM}}}}
\newcommand{\Bsdm}{\bar{s}_{_{\mathrm{DM}}}}
\newcommand{\sigmaDs}{\sigma_{\accentset{\sim}{s}}}
\newcommand{\Dsirr}{{\accentset{\sim}{s}}_{_{\mathrm{res}}}}
\newcommand{\sbkg}{s_{_{\mathrm{bkg}}}}
\newcommand{\dF}{{\accentset{\sim}{F}}}

\newcommand{\gtil}{\tilde{f}}
\newcommand{\gtilsph}{\tilde{f}^{(R)}_{\mathrm{sph}}}

\newcommand{\etairr}{\eta_{\mathrm{\hspace{0.5 pt}res}}}
\newcommand{\etaDM}{\eta_{_{\mathrm{DM}}}}

\begin{document}

\title{Decoherence as a way to measure extremely soft collisions with dark matter}
\author{C.\ Jess Riedel\footnote{E-mail: jessriedel@gmail.com}}
\affiliation{Perimeter Institute for Theoretical Physics, Waterloo, Ontario, Canada}
\author{Itay Yavin}
\affiliation{Perimeter Institute for Theoretical Physics, Waterloo, Ontario, Canada}
\affiliation{Department of Physics, McMaster University, Hamilton, Ontario, Canada}

\begin{abstract}
A new frontier in the search for dark matter (DM) is based on the idea of detecting the decoherence caused by DM scattering against a mesoscopic superposition of normal matter.  Such superpositions are uniquely sensitive to very small momentum transfers from new particles and forces, especially DM with a mass below 100 MeV.  Here we investigate what sorts of dark sectors are inaccessible with existing methods but would induce noticeable decoherence in the next generation of matter interferometers. We show that very soft, but medium range (0.1 nm - 1 $\mu$m) elastic interactions between nuclei and DM are particularly suitable. We construct toy models for such interactions, discuss existing constraints, and delineate the expected sensitivity of forthcoming experiments.  The first hints of DM in these devices would appear as small variations in the anomalous decoherence rate with a period of one sidereal day.  This is a generic signature of interstellar sources of decoherence, clearly distinguishing it from terrestrial backgrounds.   The OTIMA experiment under development in Vienna will begin to probe Earth-thermalizing DM once sidereal variations in the background decoherence rate are pushed below one part in a hundred for superposed 5-nm gold nanoparticles. The proposals by Bateman \emph{et al.}\ and Geraci \emph{et al.}\ could be similarly sensitive, although they would require at least a month of data taking.  DM that is absorbed or elastically reflected by the Earth, and so avoids a greenhouse density enhancement, would not be detectable by those three experiments.  On the other hand, the aggressive proposals of the MAQRO collaboration and Pino \emph{et al.}\ would immediately open up many orders of magnitude in DM mass, interaction range, and coupling strength, regardless of how DM behaves in bulk matter.
\end{abstract}

\maketitle
\section{Introduction}
\label{sec:introduction}

One of the basic challenges facing fundamental physics in the 21$^{\rm st}$ century is finding ways to learn more about the nature of the dark matter (DM) that makes up some 80\% of the matter in the Universe. If it is a particle, then we would like to know whether it is a boson or a fermion, to measure its mass, and to understand what interactions, other than gravity, it has with itself as well as with the particles we are made of. Even if we restrict ourselves to the case of fermionic DM, in which case galactic phase-space considerations generally require masses greater than $\mathcal{O}(100)$ eV, the allowed mass range is still very large. Given our deep ignorance about the basic properties of DM, it is important to devise experiments to cover as much of the parameter space as possible.

The most developed programs looking for DM above the keV mass range are: direct detection experiments; indirect detection observations; and direct production at colliders. Direct detection experiments look for anomalous energy deposition at the keV range from collisions with dark matter with a mass greater than a GeV and weak (and now sub-weak) cross sections \cite{supercdmscollaboration2016new,pandax-iicollaboration2016dark,angloher2016results,luxcollaboration2017results}. These experiments are insensitive for masses below a GeV because the kinetic energy of such particles in the halo is below a keV and they do not impart enough energy to be detected. Recently, promising new ideas for experiments targeting lighter DM in the keV--GeV mass range have emerged~\cite{Essig:2011nj,graham2012semiconductor,essig2013dark,lee2015modulation, hochberg2016detecting, hochberg2016superconducting, hochberg2016directional} and the first limits were announced in Ref.~\cite{Essig:2012yx}; these exploit the increased energy transfer if electrons, rather than heavier nucleons, are struck. Alongside the lower mass, the cross sections targeted by these experiments are generally considerably higher than weak scale (1--1000 pb). Detection through indirect observations searches for the annihilation or decay of dark matter into stable forms of matter, such as photons, protons, electrons and their antiparticles~\cite{essig2013dark,essig2013constraining}. These processes must be sufficiently frequent and energetic to be distinguished from background, which becomes more challenging below a DM mass of a GeV.  The most notable hints in this area arise in searches for anomalous x-ray lines in the keV range~\cite{essig2013constraining}. Finally, searches for the production of DM at colliders looking for excess events with large missing energy are sensitive to DM with a large mass range that is only limited by the available center-of-mass energy. These searches are generally limited to production cross sections somewhat larger than weak scale. These different approaches to learning more about the nature of DM are of course overlapping and act to inform and complement each other. 

A particularly challenging task is to detect DM with a mass in the MeV range and below, especially through interactions with the nucleus. The available kinetic energy of such particles in the halo is below an electron-volt and there are currently no proposals utilizing conventional means that can search for collisions involving such low energy depositions. Recently, a novel approach based on detecting the decohering effects due to DM interactions with matter was suggested by Riedel in Ref.~\cite{Riedel:2012ur}. While the proposal delineated the achievable sensitivity in cross section, it remained unclear what type of interactions can actually give rise to such a signal. In this paper we close this gap and show that it is interactions mediated by new long range forces between matter and DM that can be effectively searched for with this technique. The key insight underlying our work is this: \textsl{the big advantage decoherence has over other techniques is in detecting collisional processes with a cross section that is dominated by extremely low momentum transfers}. This compliments a recent derivation of a standard quantum limit for diffusion, showing that a given test mass is strictly more sensitive to small momentum transfers when placed in an extended spatial superposition than in any localized (even zero-temperature) state \cite{Riedel:2015xx}.

It is important to recognize that the answer to the question ``What is the nature of DM?", may not be unique. DM, just like us, may be composed of several different stable relics interacting through a variety of forces.  Laboratory experiments looking for DM are generally sensitive even to cosmological relics that only form a subcomponent of DM. Astrophysical constraints on the interactions of the dominant component of DM, which may be inapplicable to a subcomponent, should therefore not limit our vision for what is possible to look for in laboratory experiments.

We close this introduction with two comments. First, it should be duly noted that the signals we discuss in this paper are extremely weak and will require great control over the experimental apparatus. It will likely take quite a few more years before the experiments  reach the required sensitivity. Decoherence is ubiquitous and may arise from a variety of other mundane sources.  Once a signal is seen it would be necessary to run several crosschecks (discussed later) before rejecting alternative explanations and confidently attributing the signal to DM. Second, the interactions between matter and DM that we consider in this paper, while physically consistent, are not motivated by any deep principle or any other consideration. This may seem unappealing to some. Despite these two cautionary remarks, we believe that experiments utilizing large quantum superpositions represent a promising new frontier to look for new physics and new fundamental interactions. The purpose of this paper is to expose the type of new physics such experiments can probe and to serve as a proof of principle that such interactions are physically consistent.

Sec.~\ref{sec:interactions} begins with the key aspects of decoherence of quantum superpositions and discusses the concrete DM scenarios on which they have the largest advantage.  Sec.~\ref{sec:experiment} reviews the relevant experimental considerations, including the most promising interferometric devices, the properties of and constraints on the DM scenario, the decoherence process itself, and methods for isolating signal from background.  Sec.~\ref{sec:results} presents the results, summarized in Figs.~\ref{fig:sensitivity-mdm-big} and~\ref{fig:sensitivity-mdm-many}, and Sec.~\ref{sec:discussion} concludes with brief discussion.

\section{Long-range interactions}
\label{sec:interactions}

The basic challenge facing the proposal of Ref.~\cite{Riedel:2012ur} is that on the one hand simple estimates of the flux and experimental sensitivity reveal that the interaction cross section required are very large and are at the level of $10^{-20} - 10^{-26}\cm^2$. On the other hand, the many experimental searches for DM discussed in the introduction, including direct production searches which are sensitive to arbitrarily low DM masses, have already explored and excluded far smaller cross sections than that. So, the natural question is what type of interactions could be searched for with decoherence experiments that are not already strongly excluded by other searches? 

The basic objective of matter interferometer experiments is to  create and verify a very large aggregate of atoms in a spatial superposition separated by some distance $\DeltaX$, often with $\DeltaX \sim$ tens of nanometers. The experimentalists then observe the decoherence of this superposition as external particles in the environment scatter against the sample. As the  Heisenberg microscope thought experiment suggests, the type of scattering processes that destroy the superposition are ones that reveal ``which path" information about the scattering, i.e., against which of the superposed aggregates did the scattering take place? Such scattering events need to resolve the spatial superposition, and therefore need to involve momentum exchange of the order of $q \sim 2\pi \hbar/\DeltaX 
\sim 10 \eV/c$. (We set $\hbar = c = 1$ hereafter.) There lies the power of such macroscopic decoherence experiments: \textsl{they are potentially sensitive to new physics processes with far lower momentum transfers than anything achievable through the direct, indirect, or collider experiments discussed above!}

Thought of in this way it becomes clear what the type of interactions are that can be best probed by decoherence experiments. Scattering due to a long-range force, or a light mediator in the parlance of quantum field theory, is dominant at low-momentum transfers. In such a case, the large cross section needed for observability at decoherence experiments can be reconciled with the very small cross sections excluded by all the other searches. In fact, as we shall see below, the strongest constraints on this scenario are not coming from other DM searches at all, but rather from experiments looking for additional long-range forces between matter and itself. This becomes clear when thinking about the physics in terms of momentum transfer scales and the fact that if there exist a long-range force between matter and DM, then quantum field theory generally requires this force to also modify the interactions of matter with itself.

\subsection{Non-relativistic description}

The galactic halo particles making up the dark matter that we hope to detect move at relatively slow speeds, about 230 km/sec or so. 
In the non-relativistic regime relevant for the collision of such particles with matter we take the differential cross section to be of Yukawa type,
\be
\label{eqn:yukawa_xs}
\frac{d\sigma}{d\Omega} =  \frac{4\alphaM \alphaDM\mDM^2}{\left(q^2 + \mmed^2 \right)^2}
\ee
Here $\alphaM = \gM^2/4\pi$ ($\alphaDM = \gDM^2/4\pi$) is the dimensionless parameter associated with the coupling $\gM$ ($\gDM$) of the new force to normal (dark) matter, $\mDM$ is the mass of the DM particle\footnote{More generally it should be the reduced mass of the DM particle and the target particle, but for the purpose of this paper we mostly focus on the case of light DM colliding against the much heavier nucleus.}, $\mmed$ is the mass  of the particle mediating the new force. The momentum transfer is $q = 2 \mDM v \sin\theta/2$, where $\theta$ is the angle of the collision in the lab frame and $v$ is the DM incoming velocity. When the Compton wavelength of the mediator mass is comparable to the separation of the superposition, $\mmed^{-1} \sim \DeltaX$, then the cross section is largest for momentum exchange that is very small $q \sim \mmed$ yet sufficiently large to resolve the superposition. 

The strong dependence of the differential cross section in Eq.~(\ref{eqn:yukawa_xs}) on the momentum exchange and mediator mass leads to interesting behavior of the detection sensitivity as a function of the model parameters. For the purpose of calculating this sensitivity, the non-relativistic differential cross section is all we need. However, in order to better understand the physics involved and to be able to relate the current proposal to other searches for new physics, it is useful to have some examples of the possible microscopic origin of Eq.~(\ref{eqn:yukawa_xs}) coming from a more fundamental description. 

\subsection{Quantum field theory description}

The cross section in Eq.~(\ref{eqn:yukawa_xs}) is the result of a Yukawa-type potential between DM and matter,
\be
\label{eqn:yukawa_potential}
V(r) = -\frac{\gM \gDM}{4\pi} \frac{1}{r}e^{-\mmed r}.
\ee
In quantum field theory, such a potential arises from the exchange of a boson of mass $\mmed$ that couples to both DM as well as normal matter (see, e.g., Ref.~\cite{Peskin:1995ev} p.~121--122). The simplest model that realizes such an exchange is one where the mediator is a simple scalar ($\phi$)  that couples to nucleons\footnote{Fundamentally the coupling must be to the constituents that make up nucleons, i.e.\ quarks and gluons, but such fine-grained description will not be important for what follows. Our focus on nucleons as compared to electrons is not an essential one. The only requirement is that the atom as a whole is charged under the new force so that there is no screening of the Yukawa potential.} ($N$) as well as to DM ($\psi$), which we take to be a Dirac fermion. The Lagrangian density for such a model is then,
\be
\nonumber 
\mathcal{L} = \mathcal{L}_{\rm SM} &+&  \tfrac{1}{2} \left( \partial^\mu \phi \partial_\mu \phi - \mmed^2 \phi^2\right) + \bar{\psi}\left(i\gamma^\mu \partial_\mu - \mDM \right) \psi \\
\label{eqn:couplngs}
&+& \gDM \phi \bar{\psi}\psi + \gM \phi \bar{N}N
\ee
where the kinetic terms for the nucleons and all their other standard interactions are in $\mathcal{L}_{\rm SM}$. Using the Feynman prescription the lowest-order contribution to the amplitude describing $N\psi \rightarrow N\psi$ scattering is given by,
\vspace{1.0cm}
\be
\nonumber
i\mathcal{M} =& \parbox[t]{2.5cm}{\vspace{-1.6cm}
	\includegraphics[scale=0.6]{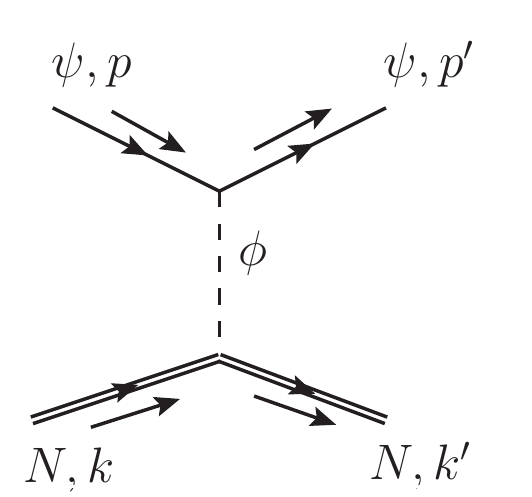} }  
\\ 
\nonumber
=& -i \gM \gDM \left(\bar{u}(p') u(p)  \frac{1}{(p-p')^2 - \mmed^2} \bar{u}(k') u(k) \right)
\nonumber
\\ 
\nonumber
&~
\\ 
\nonumber 
&\kern-10em \xrightarrow{\rm non-rel}  \frac{i \gM \gDM}{{\bf q}^2 + \mmed^2} 2\mDM \delta^{ss'} 2\mN \delta^{rr'}
\ee
where $\delta^{ss'}$ and $\delta^{rr'}$ enforce spin conservation in the non-relativistic limit and ${\bf q = p-p'}$ is the spatial component of the four-momentum transfer. This can be compared with the Born approximation of the amplitude in non-relativistic quantum mechanics to obtain the Yukawa potential, Eq.~(\ref{eqn:yukawa_potential}), through the inverse Fourier transform of the momentum exchange ${\bf q}$  (the factors of $2\mDM$ and $2\mN$ must be dropped as they relate to the relativistic normalization of states). 

\begin{figure} [tbp]
	\centering 
	\newcommand{\pbwidthfactor}{0.95}
	\includegraphics[width=\pbwidthfactor\columnwidth]{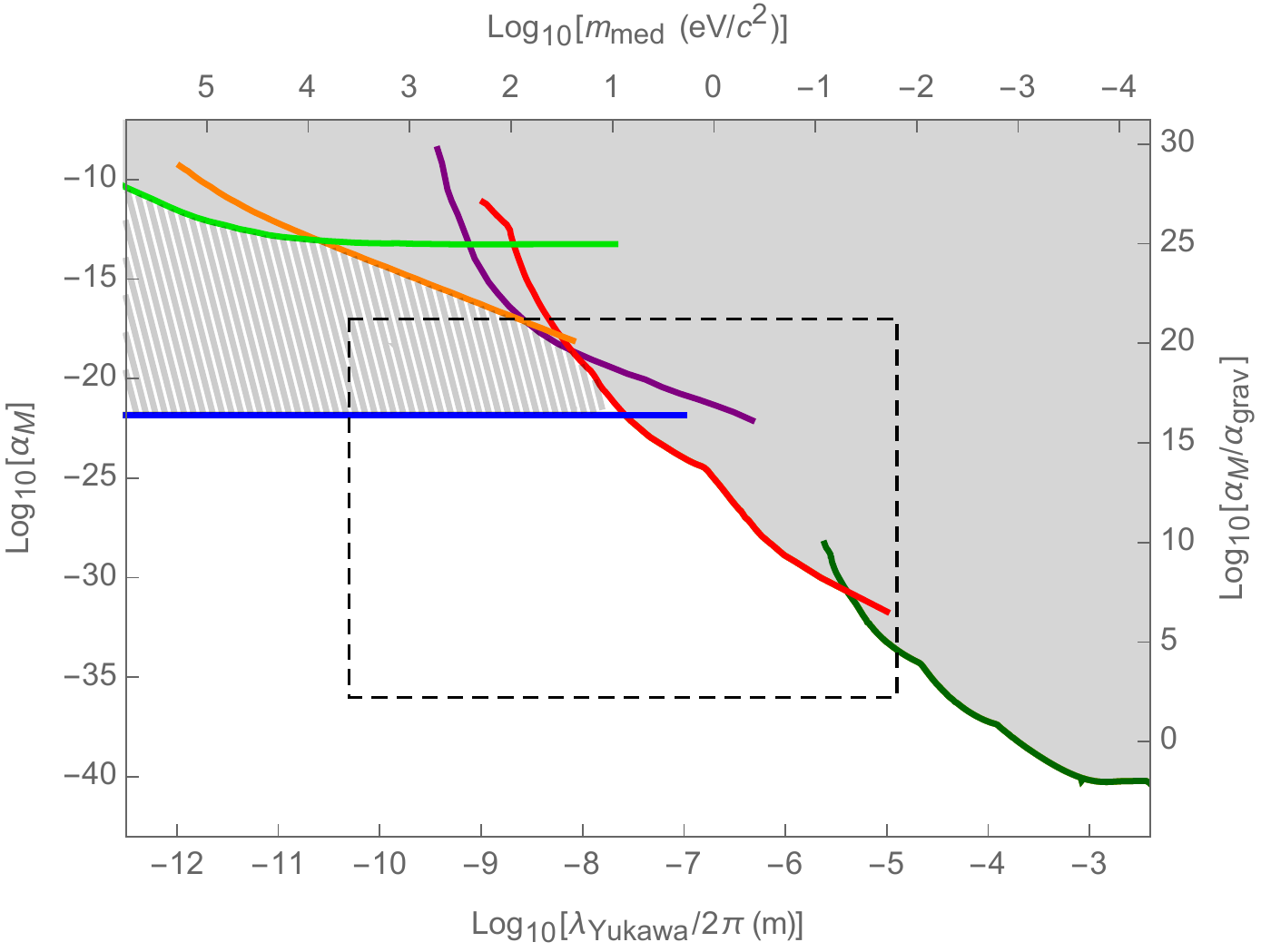}
	\caption{\textbf{Existing experimental constraints on possible new Yukawa forces.}  The region shaded solid gray has been excluded directly by neutron-lead scattering \cite{Leeb:1992xx} (light green), measurements of the neutron electric form factor \cite{Nesvizhevsky:2008xx} (orange), neutron optics experiments \cite{Leeb:1992xx} (purple), Casimir forces measurements \cite{Adelberger:2003xx} (red), and torsion balances \cite{Adelberger:2003xx} (dark green).  The region hashed gray is excluded by the strong but somewhat model-dependent constraint arising from the rate of stellar cooling \cite{Raffelt:1996xx,Raffelt:1999xx} (blue).  The dashed box encloses our region of primary interest, investigated in Figs.~\ref{fig:sensitivity-mdm-big} and \ref{fig:sensitivity-mdm-many}.}
	\label{fig:exclusions}
\end{figure}

This microscopic description of the interaction makes it clear that the same scalar exchange would also result in a potential between matter and itself,
\vspace{1.0cm}
\be
\parbox[t]{2.cm}{\vspace{-1.6cm}
	\includegraphics[scale=0.6]{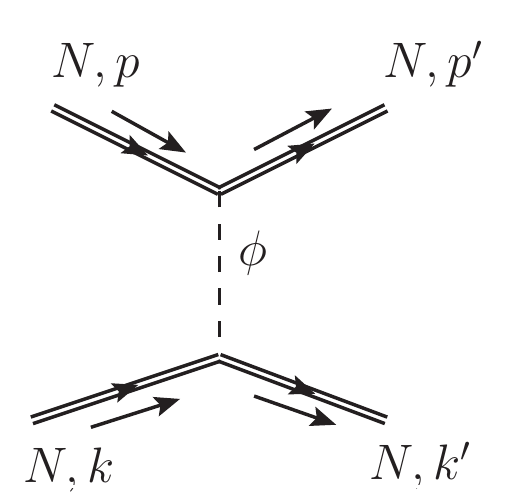} }  \quad\quad \Longrightarrow \quad\quad V_{\rm M}(r) = -\frac{\gM^2}{4\pi} \frac{1}{r}e^{-\mmed r}
\ee
and similarly there would be DM self-interactions with $\gDM$ instead of $\gM$. Transcribing the physics into quantum field theory has immediately revealed a very strong constraint on the model: new long-range forces between nucleons are extremely constrained by a variety of experiments and require the coupling $\gM$ to be extremely small, $\gM \ll 1$.  In fact, for the range of mediator masses we will consider below, $10^{4} - 10^{-2}\eV$ (corresponding to a force range of $10^{-9}-10^{-3}$ cm), the constraints require $\gM \lesssim 10^{-4} - 10^{-15}$, respectively.\footnote{These constraints are strictly stronger than limits inferred from upper bounds on the rate at which DM collisions heat interstellar hydrogen \cite{Chivukula:1990xx} for all DM masses we will consider.} See\footnote{Hardy and Lasenby have recently argued that the limits from stellar cooling can be strengthened by $\sim 10^3$ once mixing of the new scalar mediator with plasma oscillations is properly taken into account \cite{hardy2016stellar}.} Fig.~\ref{fig:exclusions}.

The self-interactions of DM through the scalar exchange are also constrained from a variety of observations such as the Bullet Cluster and the shape of galaxy clusters. The constraints, reported as the cross section per DM mass, are fairly stringent and require $\sigma/m \lesssim 0.1\cm^2/{\rm g}$~\cite{Rocha:2012jg}. This means that either the coupling to DM must also be very small, $\gDM \ll 1$, or that this is a subcomponent of DM. In fact, for the parameters we will be interested in, with DM mass in the MeV range and mediator mass in the eV range, the coupling $\alphaDM$ would have to be so small that the cross section against normal matter, Eq.~(\ref{eqn:yukawa_xs}), would be far below current sensitivity. 

Thus, the current proposal is really about a search for a subcomponent of DM with long-range interactions. As discussed in the introduction, given the complexity of the baryonic sector there is no reason to limit ourselves only to searches for the dominant component of DM. Indeed, in a recent publication the authors of Refs.~\cite{Fan:2013yva, kramer2016updated} advocated this scenario as a natural possibility for a subcomponent and showed that it can even form a dark galactic disk. The amount of DM near the Earth in this case has even larger uncertainties than usual (the dark galactic disk might even be misaligned with the baryonic disk), but in what follows we simply set $\rhoDM = 0.04\GeV/\mathrm{cm}^3$ and $\alphaDM = 1$ and our results can be easily rescaled to other values. 

One obvious question is why the coupling of $\phi$ to DM is so much stronger than to matter, $\gDM \gg \gM$. One possibility is that this is just the way it is for no deep reason, these are pure numbers undetermined by the model. After all, in the Standard Model the Higgs boson coupling to the top quark is some five orders of magnitude stronger than the Higgs coupling to electrons. Another possibility is to obtain the coupling of $\phi$ to the nucleons through mass mixing with the Higgs boson. In this case $\phi$ would inherit all the couplings of the Higgs, including its coupling to nucleons ($\sim 10^{-3}$) times the mixing angle between $\phi$ and the Higgs boson. However, for the masses and couplings we consider this scenario would require severe fine-tuning of the model parameters, no better than simply choosing $\gDM \gg \gM$.

Choosing the mediator to be a vector boson, rather than the scalar $\phi$, does not improve things in terms of model building. One possibility is to gauge B-L of the Standard Model, and endow the associated gauge boson with a mass through the spontaneous breaking of B-L. However, it still leaves open the question why $\gDM \gg \gM$. Both couplings are proportional to the B-L gauge coupling times the charge. This requires an extremely large charge ratio between DM and normal matter, which is not easy to obtain without additional assumptions on the model (e.g.\ having the DM  be some sort of composite object made of many charged particles, hence carrying a very large charge, stabilized by yet another force). 

The most natural way to obtain a small coupling to nucleons is based on the ideas of Shifman, Vainshtein, and Zakharov~\cite{Shifman:1978zn}. Suppose there exist additional heavy quarks with a coupling to the scalar,
\be
\mathcal{L}_h =  \bar{h}\left(i\gamma^\mu \partial_\mu - M_h \right) h + g_h \phi \bar{h} h
\ee
The scalar also couples to DM as in Eq.~(\ref{eqn:couplngs}), but at this point has no couplings to nucleons. The heavy quarks are very heavy, $M_h \gg \mN$. At energies below the mass of the heavy quarks, but above the scale of QCD, the heavy quarks can be integrated out and instead one obtains coupling to the gluon field strength~\cite{Shifman:1978zn},
\be
\mathcal{L} = \mathcal{L}_{\rm SM} + \mathcal{L}_{\phi} - \frac{2}{3} \frac{\alpha_s}{8\pi M_h} \phi\, G_{\mu\nu}G^{\mu\nu}
\ee
where $\mathcal{L}_\phi$ is the Lagrangian describing $\phi$ and its interaction with DM, $\alpha_s$ is the strong coupling constant, and $G_{\mu\nu}$ is the field-strength of the gluon. The coupling with the gloun field-strength generates the required coupling with nucleons at scales below the confinement scale of QCD. The coupling of $\phi$ to nucleons, $\gM$, is obtained  by the overlap of the gluon field strength with the nucleon wavefunction, which according to Ref.~\cite{Shifman:1978zn} in the chiral limit is given by
\be
\langle N | G_{\mu\nu} G^{\mu\nu} | N \rangle =  -\frac{8\pi}{9\alpha_s} \mN \bar{N} N
\ee
Thus, at low energy we obtain the Lagrangian in Eq.~(\ref{eqn:couplngs}) with
\be
\gM = \frac{2g_h}{27} \frac{\mN}{M_h}
\ee
The small ratio of nucleon mass to heavy quark mass can thus provide a natural explanation for the smallness of $\gM$ without requiring $g_h$ to be particularly small. 

\begin{figure} [tbp]
	\centering 
	\newcommand{\pbwidthfactor}{0.99}
	\includegraphics[width=\pbwidthfactor\columnwidth]{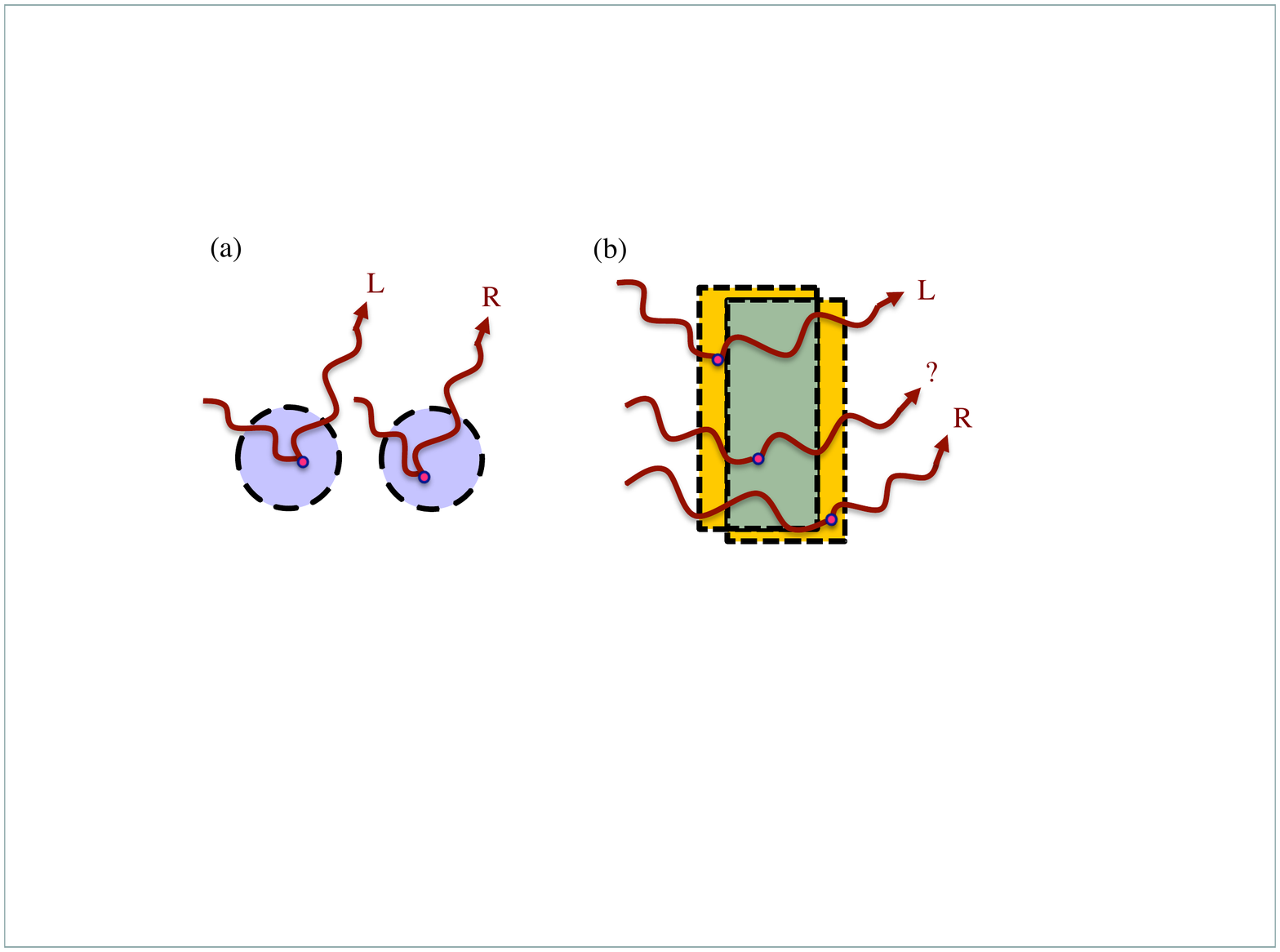}
	\caption{\textbf{Interferometers vs. resonators.}  Matter interferometers, (a), superpose objects over a distance comparable to or larger than their physical size (nanometers, generally).  In contrast, quantum nanomechanical resonators, (b), are typically much larger (microns) but are superposed over much shorter distances (femtometers). Scattered environmental particles can carry off which-path ($L$ or $R$) information, decohering the superposition.  However, a particle scattered from a region jointly shared by both parts of the superposition intuitively does not acquire which-path information, and hence does not contribute to the decoherence rate.  This suppression of decoherence can be confirmed analytically, and applies even for momentum transfers too small to be localized within the superposed object.}
	\label{fig:resonator}
\end{figure}

\newcommand{\KDTLcomp}{\mathrm{C}_{284}\mathrm{H}_{190}\mathrm{F}_{320}\mathrm{N}_{4}\mathrm{S}_{12}}
\newcommand{\KDTLname}{KDTL \cite{Gerlich:2011xx,Eibenberger:2013xx}}
\newcommand{\OTIMAname}{OTIMA \cite{Nimmrichter:2011x1,Nimmrichter:2011x2,Arndt:2014xx}}
\newcommand{\MAQROname}{MAQRO \cite{Kaltenbaek:2012x1,Kaltenbaek:2015x1,Kaltenbaek:2015x2}}
\newcommand{\vspacer}{\vspace{20px}}

\newcommand{\jX}{\phantom{.0}}
\newcommand{\jXX}{\phantom{.00}}
\newcommand{\jY}{\phantom{0}}
\newcommand{\jYY}{\phantom{00}}
\newcommand{\jYYY}{\phantom{000}}
\newcommand{\jZ}{\phantom{7.5\times}}
\newcommand{\jW}{\phantom{100\,}}
\newcommand{\jWW}{\phantom{10\,}}
\begin{table*}[tbh] 
	\def\arraystretch{1.3}
	\begin{tabular}{|c|c|c|c|c|c|c|c|}
		\hline
		{experiment} & \begin{tabular}{c}target\\composition\end{tabular} & \begin{tabular}{c}target\\radius (nm)\end{tabular} & \begin{tabular}{c}nucleon\\count\end{tabular} & \begin{tabular}{c}superposition\\separation (nm)\end{tabular} 
		& \begin{tabular}{c}exposure\\time (ms)\end{tabular} & \begin{tabular}{c}measurement\\rate (Hz)\end{tabular} & {status} \\
		\hline
		\KDTLname   						& $\KDTLcomp$ 					& $\sim\! 1$ 		& $1.0 \times 10^{4\jY}$ 	& $266\jX$ 	& $\jW \jYY 1.24$		& $10\,000\jXX$		& completed 	\\
		\OTIMAname 							& Gold (Au) 					& $\jYYY 5\jX$ 		& $6\jX \times 10^{6\jY}$ 	& $\jY 78.5$& $\jW \jY 94\jXX$ 		& $\jWW 600\jXX$	& prototype\\
		Bateman \emph{et al.}\ \cite{Bateman:2014xx}& Silicon (Si) 					& $\jYYY 5.5$ 		& $1.1 \times 10^{6\jY}$ 	& $150\jX$ 	& $\jW 140\jXX$ 		& $\jWW \jYY 0.5\jY$& proposed\\
		Geraci \emph{et al.}\ \cite{geraci2015sensing} 	& Silica (${\rm SiO}_2$) 				& $\jYYY 6.5$ 		& $1.6 \times 10^{6\jY}$ 	& $250\jX$ 	& $\jW 250\jXX$ 		& $\jWW \jYY 0.5\jY$& proposed\\
		Wan \emph{et al.}\ \cite{wan2016free} 		& Diamond (C) 					& $\jYY 95\jX$ 		& $7.5 \times 10^{9\jY}$ 	& $100\jX$ 	& $\jW \jYY 0.05$		& $\jWW \jYY 1\jXX$	& proposed\\
		\MAQROname   					    & Silica (${\rm SiO}_2$) 	& $\jY 120\jX$		& $1\jX \times 10^{10}$ 	& $100\jX$ 	& $100\,000\jXX$		& $\jWW \jYY 0.01$	& proposed\\
		Pino \emph{et al.}\ \cite{pino2016quantum}  & Niobium (Nb) 					& $1000\jX$			& $2.2 \times 10^{13}$ 	& $290\jX$ 	& $\jW 450\jXX$ 		& $\jWW \jYY 0.1\jY$& proposed\\
		\hline
	\end{tabular}
	\caption{Parameters describing some existing and proposed interferometers.  Their potential DM sensitivity is illustrated in Figs.~\ref{fig:sensitivity-mdm-big} and \ref{fig:sensitivity-mdm-many} (except for KDTL and Wan et al.).  The KDTL and OTIMA interferometers can be used to superpose multiple types of objects; we specialize to the choices above.  For the purposes of estimating decoherence rates, we model the superposed object as a sphere with the given radius, superposed over the given length and timescales.  The uniform sphere approximation is poor for the organic molecules (which are only about a dozen atoms across) superposed in KDTL, but this experiment is shown only for comparison and is not plausibly sensitive to DM. The measurement rate is used to estimate the increase in sensitivity from integrating over one month for all experiments.  KDTL and OTIMA pass a large continuous flux of particles through a sequence of gratings, yielding a high rate of individual measurements, while the other three experiments superpose one particle at a time. For all experiments, we assume a residual background of sidereal fluctuations in the decoherence rate of one part in a thousand (see Table~\ref{tab:common-parameters}), which limits the sensitivity of OTIMA. For the terrestrial experiments interfering a single nanoparticle at a time  (Bateman et al., Geraci et al., Wan et al., Pino et al.), the rate at which the entire process can be repeated had not been analyzed in detail in the relevant proposals; the measurement rates we assume above are rough conservative estimates based on the length of the entire measurement process, and private correspondence with the authors.}
	\label{tab:exp-parameters}
\end{table*}

\section{Experimental considerations}
\label{sec:experiment}

We now examine the possibility of detecting DM through the decoherence it induces in future experiments featuring massive quantum superpositions. In the absence of an independently motivated and specific DM scenario, we focus on what can be learned with minimal modification of these devices, and especially with passive data analysis.  Indeed, the unusual sensitivity of quantum superpositions to very weak Brownian jostling from any source \cite{Riedel:2015xx} makes them good systems for general exploration of many new particles and forces.\footnote{Matter interferometers may also use unitary (non-decoherent) dynamics to strengthen constraints on new Yukawa forces with ranges over a micron \cite{geraci2015sensing}.}  Relying on principles common to all large superpositions is more likely to uncover unexpected new physics than techniques that depend on the details of any particular DM model. 

As we discuss below, decoherence that varies with the sidereal day is a compelling preliminary sign of an interstellar source, and this behavior is generic for new particles interacting strongly enough with normal matter to influence terrestrial superposition experiments.  Once a candidate signal has been identified, more extensive experimental interventions would be justified in order to rule out alternative explanations.

\subsection{Matter interferometers}

We concentrate on \emph{matter interferometers}: devices that produce and confirm spatial superpositions of material objects over distances that are comparable to the object size itself.\footnote{For instance, the gold nanospheres superposed by the OTIMA experiment will be roughly 10 nm in diameter but will be superposed over distances exceeding 70 nm \cite{Nimmrichter:2011x1}.}  Some, like the KDTL (Kapitza-Dirac-Talbot-Lau) \cite{Eibenberger:2013xx} and OTIMA (macroscopic quantum resonator) \cite{Nimmrichter:2011x1,Nimmrichter:2011x2,Haslinger:2013xx,Arndt:2014xx} experiments, pass a flux of particles through several sets of gratings -- a generalization of the original two-slit experiment by Young -- and interference between different paths is observed.  Others cool and superpose individual nanoparticles in optical or magnetic traps, like the proposed MAQRO satellite \cite{Kaltenbaek:2012x1, Kaltenbaek:2015x1, Kaltenbaek:2015x2} and the terrestrial proposals by Bateman \emph{et al.}\ \cite{Bateman:2014xx}, Geraci et al. \cite{geraci2015sensing}, Pino \emph{et al.}\ \cite{pino2016quantum,romero-isart2016coherent}, and Wan \emph{et al.}\ \cite{wan2016free}.  For our main results we consider the sensitivity of these seven benchmark experiments, with parameters given in Table~\ref{tab:exp-parameters}.  In this subsection, we briefly discuss two alternative classes of devices for comparison: cold test masses and quantum (nanomechanical) resonators.

For $\mDM \lesssim 100 \MeV$, and especially when predominantly scattering forward, DM collisions act as an extremely weak Brownian bath, i.e., a source of momentum diffusion for a test mass.  If a given test mass is prepared in a thermal state of some trapping potential, the minimal detectable rate of diffusion is inversely related to its temperature, achieving a finite sensitivity at zero temperature in accordance with the uncertainty principle; this defines a standard quantum limit (SQL) \cite{Riedel:2015xx}.  The same test mass, if placed in a spatial superposition, has sensitivity that surpasses the SQL, improving quadratically with the spatial coherence length.  In this restricted sense, matter interferometers are an extension of cold classical test masses to negative effective temperatures. 

Of course, the enhancement from quantum coherence must be traded off against the limitations it imposes, so it is worth comparing interferometers to state-of-the-art cold test masses. One can estimate that the cooling of a 70-nm radius nanospheres in optical traps to $50\,\mathrm{mK}$ described in Ref~\cite{Gieseler:2012xx} has a diffusion sensitivity one to two orders of magnitude less than the (vastly less massive) superposed organic molecules of the existing KDTL interferometer.  (See also related experiments \cite{chang2010cavity,Li:2011xx,kiesel2013cavity,fonseca2016nonlinear,millen2015cavity,vovrosh2016controlling}.)   When these systems can be cooled to the ground state, they will approach the sensitivity of the interferometric proposals of Bateman \emph{et al.}\ and Geraci \emph{et al.}\ that we analyze in Figs.~\ref{fig:sensitivity-mdm-big} and \ref{fig:sensitivity-mdm-many}, but remain less sensitive than the (much less massive) superposed nanospheres in the OTIMA interferometer, and vastly less sensitive than the (comparably sized) superposed nanospheres of the MAQRO satellite and the Pino \emph{et al.}\ proposal.  This illustrates the sense in which quantum spatial coherence is a resource, like low temperatures, for detecting soft momentum transfers.

\emph{Quantum (nanomechanical) resonators} \cite{oconnell2010quantum,chan2011laser,teufel2011sideband,palomaki2013entangling,clark2017sideband} are mechanical devices placed in coherent superpositions of different vibrational modes.  They differ conceptually from interferometers in that the spatial extent of the superposition (i.e, the coherence length) of a quantum resonator is (much) smaller than the superposed object itself.  Although usually much more massive, quantum resonators typically achieve spatial coherence only on the scale of femtometers or less, which is far from the nanometers produced by the interferometers we consider. (See Fig.~\ref{fig:resonator}.)  As described below, the sensitivity penalty is generally quadratic in the coherence length.  The increased mass of a resonator is not sufficient to compensate, especially since the coherent scattering enhancement (discussed below) saturates for objects larger than the length scale of the typical momentum transfer.  However, in the future the powerful optomechanical techniques \cite{aspelmeyer2014cavity-a,aspelmeyer2014cavity-b} used in quantum resonators could conceivably produce coherence over much larger distances using suspension \cite{corbitt2007all-optical, abbott2009observation,cohadon2014suspended, weaver2016nested, mueller2015enhanced, page2016thermal} and especially levitation \cite{singh2010all-optical, guccione2013scattering-free} rather than mechanical clamping.

For these reasons we concentrate on matter interferometers as the most promising devices for probing very soft DM collisions in the medium-term future.

\subsection{Dark matter flux}

We assume a uniform mass density $\rhoDM\approx 0.04\GeV$ in the Solar System to represent a 10\% interacting DM component \cite{catena2010novel}.\footnote{Lower bounds on the mass of a fermionic DM particle \cite{viel:2013xx} are significantly relaxed when it constitutes only a portion of the total DM mass distribution.}  We set $\alphaDM \approx 1$ and allow $\mDM$ and $\mmed$ to vary in the ranges $1\keV \lesssim \mDM \lesssim  10\MeV$, $10\meV \lesssim \mmed \lesssim 10\keV$. These mediator masses $\mmed$ correspond to a force with range $20~\mathrm{pm} \lesssim \lambdamed \lesssim 20\um$. We take the velocity distribution to be thermal (Maxwellian) in the galactic rest frame, concentrated around $\vDM\approx 230\kms$, except for a cutoff at the galactic escape velocity $\vEsc \sim 550 \kms$ \cite{smith2007rave}. The Solar System moves at a velocity $\vE \approx \vDM$ in the Milky Way, roughly toward the star Vega, so that the velocity distribution in the Solar System looks like\footnote{We ignore the $\sim\! 10$\% oscillation in the apparent DM wind over a year due to the Earth's motion with respect to the sun, $\sim~\!\!30 \kms$.  This effect is discussed in Sec.~\ref{sec:evidence}.} 
\be
p({\bf v}) \propto e^{-({\bf v}+{\bf \vE})^2/\vDM^2}.
\ee
In terrestrial experiments, the DM flux is also modified by the Earth and its atmosphere.  

\begin{figure} [tb]
	\centering 
	\newcommand{\pbwidthfactor}{0.95}
	\includegraphics[width=\pbwidthfactor\columnwidth]{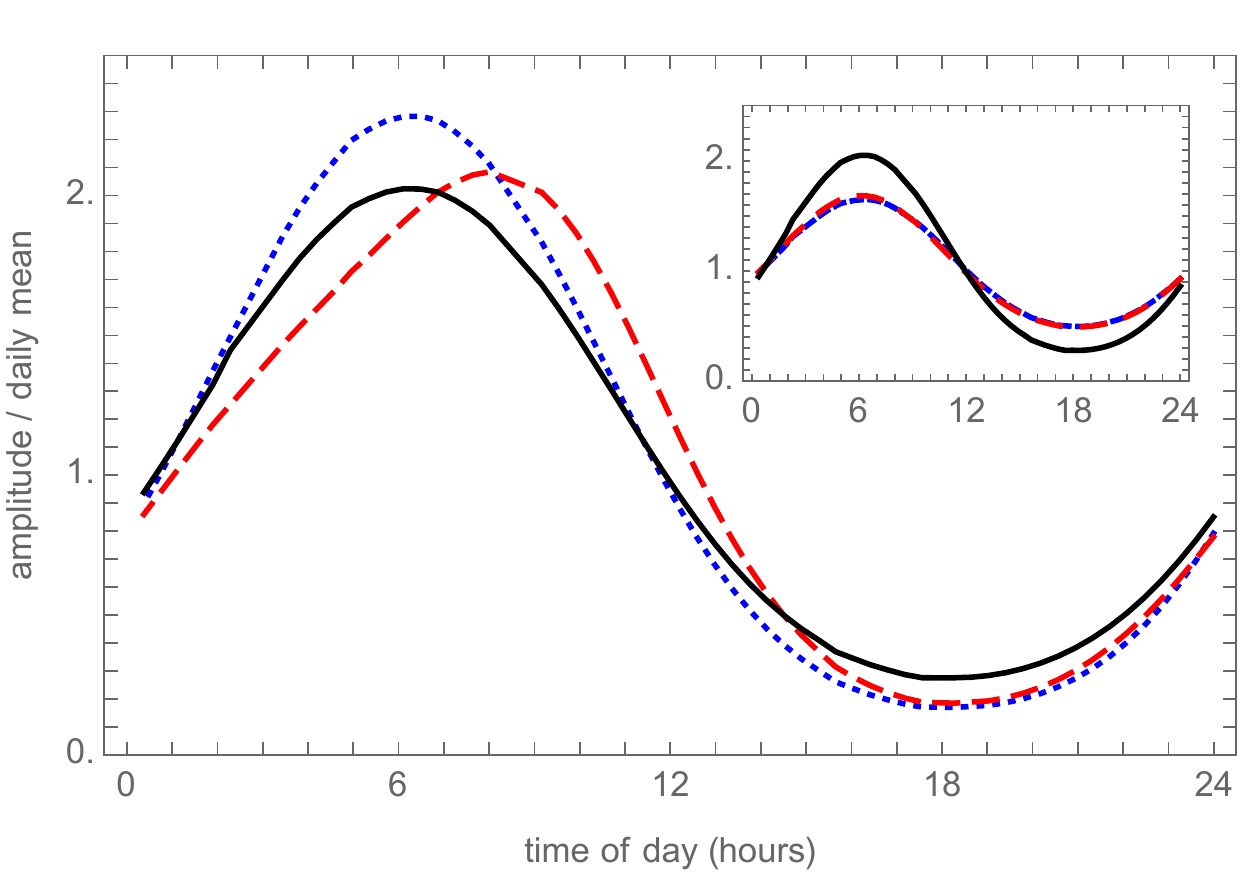}
	\caption{\textbf{Sidereal daily variation.}  Consider the OTIMA experiment at lattitude $48^\circ$N (Vienna, Austria) with the spatial coherence vector ${\bf \DeltaX}$ (i.e., the direction connecting different slits) pointed horizontal and oriented at $70^\circ$ with respect to geographic north.  Let the DM be characterized by $\mDM = 1\keV$ and $\mmed = 200\eV$ (inset: $\mDM = 1\MeV$ and $\mmed = 20 \eV$).  In accordance with the Solar Systems's velocity through the Milky Way, the apparent wind blows at $\vE\approx\vDM$ from a $+38^\circ$ declination.  The total DM flux (black, solid) varies over the course of the day due to shielding by the Earth.  The flux is largest when the DM wind blows down from overhead, and smallest 12 hours later when the experiment is inside the partial shadow cast by the Earth.  The variation in the DM-induced decoherence rate (colored lines) tracks the variation in the overall flux, both when the DM isotropizes (blue, dotted) or does not isotropize (red, dashed) in the atmosphere before reaching ground level.  Variations in the typical DM momentum with respect to ${\bf \DeltaX}$ give only modest corrections.  There is a small relative phase shift between the DM flux and the un-isotropized decoherence rate in the main plot because the dominant momentum transfer is comparable to $\DeltaX$, so the rate of decoherence is sensitive to the orientation of the DM distribution even with the energy spectrum fixed.  This is not true for the heavier DM shown in the inset, which consequently does not exhibit the asymmetry. These order-unity daily variations of the decoherence rate are a generic feature of all DM scenarios coupled strongly enough to be shielded by the Earth.  The absolute phase with respect to the solar day drifts together smoothly over the course of the year because the DM has interstellar origins and tracks the sidereal day, allowing it to be distinguished from terrestrial sources of decoherence.}
	\label{fig:daily}
\end{figure}

DM may pass completely through the Earth for sufficiently small $\alphaM$, but such weak interactions are outside the sensitivity range of most of the terrestrial\footnote{The MAQRO satellite and the Pino \emph{et al.}\ proposal might be sensitive to such weakly coupled scenarios, but shielding by the Earth is irrelevant for the MAQRO satellite because it would occupy a Lissajous orbit around one of the Sun-Earth Lagrange points L1 and L2 \cite{Kaltenbaek:2015x2}.  In the case of Pino et al., we assume the Earth effectively blocks DM originating from below the horizon, but this likely depends on details of the DM behavior in bulk material, which we do not explore.} experiments we consider.  Therefore, the Earth acts as an effective DM windscreen.  Since the apparent direction of the DM wind in the lab frame changes with the Earth's rotation throughout the day, this generically produces a large daily modulation in the total DM flux. See Fig.~\ref{fig:daily}.

We do not know how DM will behave passing through solid bulk matter, so for our main results we conservatively ignore all DM entering the Earth (from above or below) by setting the flux originating below the horizon to zero.   The possibility that DM is reflected by the Earth's surface, or is absorbed and reemitted as a thermalized gas, is considered in Appendix~\ref{sec:greenhouse}; the substantial additional sensitivity in such scenarios is illustrated in Figs.~\ref{fig:sensitivity-mdm-big} and \ref{fig:sensitivity-mdm-many}.

For much of the parameter space we consider, the DM passes easily through the atmosphere.  Even when it does interact, the forward and elastic nature of Yukawa scattering means it takes many scattering events to isotropize the direction of the DM, and even more to thermalize its energy.  The effect of the atmosphere on experimental sensitivities is detailed in Appendix~\ref{sec:shielding}. The effect of solid material less than a meter thick should be comparatively small.\footnote{The overhead mass of the atmosphere is roughly equivalent to several meters of concrete or one meter of lead, and different nuclei should not combine coherently in amorphous materials -- even when they are within a single DM de Broglie wavelength -- except for scattering in the forward (non-attenuating) direction \cite{Riedel:2012ur}.} Shielding from experimental equipment and building material is therefore ignored, although this would need to be reconsidered for experiments performed underground or near the bottom of large buildings.

\subsection{Decoherence}
\label{sec:decoh}

The essential quantum effects in an interferometer relevant to DM can be modeled by taking the superposed object to evolve in a two-dimensional subspace spanned by two macroscopically distinct states.  These can be thought of as different position eigenstates, as different slits in a grating, or as the left and right arms of a Mach-Zehnder interferometer.  If the object is initially prepared in a balanced superposition of the states, $\vert \psi_0 \rangle \propto \vert L \rangle + \vert R \rangle$, and if transitions between the two states $\vert L \rangle$ and $\vert R \rangle$ are forbidden, then the most general possible open-system dynamics produce a final state
\begin{align}
\rho = \frac{1}{2}\begin{pmatrix} 1 & \gamma \\ \gamma^* & 1 \end{pmatrix},
\end{align}
where $\gamma = \exp(-s+i \phi)$ is the \emph{decoherence factor} associated with the evolution, decomposed into the dimensionless decoherence $s$ and the phase shift $\phi$. For Markovian dynamics, $s$ increases monotonically in time and $\tau^{-1} \equiv ds/dt$ is called the \emph{decoherence rate}.\footnote{Contributions of different independent parts of an environment, such as different species of scattering particles, combine additively in the decoherence rate: $\tau^{-1} = \sum_i \tau_i^{-1}$.}

The experiment concludes by effectively making a measurement in the basis $\{ \vert \pm \rangle \propto \vert L \rangle \pm \vert R \rangle\}$, obtaining outcome $\vert + \rangle$ with probability $P_+ = (1+\mathrm{Re}\, \gamma )/2$.  By inserting an adjustable phase offset $\phi_0$ (e.g., by aligning beam splitters or shifting a grating), both the real and imaginary parts of $\gamma$ can be inferred over many trials.  The presence of a quantum superposition is demonstrated when $\gamma$ differs significantly from zero, i.e., when the decoherence is low: $s \lesssim 1$.

Since the DM interaction with other potential sources of decoherence is negligible, the decoherence factor decomposes as $\gamma = \gammaDM \gamma_{\mathrm{other}}$ where \cite{Joos:1985xx, Gallis:1990xx, Schlosshauer:2008xx, Riedel:2012ur}
\begin{gather}
\label{main-decoh-factor}
\gammaDM = e^{-\sDM+i\phiDM} = \exp [ - \int_0^T \!\! d t \,  F({\bf \DeltaX})]
\end{gather}
and
\begin{gather}
\label{decoh-rate}
F({\bf \DeltaX}) = \int \! d {\bf k} \, n({\bf k}) \frac{k}{\mDM} \int \! d  \Omega   \left[1 - \exp(i {\bf q} \cdot {\bf \DeltaX}/\hbar)\right] \frac{d\sigma_{\mathrm{T}}}{d\Omega} .
\end{gather}
Here, $n({\bf k})$ is the phase-space density of the DM distribution,  ${\bf k}$ is the incident DM momentum, ${\bf q}$ is the momentum transfer, and $d\sigma_{\mathrm{T}}/d\Omega$ is the (spin-independent) elastic scattering cross section for the entire superposed target.\footnote{We ignore the inelastic formation of DM-nucleus bound states.  Although bound states do exist when $(\mDM /\mmed)N\sqrt{\alphaDM\alphaM} \gtrsim 1$ \cite{bennett1981upper,petraki2016radiative}, the cross section for formation of such states through radiative capture of DM is suppressed by additional factors of $\vDM^{-1}N\sqrt{\alphaDM\alphaM}$ compared to elastic scattering \cite{petraki2016radiative}.}

The cross section of a target of $N$ nucleons, rather than being simply related to the single-nucleon cross section \eqref{eqn:yukawa_xs} by a multiplicative factor $N$, can receive a coherent scattering enhancement \cite{Taylor:2006xx} in the Born approximation when the nuclei are located in a region smaller than the reduced de Broglie wavelength $\lambdabar_q = \lambda_q/2\pi = 1/q$ associated with the momentum transfer ${\bf q}$. In this case, the cross section scales like $N^2$, so superposed objects become dramatically more sensitive to decoherence as they increase in size.\footnote{In the better known case of neutron scattering \cite{Squires:1978xx}, the exploitation of the coherence enhancement is limited by the detector's ability to resolve small angles in the forward direction; large enhancements are only seen for volumes $q^{-3}$ large enough to contain the $N$ neucleons, corresponding to the smaller scattering angles $\theta \sim q/k$. For the detection of decoherence, an analogous role is played by the the distance $\DeltaX$ over which quantum coherence can be maintained.  Superpositions with larger $\DeltaX$ are decohered by smaller momentum transfers $q$, corresponding to the detection of very small scattering angles.}  In fact, even for short range $\lambdamed$ and surprisingly small coupling $\alphaM$, the effective DM-scattering cross section of the target can approach its geometrical size.  This explains how  decoherence from a weakly coupled 10\% interacting subcomponent of (say) keV DM, which has a number of density of a few tens of thousand per cubic centimeter, can compete with decoherence from collisions with the residual molecules in a laboratory vacuum that is roughly two orders of magnitude denser and only three orders of magnitude slower.  The effect of coherent elastic scattering on the behavior of the decoherence rate $F({\bf \DeltaX})$ is discussed further in Appendix~\ref{sec:decoh-calc}.

\begin{table}[]
	\def\arraystretch{1.2}
	\begin{tabular}{ | r  l | c |}
		\hline
		$\vE$     & $=230\kms$ & \begin{tabular}{c}
			Earth velocity in\\ galactic rest frame \end{tabular}\\
		\hline
		$\vDM$     & $=230\kms$ & \begin{tabular}{c}
			Thermal DM velocity\\ in galactic rest frame \end{tabular} \\
		\hline
		$\rhoDM$  & $=0.04 \GeV/\mathrm{cm}^3$ \,& \begin{tabular}{c}
			Density of interacting\\ DM subcomponent \end{tabular}\\
		\hline
		\,\,$\alphaDM$  & $=1$ & \vphantom{\begin{tabular}{c} X\\X \end{tabular}}DM coupling\\
		\hline
		$\etaDM$ &$= 50\%$  & \begin{tabular}{c}Sidereal variation in \\DM decoherence rate$^*$\end{tabular}\\
		\hline
		$\etairr$ &$= 10^{-3}$  & \begin{tabular}{c}Residual sidereal\\decoherence background$^*$\end{tabular} \\
		\hline
		$\Trun$  & $=1\,\mathrm{month}$ & \vphantom{\begin{tabular}{c} X\\X \end{tabular}}Data taking timespan\\
		\hline
		$\vis$ &$= 50\%$  & \begin{tabular}{c}Interference fringe\\  visibility achieved\end{tabular} \\
		\hline
	\end{tabular}
	\caption{Parameters taken to be common to all terrestrial experiments.  These also apply to the satellite proposal MAQRO, with the exception of those marked with *. Being far from Earth, MAQRO will not observe a sidereal variation due to shielding of the DM wind.}
	\label{tab:common-parameters}
\end{table}
	
\subsection{Signal}

The signal induced by DM in matter interferometers is \emph{anomalous decoherence}, i.e., decoherence in excess of what is expected from conventional sources like stray gas molecules or background blackbody radiation.  Of course it is difficult to fully characterize or completely eliminate conventional sources of decoherence, and the mere observation of unexplained decoherence, like any unexplained noise, is certainly not convincing evidence for new particles or forces.  However, a small excess of DM-induced decoherence can be distinguished from larger backgrounds through several methods.  Most notably, the terrestrial DM flux is expected to vary with the Earth's rotation over the \emph{sidereal} day (23 hours, 56 minutes, 4 seconds), leading to large oscillations in the DM-induced decoherence rate.  The variation is due to shielding of the DM wind by the Earth, and is generically true whenever DM interactions are strong enough to be relevant to a terrestrial interferometer. 
This characteristic period can distinguish interstellar sources of decoherence from mundane terrestrial ones, such as vibrations or temperature-dependent blackbody radiation, even when those sources vary with the  \emph{solar} day (24 hours).  Additional background-rejection methods are discussed in the next section.

As shown in Fig.~\ref{fig:daily}, the magnitude of the daily variation of the DM-induced decoherence rate is comparable to the mean rate: $\dF({\bf \DeltaX}) / \bar{F}({\bf \DeltaX}) \equiv \etaDM \sim 50\%$.   We take the daily variation of the DM decoherence seen by each shot in the experiment to be
\begin{align}
\label{eq:sidereal-decoh}
\Dsdm = \etaDM \sDM = \etaDM \mathrm{Re}\int_0^T\!\! dt\, F({\bf \DeltaX}).
\end{align}
A statistical estimator for $\Dsdm$ can be computed from observed frequencies in a simple bin-counting experiment.  We consider the event rates for two different outcomes ($+$ vs.\ $-$, i.e., ``peak'' vs. ``trough'') during two different halves of the day (``morning'' vs.\ ``evening'').  The number of events observed in each of these four bins is an independent Poissonian variable, and in Appendix~\ref{sec:statistics} we derive the asymptotic estimator error 
\begin{align}
\label{eq:decoh-error-main}
\sigmaDs = 2 \sqrt{(\vis^{-2}-1)/\barNum}
\end{align}
for visibility $\vis \approx \Re (\gamma_{\mathrm{other}}) $ and total expected number of events $\barNum$ (summed over all four bins). This assumes only that the DM decoherence is small compared to other sources of decoherence ($\Dsdm \ll 1-\vis \le 1$), and that the fractional daily variation in the overall count rate is small  ($\DeltaNum \ll \barNum$).  Thus, an interferometer's sensitivity to anomalous decoherence exhibits Poissonian scaling ($\propto \barNum^{-1/2}$), but fails if any conventional decoherence sources drive $\vis \to 0$.

Gathering additional data over multiple weeks to statistically increase sensitivity to small sidereal DM decoherence rates does not require an assumption that other, conventional sources of decoherence are absolutely stable on those long timescales.  Only daily variations of mundane sources (which are generally a small fraction of the overall decoherence rate) have the ability to mimic DM in the short term, and these can be rejected in the long term by looking for the expected phase drift of the sidereal signal over the year.  The sensitivity of this passive strategy is ultimately limited only by  \emph{seasonal} variation in the size of \emph{daily} fluctuations in the decoherence rate from terrestrial sources.  Let us call this the \emph{sidereal decoherence background}.   If a distinctive sidereal signal were observed, active techniques could then be brought to bear to eliminate alternative explanations.

Interferometer experiments are concerned chiefly with the constant part of the decoherence background, and estimates of the expected amplitude of fluctuations in the decoherence background are not currently available.  For the purposes of estimating the limits of DM sensitivity in Figs.~\ref{fig:sensitivity-mdm-big} and \ref{fig:sensitivity-mdm-many}, we assume each experiment can control the sidereal fluctuations in decoherence from conventional sources\footnote{One can distinguish ``true'' decoherence, due to the entanglement of the target with the environment, from mere dephasing, due to an unknown classically noisy phase shift.  Both reduce fringe visibility, but the latter can in principle be undone by identifying the phase shift and subtracting it off.  For interferometers recording a continuous fringe pattern, this can be accomplished by looking at the second-order correlation function of particle impacts, and has been demonstrated experimentally \cite{rembold2014correction,gunther2015multifrequency,rembold2017secondorder}.  Conventional sources of dephasing with a sidereal period cannot be distinguished from a possible oscillatory phase shift due to the force of the DM wind, but they both can be subtracted off together to allow observation of DM decoherence.} to one part in a thousand, $\etairr \approx 10^{-3}$.  In practice, this could be achieved by either suppressing all daily noise below this threshold, or by taking data at different times of the year to distinguish the sidereal and solar days.  Our results can be scaled for alternative background rates as discussed in Sec.~\ref{sec:results}.

\subsection{Background rejection}
\label{sec:evidence}

Once a candidate sidereal signal has been found, there are several distinct strategies for gathering more evidence that the source is astrophysical, and for characterizing the cross section and flux.  Here we list the most notable methods.

\begin{figure*} [ptb]
	\centering 
	\newcommand{\pbwidthfactor}{0.95}
	\includegraphics[width=\pbwidthfactor\textwidth]{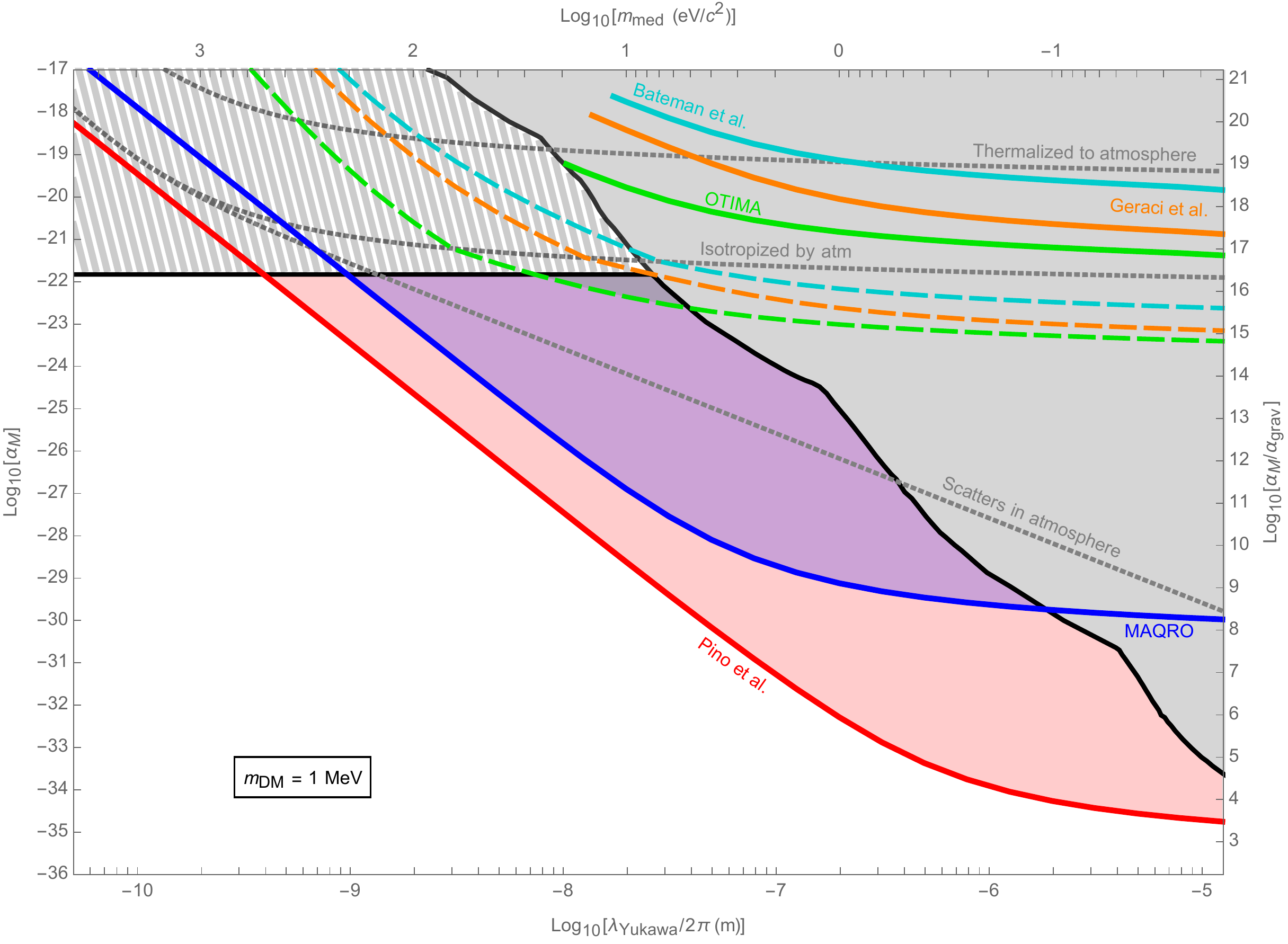}
	\caption{\textbf{Interferometer sensitivity to MeV dark matter.} Each point in the $\mmed,\alphaM$-plane corresponds to a DM scenario with fixed $\mDM = 1\MeV$, $\alphaDM=1$.  The characteristic length scale $\lambdamed = \mmed^{-1}$ of the Yukawa potential is given by the inverse mass of the associated mediator. The solid black line bounds the gray and hashed-gray regions that have already been excluded, as detailed in Fig.~\ref{fig:exclusions}.  Three gray dotted curves depict the progressively larger minimum couplings $\alphaM$ for which DM incident on the atmosphere scatters at least once, is isotropized, and thermalizes before it reaches the ground.  The first two coincide when the interaction is short range, $\lambdamed \ll \lambdaDM \sim 0.26\nm$.  Under the conservative assumption that all DM incident on the Earth's crust is absorbed, the solid colored lines denote the limit of sensitivity for five proposed interferometer experiments: OTIMA~\cite{Nimmrichter:2011x1,Nimmrichter:2011x2,Arndt:2014xx}, Bateman et al.~\cite{Bateman:2014xx}, Geraci et al.~\cite{geraci2015sensing}, the MAQRO satellite~\cite{Kaltenbaek:2012x1, Kaltenbaek:2015x1, Kaltenbaek:2015x2}, and Pino et al.~\cite{pino2016quantum}. (Because of atmospheric shielding, the experiments could only detect DM where the sensitivity line dips below the isotropization curve, as discussed in Appendix~\ref{sec:shielding}.)   Non-excluded regions for which DM would induce detectable decoherence in the experiments are shaded in the corresponding color, and account for atmospheric effects.  If the surface of the Earth isotropically reflects incident DM, then the sensitivity lines would shift slightly downward (more sensitive) by roughly a factor of $\log_{10}(2) \approx 0.3$ and would not need to pass below the isotropization curve.  The sensitivity lines end when the Born approximation breaks down; as described in Appendix \ref{sec:born}, this indicates that the cross section has become of the order of the geometric cross section, and the sensitivity would generally saturate for arbitrarily large $\alphaM$, although there can be a sensitivity enhancement for resonant scattering with attractive potentials. The dashed colored lines denote the increased sensitivity if there is a DM greenhouse effect, i.e., if rather than being absorbed or reflected, DM incident on the Earth's surface thermalizes to the temperature of the upper crust and is re-emitted (see Appendix \ref{sec:greenhouse}).  We do not include greenhouse lines for the MAQRO satellite or the Pino \emph{et al.}\ proposal; the former operates away from the Earth, and the latter would need to account for scattering lengths that could exceed the dimensions of the atmosphere or the Earth.}
	\label{fig:sensitivity-mdm-big}
\end{figure*}

\begin{figure*} [tbhp]
	\centering 
	\newcommand{\pbwidthfactor}{0.99}
	\includegraphics[width=\pbwidthfactor\textwidth]{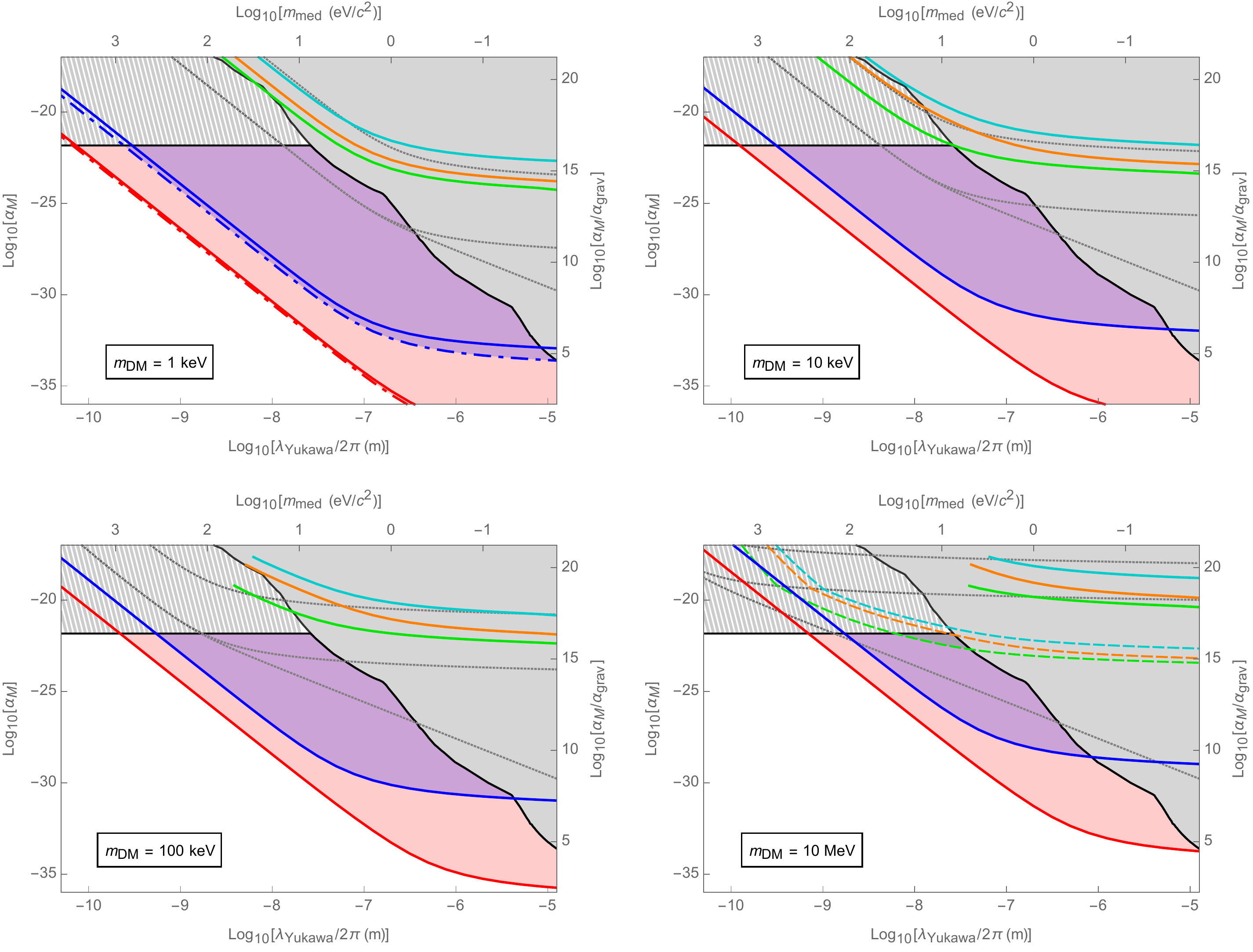}
	\caption{\textbf{Interferometer sensitivity for other DM masses.} The same quantities displayed in Fig.~\ref{fig:sensitivity-mdm-big} with $\mDM = 1\keV$, $10\keV$, $100\keV$, and $10\MeV$.  
		The dot-dashed lines for MAQRO and Pino \emph{et al.}\ when $\mDM=1\keV$ bound the regions with a detectable phase shift from the DM wind \cite{Riedel:2012ur} that exceeds the decoherence rate.  (See Appendix~\ref{sec:statistics}.)   Due to scattering in the atmosphere at this mass, the DM distribution seen by the other four experiments would always be isotropic, implying no net DM wind and hence no phase shift.  The greenhouse scenario (dashed colored lines) only increases the sensitivity when the Earth is cooler than the temperature of incident DM, translating to $\mDM \gtrsim 120 \keV$.  For smaller masses, the dominant source of decoherence is just the unthermalized component of the DM distribution, so greenhouse effects have little effect on sensitivity.}
	\label{fig:sensitivity-mdm-many}
\end{figure*}

First, if the length scale $1/q$ associated with the typical momentum transfer is comparable to or longer than the spatial coherence $\DeltaX = \abs{\bf \DeltaX}$, the DM decoherence is modulated by order unity when the direction ${\bf \DeltaX}$ is changed, i.e., by physically rotating the interferometer.\footnote{Ref.~\cite{Riedel:2012ur} has a plot of this effect.}  This makes the superposition a \emph{directional} target.  By varying the orientation independently of the time of day, one can roughly identify a celestial direction associated with extremal decoherence, giving an important additional constraint (up to an unknown polarity) on any candidate direction for the DM wind.  However, this mechanism does not work if $q \DeltaX \gg 1$ since then each scattering event will cause complete decoherence regardless of orientation, or if the local DM environment is isotropic due to prior scattering in the atmosphere (see Appendix~\ref{sec:shielding}).

Second, for DM couplings not too far below the ``isotropized by atm'' line in Figs.~\ref{fig:sensitivity-mdm-big} and~\ref{fig:sensitivity-mdm-many}, it is possible to directly modulate the DM flux by, e.g., adding and removing lead shielding, or by moving the detector underground. Observing this effect would be particularly striking evidence to rule out many mundane sources of anomalous decoherence.  This mechanism fails if the DM is too weakly interacting or if the DM is locally isotropic.

Third, the parameters describing the interferometer (e.g., equipment temperature, vacuum quality, target size, target speed, target composition, interferometric arm length) can be varied to see if the observed dependence of the decoherence rate is consistent with unobserved soft scattering events.  For instance, changing the isotopic composition of the target would not much change mundane (electromagnetic) sources of anomalous decoherence, but would significantly change decoherence from DM that probes the dark charge of the nucleus.  Likewise, slightly varying the temperature of the experimental equipment and measuring any change in decoherence rate can upper bound temperature-dependent noise in the experiment, which could rule out sidereal temperature fluctuations as an explanation of any anomalous decoherence that is observed.

Fourth, the time dependence of the DM distribution has significant structure.  Earth's orbital speed around the Sun, $\sim$\,$30\kms$,  modulates the average speed set by the Sun's path in the galactic rest frame, $\sim$\,$230 \kms$.  This introduces a slight annual modulation of the mean decoherence rate, which is analogous to the signal sought by several weakly interacting massive particle (WIMP) experiments \cite{cherwinka2011search,aalseth2011search,herrero-garcia2012annual}.  It also would adjust the amplitude of the daily variations (since the apparent DM wind changes), an effect which is not found with WIMP DM that cannot be shielded by the Earth. The phase of these annual and daily modulations is roughly determined by the direction the Solar System travels in the Milky Way (with adjustments due to gravitational focusing by the Sun \cite{lee2014effect}).  With sufficient sensitivity in the future it is even conceivable to observe monthly fluctuations due to the Moon \cite{britto2015monthly}.

Fifth, the magnitude of the expected sidereal variation can be changed by moving the experiment to different geographic locations.  Locations near the equator will see larger daily changes in the DM wind than locations near the poles.  The phase of the daily signal should change predictably with longitude.

Finally, we emphasize that the observation of quantum interference is good evidence that \emph{all} sources of decoherence are small ($s \lesssim 1$) on the relevant timescale.  Upper bounds on the anomalous decoherence rate immediately and robustly constrain DM, without requiring any of the active techniques described above.

\section{Results}
\label{sec:results}

Neither the KDTL experiment nor the Wan \emph{et al.}\ proposal would be sensitive to the DM we consider.  
The sensitivities of the other five benchmark experiments from Table.~\ref{tab:exp-parameters} are depicted in Figs.~\ref{fig:sensitivity-mdm-big} and~\ref{fig:sensitivity-mdm-many}.  The decoherence sensitivity curves are defined as the critical matter-DM coupling $\alphaMsense$ for which the magnitude of the sidereal variation $\Dsdm$ [Eq.~\eqref{eq:sidereal-decoh}] begins to exceeds the residual background $\Dsirr = \etairr \sbkg = \etairr \ln (\vis^{-1})$ combined with the estimator error $\sigmaDs$ [Eq.~\eqref{eq:decoh-error-main}]:
\begin{align}
\Dsdm(\alphaMsense) \equiv \sigmaDs + \Dsirr.
\end{align}
We take the daily variation of the DM decoherence rate\footnote{The MAQRO satellite is an exception for which we use the absolute DM decoherence rate $F$, not the variation $\dF$.  MAQRO would not operate in the vicinity of the Earth and therefore would not expect to see a sidereal variation in the flux, necessitating other methods determining whether anomalous deocherence were due to DM.  Unlike terrestrial experiments which often aim to just achieve interference ($\vis>0$), the MAQRO experiment will explicitly test modifications to quantum mechanics that induce objective wavefunction collapse by looking for very small changes to the interference pattern.  A preliminary analysis of the mundane sources of decoherence in MAQRO has been done \cite{Kaltenbaek:2012x1, Kaltenbaek:2015x1}, and the background in the pristine satellite environment is small enough to not only achieve interference but to increase sensitivity beyond this through integration over at least 20 months of data taking.  Therefore, observing anomalous decoherence in the first month would be very notable on its own.  In the absence of a strong sidereal variation from Earth shielding, other effects like annual variations and directional dependence, as discussed in Sec.\ref{sec:evidence}, would be needed to determine whether anomalous decoherence could be attributed DM.  Indeed, MAQRO aims to test objective wavefuction collapse, and must be capable of ruling out astrophysical sources (like DM) if any anomalous decoherence is observed since these plausibly have higher \emph{a priori} probability than a fundamental breakdown of quantum mechanics.} to be $\etaDM \approx 50\%$, the residual sidereal daily background to be $\etairr \approx 10^{-3}$ of the total decoherence from all conventional sources, the data-taking run to be $\Trun = 1$~month, and the fringe visibility achieved to be $\vis \approx 50\%$.

We neglect atmospheric shielding when plotting the sensitivity curve, which allows us to include less sensitive experiments -- that are not sensitive to any DM that can pass through the atmosphere -- for comparison.  However, limitations imposed by the atmosphere are taken into account by the shaded regions denoting un-excluded DM parameter space that would be probed.

DM elastic scattering with individual molecules in the atmosphere can be reasonably modeled, but interactions in the Earth could be very complex, especially if there are other, inelastic modes.  Our main sensitivity curves in Fig~\ref{fig:sensitivity-mdm-big} and \ref{fig:sensitivity-mdm-many} conservatively ignore any DM that thermalizes in the atmosphere or is incident on the Earth's surface.  In this case, only the ambitious proposals by Pino \emph{et al.}\ and the MAQRO collaboration could detect DM.  However, as illustrated by the dashed lines in those figures, the OTIMA experiment
would be sensitive to a narrow class of DM that thermalizes in the Earth's crust and is re-emitted.  This is due to a strong greenhouse effect for such DM, which is discussed further in Appendix~\ref{sec:greenhouse}.  

An intermediate scenario, in which DM is isostropically reflected from the surface of the Earth without thermalizing, can also be considered.  The primary difference here would be that even when DM isotropizes in the atmosphere its flux would not be greatly suppressed at the surface of the Earth, and so could better decohere some superpositions.  However, for the experiments we consider, the only part of parameter space in which this increases sensitivity by more than about a factor of two is already excluded by the stellar heating bound.  Therefore it is not plotted separately, but is discussed in Appendix~\ref{sec:greenhouse}.

Our calculation of the decoherence rate for different experiments and DM parameters stretches over several qualitatively distinct regimes defined by the ratios between four characteristic length scales: the range of the dark force $\lambdamed = 1/\mmed$,  the typical DM reduced de Broglie wavelength $\lambdaDM = 1/(\mDM \vDM)$, the radius of the target $R$, and the spatial extent of the superposition $\DeltaX$.   The characteristic inverse momentum transfer is given by\footnote{The upper bound from the length scale $R^{-1}$ on the characteristic momentum transfer arises because the scattering is always dominated by the coherent scattering component in the regime we consider, and any momentum transfers that probe the interior of the target cannot be coherent over it.  The implication, when this occurs, is that only very forward scattering contributes.}
\begin{align}
\label{eq:momtransfer}
q^{-1} &\sim \max(\lambdamed,R,\lambdaDM).
\end{align}

The sensitivity curves, $\alphaMsense(\mmed)$, are calculated in Appendix~\ref{sec:decoh-calc} using \eqref{decoh-rate}. An order-of-magnitude estimate for the sensitivity curve is given in several limiting cases by\footnote{This neglects the contribution from incoherent scattering, but that is almost always subdominant for the experiments we consider.}
\begin{align}
\frac{1}{\alphaMsense} &\sim \chistat T N^2 \alphaDM \rhoDM \lambdaDM \Sigma^2 
\end{align}
where
\begin{align}
\Sigma = \begin{cases} 
\lambdamed^2/R, & \text{if}\ \DeltaX \gg R \gg \max(\lambdamed,\lambdaDM)\\
\lambdamed, & \text{if}\ \DeltaX \gg \lambdamed \gg \max(R, \lambdaDM) \\
\DeltaX, & \text{if}\ \lambdamed \gg \max(\DeltaX, R, \lambdaDM)\\
\DeltaX \lambdamed^2 /\lambdaDM^2, & \text{if}\ \lambdaDM \gg \max(\DeltaX, \lambdamed, R)
\end{cases}
\end{align}
is a characteristic length scale and 
\begin{align}\begin{split}
\chistat &= \left[2 \sqrt{\frac{\vis^{-2}-1}{\Trun \Gammacount}} + \etairr \ln (\vis^{-1})\right]^{-1} \\
&\sim \min(\sqrt{\Trun \Gammacount},\etairr^{-1})
\end{split}\end{align}
is the statistical enhancement. These limiting cases give reasonable estimates for the large-$\mmed$ and small-$\mmed$ regions of Figs.~\ref{fig:sensitivity-mdm-big} and \ref{fig:sensitivity-mdm-many}.

\section{Discussion}
\label{sec:discussion}

Future interferometers can maximize their sensitivity by increasing the mass, spatial extent, and exposure time of the superpositions produced.  Sensitivity increases quadratically with mass (in the coherence scattering regime), quadratically with spatial extent $\DeltaX$ (until this is larger than typical momentum transfer), and linearly with exposure time $T$.  In particular, the Wan \emph{et al.}\ proposal is hampered mostly by its unusually short exposure time, and modestly more aggressive parameters (an increase in radius by $\sim\!20\%$, or a tripling of exposure time) would make it sensitive to unexcluded DM with $\mDM \approx 1\keV$. (This threshold behavior is related to the saturation of the cross section during the breakdown of the Born approximation, as discussed in Appendix~\ref{sec:born}.) 

The fast data-gathering rate of the OTIMA interferometer means its sensitivity is limited by the sidereal decoherence background from conventional sources in the laboratory, the magnitude of which is currently unknown and assumed here to be $\etairr \approx 10^{-3}$.  Better understanding of this background would be very valuable, and the DM reach of OTIMA would increase linearly with its suppression.  In contrast, the proposals by Pino \emph{et al.}\ and the MAQRO collaboration are limited by the amount of data that can be collected in a reasonable time (taken here to be one month), and so are not particularly sensitive to this assumption.  The proposals by Bateman \emph{et al.}\ and Geraci \emph{et al.}\ are intermediate between these cases.

Let us briefly point out two other possibilities, albeit with little \emph{a priori} motivation, that would produce an enhanced DM signal.  First, relatively strong matter-DM interactions might lead to DM clumping in the Solar System \cite{damour1999new,peter2008particle,adler2009flyby}, although this is not thoroughly understood and gravitational three-body interactions are not effective \cite{edsjo2010comments}. Direct constraints on the DM mass density in the vicinity of the Earth are weak, being compatible with an increase of $10^5$ above the interstellar average if distributed smoothly through the Solar System \cite{iorio2006solar, khriplovich2007density, frere2008bound, pitjev2013constraints} and $10^{13}$ if concentrated within the orbit of the Moon \cite{adler2008placing}. Second, depending on the exact shape and size of the superposed target, resonance scattering -- which occurs outside the region of validity for the Born approximation and would be especially important for $\mDM \gtrsim 10\MeV$ -- could greatly enhance the scattering cross section.  Resonant behavior is illustrated in Ref~\cite{bateman2015existence} for a related low-mass DM search proposal using matter interferometers.\footnote{Note that the DM candidate considered in Ref.~\cite{bateman2015existence} appears to violate constraints on the annihilation rate arising from power injection into the cosmic microwave background because thermal DM with this little mass can reionize hydrogen by annihilation during matter-radiation equality \cite{Lin:2011gj,madhavacheril2014current}.  We thank Gordan Krnjaic for bringing this to our attention.}

To conclude: the impressive metrological power of matter interferometers complements their exciting but speculative role testing the foundations of quantum mechanics \cite{Nimmrichter:2011x2,Arndt:2014xx,Kaltenbaek:2015x2,pino2016quantum,romero-isart2016coherent}.  Even in the absence of dark matter, they put model-independent limits on anomalous sources of weak diffusion, especially from any interstellar sources shielded by the Earth.

\section*{Acknowledgments}
We thank Asimina Arvanitaki, Sougato Bose, Sandra Eibenberger, Andrew Geraci, Klaus Hornberger, Rainer Kaltenbaek, Robert Lasenby, Michael Niemack, Oriol Romero-Isart, and Hendrik Ulbricht  for very helpful discussion.  Research at Perimeter Institute is supported by the Government of Canada through the Department of Innovation, Science and Economic Development Canada and by the Province of Ontario through the Ministry of Research, Innovation and Science. IY acknowledges support by funds from the NSERC of Canada and the Early Research Awards program of Ontario.

\appendix

\section{Velocity dependence}
\label{sec:velocity}

The rate for Yukawa-type scattering processes grows strongly as velocity decreases.  In particular, the momentum-transfer cross section
\begin{align}\begin{split}
\sigma_{\mathrm{tr}} &= \int d\Omega (1-\cos \theta) \frac{d\sigma}{d\Omega} \\
&\propto \int d\Omega  \frac{1-\cos \theta}{\left(\betafac^2(1-\cos \theta)/2 + 1\right)^2},
\end{split}\end{align}
with
$\betafac \equiv 2 \mDM \vDM/ \mmed$,
grows like $\sigma_{\mathrm{tr}} \propto 1/\vDM^4$ for small $\vDM$. This is the relevant cross section for computing various DM astrophysical parameters as well as the DM-stopping power of the atmosphere. 

It might be guessed that the decoherence rate observed in superposition experiments would grow similarly as $\vDM$ decreases, since the degree to which any scattering event decoheres a superposition increases with the momentum transferred.  Here we point out that the decoherence rate does not share this behavior, in contrast to atmospheric and astrophysical dependencies.  Rather, the effective cross section for decoherence behaves more like the total cross section than the momentum-transfer cross section.  This velocity dependence is important for understanding decoherence in scenarios where DM is trapped and thermalized by the atmosphere, or where the DM incident on the atmosphere is not near the typical galactic velocity $\sim 230 \kms$.  Below we assume that the mediator mass is much less than the typical incident DM momentum, $\mmed \ll \vert {\bf \pin} \vert$, so that we are in the forward scattering regime.

For general superposition separation ${\bf \DeltaX}$, the decoherence per DM particle goes like
\be
\sigma_{\mathrm{decoh}} = \int d\Omega \left[1 - \cos( {\bf q}\cdot {\bf \DeltaX})\right] \frac{d\sigma}{d\Omega}.
\ee
Formally, this is neither the total cross section nor the momentum-transport cross section. When the superposition size $\DeltaX$ is very large, each collision causes full decoherence, the phase ${\bf q}\cdot {\bf \DeltaX}$ oscillates rapidly, and the decoherence rate just goes like the total cross section $\sigma_0 \propto \vDM^{-2}$. On the other hand, when the superposition size is very small, we can expand $\cos( {\bf q}\cdot {\bf \DeltaX})$ to second order in $\DeltaX$ and we find that the relevant quantity is the momentum-transport cross section \emph{weighted} by the incident momentum squared,
\be
\frac{1}{2}({\bf q} \cdot {\bf \DeltaX})^2 \frac{d\sigma}{d\Omega} \propto m^2  \vDM^2  (1-\cos \theta) \frac{d\sigma}{d\Omega}.
\ee
Therefore, in both cases the decoherence per DM particle goes like $1/\vDM^2$, not $1/\vDM^4$.  Since, for fixed DM density $\rhoDM$, the total flux is proportional to $\vDM$, the overall decoherence rate due to DM increases only like $1/\vDM$ in the forward scattering region.  This behavior is confirmed in numerical integrals for the more general case\footnote{The coherent scattering enhancement complicates this somewhat.  In the parts of the DM parameter space where the enhancement is only partial, we get an additional boost of $1/\vDM^3$.  However, the coherence enhancement eventually saturates.}.

Unlike for decoherence, the atmospheric shielding does in fact go like $1/\vDM^4$, so this soon dominates for small velocities.  That is, slower DM is generally blocked much more effectively by the atmosphere, and this is not well compensated by the increased ability of slow DM to decohere superpositions. The fundamental difference between shielding and decoherence is that, as the incident momentum of the DM decreases, the atmosphere has to do less and less total work to isotropize it while the amount of spatial discriminating power (momentum transfer) necessary to decohere a given mass superposition does not change.

\section{Atmospheric and Earth shielding}
\label{sec:shielding}

In this paper we consider some DM-nucleon cross sections large enough that the DM may collide multiple times with molecules in the atmosphere.  Here we estimate the effectiveness of the atmosphere at modifying the DM flux and potentially shielding terrestrial experiments from it.  The main results of this paper, illustrated as solid colored lines in Figs.~\ref{fig:sensitivity-mdm-big} and \ref{fig:sensitivity-mdm-many}, make the following conservative assumptions: DM thermalized in the atmosphere is ignored, DM reaching the ground is fully absorbed, and no DM is emitted from the Earth.  (We discuss relaxing these assumptions in the next section.)

For fixed DM particle mass $\mDM$, we calculate three curves in the $\mmed$-$\alphaM$ plane delineating whether the DM (a) scatters at least once in the atmosphere, (b) scatters sufficiently to isotropize, or (c) scatters sufficiently to thermalize.  One can see that (a) is necessary but not sufficient for (b) since the Yukawa scattering \eqref{eqn:yukawa_xs} is generally dominated by the forward direction.  Likewise, (b) is necessary but not sufficient for (c) since, for non-relativistic elastic scattering, the energy transfer is suppressed by the mass ratio $\mDM/m_{\mathrm{atm}}$, where $m_{\mathrm{atm}} \sim m_{\mathrm{N}_2} \approx 26 \GeV/c^2$ is the typical mass of the (mostly nitrogen) molecules in the atmosphere.

\emph{Scattering in atmosphere.}  The mean free path of a DM particle as a fraction of the overhead atmospheric mass is
\be
\label{eq:iso}
\zeta_{\mathrm{scatt}} \sim \frac{ m_\mathrm{atm} g_{\mathrm{E}}}{ \sigma_{\mathrm{atm}} p_{\mathrm{atm}}}.
\ee
Here, $p_{\mathrm{atm}}$ is atmospheric pressure at the detector, $g_{\mathrm{E}}$ is the gravitational acceleration, and $\sigma_{\mathrm{atm}}$ is the average scattering cross section with atmospheric molecules.  The scattering curves in Figs.~\ref{fig:sensitivity-mdm-big} and \ref{fig:sensitivity-mdm-many} are defined by $\zeta_{\mathrm{scatt}} = 1$, so that values of $\alphaM$ below this curve correspond to DM which passes unimpeded through the atmosphere.  

\emph{Isotropization in atmosphere.}  When DM scatters in the atmosphere, we model its trajectory to have an initial adiabatic portion where the energy is fixed while the \emph{direction} of the momentum follows a random walk on the unit sphere.  The variance associated with the angular distribution following a single collision is
\begin{align}
\sigma_\theta^2 \equiv \langle \sin^2 \theta \rangle \approx 4 \frac{1+\betafac^2}{\betafac^6}\left[(2+\betafac^2)\ln (1+\betafac^2)-\betafac^2\right]
\end{align}
where $q = \vert {\bf k'- k} \vert = 2 k \sin (\theta/2)$, $\betafac \equiv 2 \mDM \vDM/ \mmed \approx 2 k/\mmed$, and the distribution of angular step sizes is given by the Yukawa cross section \eqref{eqn:yukawa_xs}. For small steps (forward scattering), this takes the form of diffusion on the unit sphere and proceeds mathematically similarly to on the plane~\cite{Roberts:1960xx}. The incident DM will be isotropized after roughly $N_{\mathrm{iso}}$ collisions if $\sigma_\theta N_{\mathrm{iso}}^{1/2} \sim \pi$, and this will be associated with traversing a fraction
\begin{align}
\label{eq:iso-depth}
\zeta_{\mathrm{iso}} \sim \frac{ N_{\mathrm{iso}} \zeta_{\mathrm{scatt}}}{\sqrt{3}} \sim \frac{1}{\sqrt{3}} \left(\frac{\sigma_\theta}{\pi}\right)^2\frac{ m_\mathrm{atm} g_{\mathrm{E}}}{ \sigma_{\mathrm{atm}} p_{\mathrm{atm}}}
\end{align}
of the atmospheric mass\footnote{The factor of $1/\sqrt{3}$ arises from the standard deviation in a particular direction for a random step in three dimensions.}.  The isotropization curve is defined by $\zeta_{\mathrm{iso}} = 1$, so that values of $\alphaM$ below this curve correspond to DM that reaches the surface traveling with approximately the same momentum vector as when it entered the atmosphere.

On length scales larger than the isotropization depth, but before thermalization takes place, the DM trajectory can be modeled as a random walk in fractional atmospheric penetration depth with step size $\zeta_{\mathrm{iso}}$.  This walk takes place on the interval $[0,1]$, beginning at the initial penetration depth $\zeta_{\mathrm{iso}}$, with absorbing barriers on either end (i.e., the Earth's surface\footnote{See the next section for discussion of reflection from the Earth's surface.} and the top of the atmosphere\footnote{For DM masses below $\sim \mDM^{\mathrm{esc}} \approx 37 \MeV$, mean thermal velocities at $300\K$ are above gravitational escape velocity, so Jeans escape is immediate for DM reaching the exosphere, even when accounting for thermalization.  For masses above $\mDM^{\mathrm{esc}}$, the process is slower, allowing for a somewhat larger steady-state density of DM in the atmosphere than would be suggested by the model with the exosphere as a random-walk absorber.  However, we do not concentrate such large masses here since they are not particularly amenable to detection through decoherence, especially for $\mDM \gtrsim 100\MeV$.}).  A random walk starting $n$ and $m$ steps from two absorbing barriers is known to lead to absorption in the barriers with probabilities $m/(n+m)$ and $n/(n+m)$, respectively.  Therefore, the probability of a given DM particle reaching ground level is $\zeta_{\mathrm{iso}}$ (when $\zeta_{\mathrm{iso}} \le 1$).

\emph{Thermalization in atmosphere.}  After a sufficient number of steps, DM still in the atmosphere will thermalize.  The temperature of the interstellar DM flux is about $\mDM \vDM^2/3$, so for the characteristic speed $\vDM = 230 \kms$, DM particles with mass above $\sim\!120\keV$ will be hotter than the atmosphere.  More specifically, the DM is hotter than the atmosphere by factor of roughly\footnote{The majority of the atmosphere by mass is between $240\K$ and $300\K$, and 99\% is hotter than $210\K$.}
\be
\frac{\mDM (\vDM^2/3)}{270 \K} \sim 8  \left(\frac{\mDM}{1 \MeV} \right)\left(\frac{\vDM}{230 \kms} \right)^2.
\ee

For $\mDM \ll m_{\mathrm{atm}}$ and a momentum transfer $q$, the expected energy deposition for a collision is $\Delta E = q^2/(2 m_{\mathrm{atm}})$. Therefore the DM will thermalize to $\sim~\!\!\!\!\! 270 \K$ after taking approximately $\mDM/m_{\mathrm{atm}}$ isotropic steps.\footnote{For $\mDM = 1\MeV$, the mass ratio is $\mDM/m_{\mathrm{N}_2} \approx 4\times 10^{-5}$, which ensures that there are of order $10^4$ isotropic steps before thermalization becomes significant.  Random-walking DM that avoids exiting the atmosphere will therefore pass through a fraction of order $10^2 \zeta_{\mathrm{iso}}$ of the atmospheric mass.}  That is, the DM particle thermalizes after passing through a fraction 
\be
\zeta_{\mathrm{therm}} \sim \sqrt{\frac{m_{\mathrm{atm}}}{\mDM}}\, \zeta_{\mathrm{iso}}
\ee
of the atmospheric mass; the thermalization curve is defined by $\zeta_{\mathrm{therm}} = 1$.  DM particles from models corresponding to points below this curve will not thermalize before reaching the ground.

Even if the DM thermalizes quickly after entering the atmosphere, its probability of reaching the ground is still given by $\zeta_{\mathrm{iso}}$, as computed with the initial velocity $\vDM \approx 230\kms$.  This is because the probability of reaching one absorbing barrier in a random walk is determined by the initial relative distance from the two barriers, not the step size of the random walk (for small steps).  Thus, the total DM flux at ground level is always reduced roughly by the fraction $\zeta_{\mathrm{iso}}$. This is the probability for a single incident DM particle to reach ground level without first exiting the atmosphere, and does not strongly depend on $\zeta_{\mathrm{therm}}$.  For $\zeta_{\mathrm{therm}} \gg 1$, the reduced DM flux still has essentially the same energy distribution as outside the Earth's atmosphere, while for $\zeta_{\mathrm{therm}} \ll 1$ the DM energy distribution at ground level will be thermalized to the temperature of the atmosphere.

\emph{Experimental implications.}  Now we discuss how these three thresholds interact with the ability of superposition experiments to detect DM.  For a given experiment, the sensitivity curve is defined as the minimum coupling $\alphaMsense$ sufficient to induce detectable decoherence in the relevant experiment in the absence of any atmospheric shielding.  Therefore, any region above the sensitivity curve but below the isotropization curve -- regions for which DM flux at ground level is essentially identical as the top of the atmosphere -- will be detectable by the experiment. Additionally, the experiment sees nothing for all values of $\mmed$ where the experimental sensitivity curve is above the isotropization cut off because the atmosphere can only reduce the flux.

But now fix $\mmed$ such that the isotropization curve lies above the sensitivity curve, and consider two values of $\alphaM$: the lower $\alphaM$ lies on the isotropization curve, and the upper $\alphaM$ is one order of magnitude larger.  The flux seen by terrestrial experiments in DM scenarios corresponding to the upper point is reduced by about one order of magnitude compared to a coupling on the isotropization curve.  But the total cross section has also increased by an order of magnitude to compensate, so that the rate of decoherence in the experiment is actually unchanged.  Therefore, this type of DM should be about as visible as the type lying on the isotropization curve.  This argument applies for even large values of $\alphaM$ so long as they are below the thermalization curve, i.e., so long as the rare DM particle reaching the surface has not yet thermalized.

For values of $\alphaM$ where the isotropization curve and sensitivity curve are comparable (i.e., where the experiments are at the limits of their sensitivity) the values of $\alphaM$ above the isotropization curve will likewise strain detection.  This region is not precisely defined since the istropization curve \eqref{eq:iso} has only been specified up to a factor of order unity.

Of course, terrestrial experiments will not be able to function as \emph{directional} detectors of DM for $\alphaM$ above the isotropization curve, because directional information about the DM flux will be washed out.  Likewise, adding or removing shielding materials (e.g., lead or concrete) is ineffective at modulating the flux.  But in this scenario it will still be possible to look for sidereal oscillations in the decoherence rate from the change in the total number flux and (if below the thermalization curve) mean energy illustrated in Fig.~\ref{fig:daily}.

If DM interacts sufficiently strongly to thermalize in the atmosphere before it reaches the ground, the suppression of the total number flux relative to the top of the atmosphere is still $\zeta_{\mathrm{iso}} \propto \alphaM$.  (As mentioned above, the probability of a random walk reaching one absorbing barrier is determined by the initial placement, not by the step size.)  However, as described in Appendix~\ref{sec:velocity}, the rate of decoherence for a given number flux changes, due to shifts in both the chance of any collision in the detector and the degree of decoherence per collision. For $\mDM \gtrsim 120 \keV$, the DM slows down as it thermalizes and its effective decoherence cross section increases, while the reverse happens for $\mDM \lesssim 120 \keV$.  If the stellar cooling bound can be avoided, Figs.~\ref{fig:sensitivity-mdm-big} and~\ref{fig:sensitivity-mdm-many} show that several of the experiments are potentially sensitive to DM that thermalizes in the atmosphere.  But no DM thermalizing in the atmosphere is compatible with all constraints.

\section{Earth reflection and greenhouse effects}
\label{sec:greenhouse}

In this section we relax some conservative assumptions made about how DM interacts with the atmosphere and the Earth to see how this could potentially increase the DM parameter space visible to forthcoming experiments.\footnote{This is distinct from other scenarios that might, for one reason or another, increase the total DM density in the vicinity of the Earth above the interstellar average $\sim 0.4 \GeV/\mathrm{cm}^3$, e.g., \cite{Adler:2009xx}.}

\emph{Reflection by Earth.}  For our main results, Figs.~\ref{fig:sensitivity-mdm-big} and \ref{fig:sensitivity-mdm-many}, we assumed that DM that passes through the atmosphere and reaches the ground is fully absorbed. Unlike interactions in the atmosphere, which can be reasonably treated as elastic scattering events with isolated free molecules, energy deposition in solid material is a thorny topic, depending sensitively on material composition.  However, given that the large mass ratio of DM with nucleons can suppress energy transfer,\footnote{On the other hand, light electrons, band transitions, orbital angular momentum, etc., probably give many channels for more efficient energy transfer.} it is possible that the ground acts effectively as a Lambertian (isotropic) DM reflector.  Indeed,  we expect the weakly interacting DM to penetrate into the ground, yet it may (with order unit probability) emerge from the surface with approximately the same energy.

When DM passes unimpeded through the atmosphere, this would only roughly double the flux seen by the detector; the Earth-reflected DM would travel up and out of the atmosphere.  (DM reflected by the ground would almost certainly be isotropized, so this would also reduce the size of the directional signature by a factor of two.)  However, if interactions are strong enough for the DM to isotropize in the atmosphere, the random-walk model discussed in the previous section must be changed.  In this case --- one absorbing barrier with a given flux, and one reflecting barrier --- the steady-state solution is given by an equal isotropic DM flux at all altitudes.  The experimental sensitivity curve need not dip below the isotropization curve to imply observable decoherence; all directional information will be washed out, but the atmosphere does not reduce the flux.  This is not depicted in our main plots because it is only significant for DM excluded by the stellar-cooling bound.

\emph{Thermalization by Earth.} Rather than reflecting from the ground, DM  may thermalize to the temperature of the crust.\footnote{It is also possible that DM thermalizes to the hotter temperatures deeper in the Earth's interior and then passes through the (comparatively thin) crusts without much cooling.}  When $\mDM \gtrsim \mDMesc \approx 37\MeV$, the Earth-thermalized DM simply sinks into the interior of the planet, but otherwise its typical velocity at 300 K is sufficient to bring it back to the surface.   In this case, there will be two dominant populations of DM particles in the atmosphere: those coming directly from outside the atmosphere with characteristic speed $\vDM$, and those that thermalized inside the Earth and leaked back into the atmosphere.  Appealing again to the random-walk model, one can check that the steady-state solution (with equal number flux entering and exiting the Earth) results in an Earth-thermalized number flux at the surface that is enhanced by $\zeta_{\mathrm{iso}}/\zeta_{\mathrm{iso}}^{(300 \K)}$ relative to the interstellar flux at the top of the atmosphere. In the forward scattering regime ($m \ll k$), this reduces to $\vDM^4/(\vDM^{(300 \K)})^4 = \TDM^2/(300 \K)^2$.

For $\mDM \gtrsim 120\keV$, the DM is cooling as it thermalizes to 300 K, so this can significantly increase the flux seen by the experiments.\footnote{When $\mDM \lesssim 120\keV$, the decoherence rate is still at least as large as the strictly more conservative scenario of an absorbing surface for the Earth.} This is a DM greenhouse effect. Cooler DM also has a larger effective decoherence cross section as discussed in Appendix~\ref{sec:velocity}, further increasing the potential signal to several orders of magnitude larger than the conservative Earth-absorbing scenario. The resulting enhancement to the sensitivity of the OTIMA interferometer and the proposals by Bateman \emph{et al.}\ and Geraci \emph{et al.}\ is plotted in Figs.~\ref{fig:sensitivity-mdm-big} and~\ref{fig:sensitivity-mdm-many}.

The flux of unthermalized DM penetrating to ground level closely tracks the daily fluctuation in the interstellar flux at the top of the atmosphere because the time to diffuse through the atmosphere is much less than 24 hours for almost all the parameter range we consider.  Note that if the isotropization length in bulk matter is dramatically shorter for thermalized than unthermalized DM it's possible unthermalized DM entering the Earth and then subsequently thermalizing would take much longer than a day on average to escape. In this case, the interior of the planet would serve as a large reservoir of thermalized DM that smooths out the fluctuations in the flux at the surface.  This would suppress the sidereal variations of the Earth-thermalized component, making it harder to positively attribute anomalous decoherence to DM.

\section{Statistics}
\label{sec:statistics}

In previous work \cite{Riedel:2012ur} a superposition experiment's sensitivity to DM was identified as the coupling strength $\alphaM$ necessary to produce an $e$-fold of decoherence ($\gammaDM = 1/e$).  Here, we strengthen our reach by exploiting oscillations in the DM flux expected from the Earth's rotation over the sidereal day (23 hours, 56 minutes, 4 seconds), which produce large ($\sim50\%$) oscillations in the decoherence rate from DM.  For simplicity, we follow Sec.~\ref{sec:decoh} and assume a toy model with a two-dimensional Hilbert space that generates binary data, i.e., ``$+$'' or ``$-$'' measurement outcomes; this can be obtained from the continuous fringe pattern seen in most interferometers by simply partitioning the data into peak ($+$) and trough ($-$) bins, with modest loss of information.

To estimate the maximum sensitivity, we decompose the decoherence factor\footnote{Of course, a noisy phase in the decoherence factor is equivalent to a constant degree of decoherence. Phase noise from mundane terrestrial sources that varies on a daily basis is distinguished from DM decoherence using the same appeal to the sidereal drift.} as $\gamma = \gammaOther \gammaDM$ and assume the decoherence due to DM is relatively weak, $\gammaDM = e^{-\sDM} \approx 1-\sDM$.  (A possible phase shift from the DM wind is discussed separately below.) We break the DM decoherence $\sDM$ into a constant part $\Bsdm$ and a daily variation
\be
\Dsdm = \etaDM \sDM = \etaDM \Re \int_0^T \!\! d t \,  F({\bf \DeltaX}),
\ee
where $\etaDM \sim 50\%$ (see Fig.~\ref{fig:daily}) is the relative amplitude of the daily varying part of the decoherence. We bundle the constant part in with the non-DM decoherence rate, $\vis = \Re \gammaOther e^{-\Bsdm}$.  (In other words, here we are not attempting to distinguish mundane terrestrial decoherence from the constant component of DM decoherence.)  In this idealized case, we can identify $\vis \in [0,1]$ as the (24-hour-averaged) inference fringe \emph{visibility} (also known as the \emph{contrast}) \cite{Hornberger:2009xx}.

Let us further simplify our analysis by crudely separating our events ($\{+,-\}$ measurement outcomes for individual particles passing through the interferometer) into two time periods of the day: the morning and the evening. The probabilities of getting outcomes $\{+,-\}$ during these periods are
\begin{align}
P^{\mrn}_{\pm} &= \frac{1}{2} \left[1\pm\vis\left(1+\frac{\Dsdm}{2}\right)\right],\\
P^{\eve}_{\pm} &= \frac{1}{2} \left[1\pm\vis\left(1-\frac{\Dsdm}{2}\right)\right],
\end{align}
respectively (since $\Dsdm \ll 1$).  
Suppose that, when accounting for the detection efficiency of the interferometer, the total expected number of detection events over a multi-day experimental run is $\barNum$ with a possibly nonzero expected difference $\DeltaNum$ between morning and evening due to a daily varying background.
Then the four counts ($+$/$-$ for morning/evening) are statistically independent random variables, each distributed as a Poisson process with the respective means
\begin{align}
\Num^{\mrn}_{\pm} &= \frac{\barNum+\DeltaNum/2}{4} \left[1\pm\vis\left(1+\frac{\Dsdm}{2}\right)\right],\\
\Num^{\eve}_{\pm} &= \frac{\barNum-\DeltaNum/2}{4} \left[1\pm\vis\left(1-\frac{\Dsdm}{2}\right)\right].
\end{align}
The maximum-likelihood estimator is obtained by solving for $\Dsdm$, 
\begin{align}
\Dsdm = 2\frac{\Num^{\mrn}_{+}\Num^{\eve}_{-}-\Num^{\mrn}_{-} \Num^{\eve}_{+}}{\Num^{\mrn}_{+}\Num^{\eve}_{+}-\Num^{\mrn}_{-}\Num^{\eve}_{-}},
\end{align}
and the estimator variance is  
\begin{align}\begin{split}
\label{eq:decoh-error}
\sigmaDs^2 &= \sum_{X} \sigma^2_{X} \frac{\partial \Dsdm}{\partial X}  \\
&= \frac{\barNum(4(4+\Dsdm^2)-\vis^2 (4-\Dsdm^2)^2) +8 \DeltaNum \Dsdm}{\vis^2(4 \barNum^2-\DeltaNum^2 )} \\
&\approx 4 \frac{\vis^{-2}-1}{\barNum},
\end{split}\end{align}
where the sum is taken over $X= \Num^{\mrn}_{\pm}$, $\Num^{\eve}_{\pm}$, and where we have used $\sigma^2_{X} = \bar{X}$ for Poisson processes.  The last line follows from assuming that the DM decoherence is small compared to other sources of decoherence ($\Dsdm \ll 1-\vis \le 1$), and that the fractional daily variation in overall count rate is small ($\DeltaNum \ll \barNum$). The fringe visibility for successful interferometers is typically of order unity (e.g., \cite{Arndt:1999xx,Eibenberger:2013xx}), so we set $\vis=50\%$ for the purposes of our main results.\footnote{The approximation on the last line of Eq.~\eqref{eq:decoh-error} vanishes for very high visibility, $\vis \to1$, but in fact the exact quantity actually approaches $3\Dsdm^2/\barNum$ once our assumption $1-\vis \gg \Dsdm$ is violated.}

The error on the amount of daily-varying DM decoherence for a single particle passing through the interferometer therefore goes like $\sim \sqrt{12/\barNum}$, the expected scaling for Poissonian statistics.  It is also reasonable that when there is effective decoherence from mundane sources, $\vis$ is driven exponentially to zero and the experiment becomes completely insensitive to decoherence from DM.  In other words, \emph{achieving quantum coherence is critical} to the success of this technique.

\emph{Phase shift.}  As shown in the plot for $\mDM = 1\keV$ in Fig.~\ref{fig:sensitivity-mdm-many}, there are regions of parameter space for which the phase shift from the DM wind \cite{Riedel:2012ur} is larger than the decoherence, that is, for which $\sDM > \phiDM$ where
\begin{align}
\phiDM = \Im \int_0^T\! dt F( \DeltaX)
\end{align}
and $\gammaDM = e^{-\sDM + i \phiDM} \approx 1 -\sDM + i \phiDM$.  Here we briefly describe how to detect it.

First note that searching for a phase is properly understood as a null experiment for experimental configurations that truly operate in a two-dimensional Hilbert space like the Mach-Zehnder toy model (spanned, e.g., by the states corresponding to the left and right arms).  Recall that probabilities for the two measurement outcomes are $P_{\pm} = (1 \pm \Re \gamma)/2$, where the real part of the total decoherence factor is
\begin{align}\begin{split}
\Re \gamma &= \Re (\gammaDM \gamma_{\mathrm{other}}) \\
&= \Re \gammaDM \Re \gamma_{\mathrm{other}}  - \Im \gammaDM  \Im \gamma_{\mathrm{other}} \\
&\approx (1-\sDM) \Re \gamma_{\mathrm{other}}  - \phiDM \Im \gamma_{\mathrm{other}}.
\end{split}\end{align}
The normal process of aligning the interferometer applies the correct phase to $\gamma_{\mathrm{other}}$ necessary to zero out $\Im( \gamma_{\mathrm{other}})$, thereby maximizing the fringe visibility $\vis = (P_+ - P_-)/(P_+ + P_-)$.  To search for a small phase shift varying with the sidereal day, the interferometer should instead be null-aligned by adding a constant $\pi/2$ phase shift, zeroing out $\Re(\gamma_{\mathrm{other}})$ and maximizing $\Im(\gamma_{\mathrm{other}})$. (This would require a separate data-taking session in an actual Mach-Zehnder interferometer.)  The small varying phase shift from the DM wind, $\phiDM$, would then give a small discrepancy between $P_+$ and $P_-$, which would flip sign with the sidereal day.  The above statistical procedure could then be applied to the sidereal varying component $\DphiDM$ with the replacement $\vis \to \Im \gamma_{\mathrm{other}}$.

Of course, the measurement output from the interferometers we consider is actually a continuous real value (the spatial position of the measured particle) rather than a true binary outcome $\{+,-\}$.  Information is thrown away to reduce it to the above two-dimensional toy model by simply partitioning that real value into equal-sized bins.  Thus, searching for a phase shift from the DM wind would be done during data analysis rather than necessitating a separate experimental run.  (This is accomplished crudely for the toy model by just shifting the bins by $\pi/2$, but optimal methods would of course take into account the fine-grained position.)  Likewise, full temporal information can improve on the simple morning/evening binning, which can aid in rejection of background noisy phase shifts that \cite{rembold2014correction,gunther2015multifrequency,rembold2017secondorder} that could otherwise obscure the faint decoherence signal from DM.

\section{Decoherence rate}
\label{sec:decoh-calc}

In this section we calculate the decoherence rate \eqref{decoh-rate} and express it in terms of three integrals to be computed numerically.  We operate in the Born approximation, the applicability of which is discussed in Appendix~\ref{sec:born}, and assume the target is rigid, as justified in Appendix~\ref{sec:debye}. For a target superposed over spatial extent ${\bf \DeltaX}$ for time $T$, the off-diagonal terms in the target density matrix acquire a multiplicative decoherence factor $\gammaDM = \exp [ - \int_0^T \!\! d t \,  F({\bf \DeltaX})]$ due to the DM flux.  Here 
\begin{align}\begin{split}
\label{decoh-rate-app}
F({\bf \DeltaX}) &= \frac{\rhoDM}{\mDM} \int \! d {\bf k} \, n({\bf k}) \frac{k}{\mDM} \int \! d  \Omega \, I({\bf q})\\
&\qquad \times  \left[1 - \exp(i {\bf q} \cdot {\bf \DeltaX}/\hbar)\right] \frac{d\sigma}{d\Omega}(q), 
\end{split}\end{align}
where the DM momentum distribution is
\begin{gather}
\label{eq:dmphasespacedensity}
n(\pin) = \Znorm \left( \frac{e^{-(\pin/\mDM-{\bf \vE})^2/\vDM^2}}{\mDM^3 \vDM^3} \right)
\end{gather}
(for $\pin/\mDM < \vEsc$, with $\Znorm \approx 0.192$ a normalization factor), the single-nucleon spin-independent Yukawa elastic scattering cross section is
\begin{gather}
\frac{d\sigma}{d\Omega}(q) =  \frac{4\alphaM \alphaDM \mDM^2}{\left(q^2 + \mmed^2 \right)^2} ,
\end{gather}
and the structure factor is
\begin{gather}
\begin{split}
\label{eq:struc-fac}
I({\bf q}) &= \left\langle \left\vert \sum_{i=1}^{N} e^{-i {\bf x}_i \cdot {\bf q}} \right\vert^2 \right\rangle .
\end{split}
\end{gather}
The real part of $F({\bf \DeltaX})$ gives the decoherence rate, and the imaginary part gives the rate of phase shift.  The DM velocity distribution is Maxwellian (thermal) and isotropic in the galactic rest frame, but is displaced, in the lab frame, by the Earth's velocity through the Milky Way ${\bf \vE}$.  The structure factor $I({\bf q})$ comes from summing the contributions (in the Born approximation) from each particle in the superposed target charged under the mediating force \cite{Squires:1978xx}.  The angle brackets in Eq.~\eqref{eq:struc-fac} denote a thermal average over configurations of the target, and the ${\bf x}_i$ are the positions of the scattering charges (nucleons) within it.  For simplicity we assume all protons and neutrons are equally charged with individual couplings $\alphaM$, so $N$ is the number of nucleons in the target.  (Many other scenarios, like a $\mathrm{B}\!-\!\mathrm{L}$ charge, lead to simple order-unity adjustments.) 

When $\DeltaX$ is smaller than the size of a homogeneous target, we  expect that the volume overlap (i.e., the spatial points that are jointly shared by the two configurations of the superposition) does not contribute to the decoherence.  Intuitively this is because which-path information in not encoded into any DM that scatters from this volume, since it is common to both possible configurations; see Fig.~\ref{fig:resonator}.  One can check that Eq.~\eqref{decoh-rate-app} exhibits this behavior.  For instance, a cubic target of volume $L^3$ superposed over a distance $\DeltaX < L$ (in a direction parallel to one edge) has an identical decoherence rate $F({\bf \DeltaX})$ as a $L^2 \times \DeltaX$ slab superposed over a distance $L$ (in the dimension with thickness $\DeltaX$), regardless of the form of the single-nucleon scattering cross section.

In the limit where the target is much smaller than the length scale $1/q$ associated with the typical momentum transfer, the exponent in Eq.~\eqref{eq:struc-fac} is constant over the sum.  In this case, all DM-scattering out states add coherently and the decoherence rate is larger by a factor $N^2$ compared to the case of a single charge ($N=1$), i.e., a factor $N$ larger than would be expected from incoherent scattering.  Since the next generation of matter interferometers will superpose objects larger than $N=10^6$, this is a dramatic increase in sensitivity.

When the momentum $q$ is large enough to begin probing the internal structure of the target, we expect the coherence enhancement to fade.  This is seen explicitly by expanding the structure factor as 
\begin{align}\begin{split}
\label{eq:struct-fac-decomp}
I({\bf q}) &= \left\langle  \sum_{i=1}^{N} \sum_{j=1}^{N} e^{i ({\bf x}_i - {\bf x}_j)\cdot {\bf q}}  \right\rangle \\
&= N + A^2 \sum_{i=1}^{N_a} \sum_{\substack{j=1\\j\neq i}}^{N_a} \left\langle e^{i ({\bf y}_i - {\bf y}_j)\cdot {\bf q}}  \right\rangle
\\
&\approx N +  A^2 N_a(N_a-1) \gtil({\bf q})^2
\end{split}\end{align}
where $A$ is the atomic number of the nuclei, $y_i$ are the nuclei centers, and 
\begin{align}
\label{eq:nucleitransform}
\gtil({\bf q}) = \sum_{i=1}^{N_a} \left\langle e^{i {\bf y}_i \cdot {\bf q}} \right\rangle
\end{align}
is the Fourier transform of the thermal distribution of nuclei positions.  (We always take $\mDM < 1 \GeV$ and $\vDM < \vEsc \sim 0.002~c$, so $\lambdabar_q \gtrsim 100 \fm$ is always large enough to be fully coherent over a nucleus.)  To get the third line of Eq.~\eqref{eq:struct-fac-decomp}, we neglected the edge effects and pair correlation corrections that are small for the large targets we consider.  (The effects of pair correlations, which become very small for large $N$, are illustrated in the appendix of Ref.~\cite{Riedel:2012ur}.)  

When the typical length scale $q^{-1}$ associated with the momentum transfer is large compared to the atomic spacing \AA, we can take the continuum limit of Eq.~\eqref{eq:nucleitransform}.  A uniform spherical target of radius $R$ simplifies $\gtil({\bf q})$ to
\begin{align}
\label{eq:sphere-dis}
\gtilsph({\bf q}) \equiv 3 \frac{\sin(qR)-qR \cos(qR)}{(qR)^3}.
\end{align} 
For $R \gg 1/q$, we have $\gtil({\bf q})^2 \sim 1/(qR)^4 \to 0$, showing that only the incoherent contribution $N$ remains on the last line of Eq.~\eqref{eq:struct-fac-decomp} for large $q$, as expected.  Another way to see this is to just note that, for $q$ large compared to the inter atomic distances $\vert{\bf y}_i - {\bf y}_j\vert$, the thermal average in the second line vanishes. 

The three key dimensionless quantities that control the behavior of the integrals in Eq.~\eqref{decoh-rate-app} are
\begin{align}
\label{eq:dimensionless}
\alphafac &= 2 \mDM \vDM \DeltaX,\\
\betafac &= 2 \mDM \vDM /\mmed ,\\
\sigmafac &= 2 \mDM \vDM R .
\end{align}
Some algebra and integration over the azimuthal angles gives
\begin{widetext}
\begin{align}\begin{split}
\label{eq:mainintegral}
F({\bf \DeltaX}) &= \left(N \frac{4\alphaM \alphaDM \rhoDM \mDM \vDM}{ \mmed^4}\right) \cdot \left[4(2\pi)^2 \Znorm e^{-\vE/\vDM}\right] \\
&\qquad  \times \int_0^{\vEsc/\vDM} \! d s \, e^{-s^2} s^3 \int_{-1}^{1} \! d  \Cndx \, \exp\left(2 s \Cchi \Cndx \frac{\vE}{\vDM}\right) I_0\left(2 s \Schi \Sndx \frac{\vE}{\vDM} \right)  \\
&\qquad  \times \int_{0}^{1} \! d  \Cnq \,  \frac{1-\exp\left(i \alphafac s \Cnq^2 \Cndx\right) J_0 \left(\alphafac s \Cnq \Snq \Sndx \right)}{\left(\betafac^2 s^2 \Cnq^2 +1\right)^2} \Cnq \left[1 +  A (N_a-1) \gtil \left(\sigmafac s \Cnq \right)\right] 
\end{split}\end{align}
where $s=k/(\mDM \vDM)$ is the normalized DM momentum, $J_n$ ($I_n$) is the unmodified (modified) Bessel functions of the first kind, and $C_{\bf n m}$ ($S_{\bf n m}$) is the cosine (sine) of the angle between the vectors ${\bf n}$ and ${\bf m}$.  To produce our main plots, Eq.~\eqref{eq:mainintegral} is integrated numerically.

The Earth and its atmosphere modify the flux seen by terrestrial experiments, as discussed in Appendix~\ref{sec:shielding}.  For our main plots, Figs.~\ref{fig:sensitivity-mdm-big} and \ref{fig:sensitivity-mdm-many}, we assume the Earth's surface reflects isotropically (which applies to all experiments except MAQRO). The most ambitious experiment on Earth, Pino et al., generally sees an anisotropic distribution for downward traveling DM (originating from space) and an isotropic distribution for upward traveling DM (having been reflected by the Earth).  The other three terrestrial experiments (OTIMA, Geraci et al., Bateman et a.) are only plausibly sensitive to DM that interacts strongly enough to isotropize in the atmosphere, so this completely smooths out the directional distribution. Isotropization can eliminate the directional signal but affects that overall decoherence rate by less than a factor of two.

\emph{Limiting behavior.}  It is useful to understand the limiting behavior of this integral for extreme values of $\alphafac$, $\betafac$, and $\sigmafac$. For simplicity, take $\vEsc/\vDM \to \infty$ (so $\Znorm \to \pi^{-3/2}$), take $\vE/\vDM \to 1$, let the DM wind be parallel to the superposition displacement ($\Cchi = 1$), and approximate $N_a-1 \approx N_a$.  Then the coherent-scattering part of the decoherence rate simplifies to 
\begin{align}\begin{split}
F^{(N^2)}_R({\bf \DeltaX}) &= 
\left(\frac{2^6 \sqrt{\pi}}{e} \right)
\left( \frac{N^2\alphaM \alphaDM \rhoDM \mDM \vDM}{ \mmed^4}  \right) 
\int_0^{\infty} \! d s \, e^{-s^2} s^3 \\
&\qquad  \times\int_{-1}^{1} \! d  \Cndx  \exp\left(2 s  \Cndx\right)  \int_{0}^{1} \! d  \Cnq \,  \frac{1-\cos\left( \alphafac s \Cnq^2 \Cndx\right) }{\left(\betafac^2 s^2 \Cnq^2 +1\right)^2} \Cnq  \gtil \left(\sigmafac s \Cnq \right)
\end{split}\end{align}
\end{widetext}

We concentrate on the behavior of the integral over $\Cnq$ using heuristic arguments.  Because $\Cnq \in [0,1]$, and because of the overall suppression by $e^{-s^2}$ for $s \gg 1$, the factors $(\betafac^2 s^2 \Cnq^2 +1)^2$ and $\gtil \left(\sigmafac s \Cnq \right)$ are ignorable when $\betafac \ll 1$ and $\sigmafac \ll 1$, respectively.  Furthermore, if $\betafac \gg \sigmafac$ then $\gtil (\sigmafac s \Cnq )$ can be ignored, and similarly for $(\betafac^2 s^2 \Cnq^2 +1)^2$ if $\sigmafac \gg \betafac$.  Lastly, we can approximate $1-\cos(\alphafac s \Cnq^2 \Cndx)$ by $1$ if $\alphafac \gg \mathrm{max}(\betafac,\sigmafac,1)$ (because the rapidly oscillating cosine term averages to zero) and by $(\alphafac s \Cnq^2 \Cndx)^2/2$ if $\alphafac \ll \mathrm{max}(\betafac,\sigmafac,1)$.

This gives limiting behavior in six distinct regimes.  In each regime, we can compute the leading behavior of the integrals. Letting $\Omega = \mathrm{max}(\betafac,\sigmafac,1)$ and $\Phi=\mathrm{min}(\alphafac,\Omega)$,  we get the estimate
\begin{align}
F^{(N^2)}_R({\bf \DeltaX}) \approx N^2 \frac{4\pi \alphaM \alphaDM \rhoDM \mDM \vDM}{ \mmed^4} \Ycoeff_{\Omega,\Phi} \, \frac{\Phi^2}{\Omega^4},
\end{align}
where we have set the mean apparent DM velocity to the speed of the Earth in the galactic rest frame, $\vE/\vDM \approx 1$, and where\footnote{Here, $\mathrm{erf}(x)$ denotes the error function.}
\begin{align}\begin{split}
\Ycoeff_{\betafac,\betafac} &= 4 \,\mathrm{erf}(1) \approx 3.3708,\\
\Ycoeff_{\sigmafac,\sigmafac} &= 18   \,\mathrm{erf}(1) \approx 15.1686,\\
\Ycoeff_{1,1} &= \frac{4}{e \sqrt{\pi}}+6 \, \mathrm{erf}(1) \approx 5.88642,\\
\Ycoeff_{\betafac,\alphafac} &\approx 1.61279,\\
\Ycoeff_{\sigmafac,\alphafac} &\approx 9.92504 ,\\
\Ycoeff_{1,\alphafac}  &= \frac{19}{12 e \sqrt{\pi}}+\frac{55}{24}  \,\mathrm{erf}(1)  \approx 2.25982
\end{split}\end{align}
are the dimensionless coefficients for the six limiting regimes.

From this we can calculate the critical sensitivity by setting $ F^{(N^2)}_R T = 1$ (where $T$ is the exposure time given for each experiment in Table~\ref{tab:exp-parameters}), and solving for $\alphaM$. The sensitivity of an experiment is defined by the  critical value $\alphaMsense$ at which the sidereal variation in the SM decoherence is larger than the residual decoherence background and the statistical estimator error:
\begin{align}
\Dsdm(\alphaMsense) \equiv \sigmaDs + \Dsirr.
\end{align}
Here,
\begin{align}
\Dsirr = \etairr \sbkg = \etairr \ln (\vis^{-1})
\end{align}
is the background, 
\begin{align}
\sigmaDs = 2 \sqrt{\frac{\vis^{-2}-1}{\barNum}}
\end{align}
is the error derived in Appendix~\ref{sec:statistics}, $\vis$ is the interference fringe visibility, $B_0 = \Trun \Gammacount$ is the expected number of counts, $\Trun =$ 1 month is the presumed length of the data-taking run, and $\Gammacount$ is the count rate.  Values of the parameters can be found in Tables~\ref{tab:exp-parameters} and ~\ref{tab:common-parameters}.   

This can be bundled into a statistical factor $\chistat \equiv (\sigmaDs + \Dsirr)^{-1}$ to estimate our sensitivity:
\begin{align}\begin{split}
\frac{1}{\alphaMsense} &=  T \etaDM \chistat N^2 \frac{4 \pi \alphaDM \rhoDM \mDM \vDM}{ \mmed^4}  \Ycoeff_{\Phi,\Omega} \, \frac{\Phi^2}{\Omega^4}\\ 
\end{split}\end{align}
which well approximates the sensitivity line plotted for each experiment in Figs.~\ref{fig:sensitivity-mdm-big} and \ref{fig:sensitivity-mdm-many}.

\section{Born approximation}
\label{sec:born}

For sufficiently large values of the coupling $\alphaM$, the Born approximation will break down.  The approximation is based on the assumption that the potential \eqref{eqn:yukawa_potential} is weak -- or, more intuitively, translucent -- so that the incident plane wave $e^{i {\bf x} \cdot {\bf k}}$ is only slightly perturbed as it propagates through the potential and therefore approximates the exact time-independent solution.  For sufficiently large $\alphaM$, the potential becomes opaque and we expect the total cross section to saturate at a maximum value set by the spatial range of the interaction and the size of the superposed target, i.e., of order of the geometric cross section.

To illustrate this breakdown, we recall the toy case of a spherical square-hump (or square-well) potential of strength $V_0$ and radius $R$,
\be
\label{eq:sphericalsquarewell}
V(r) = \begin{cases} 
    V_0, & \text{if}\ r < R \\
	0, & \text{if}\ r > R
	\end{cases}.
\ee
The following behavior is well known (e.g., p.\ 519 of Ref.~\cite{landau1991quantum}). For small $V_0^2$, the Born approximation applies and the total scattering cross section is
\begin{align}
\label{eq:born-landau}
\sigma = \frac{2 \pi m^2 V_0^2 R^4}{k^2}\left(1-\frac{1}{(2ka)^2} + \frac{\sin 4ka}{(2ka)^3} - \frac{\sin^2 2ka}{(2ka)^4}\right),
\end{align}
which grows quadratically with the strength of the potential, $\sigma \propto V_0^2 \propto N^2$. This is the coherent scattering regime, and it corresponds to a quadratic (not linear dependence) on the total charge.  But the Born approximation breaks down as the interaction strength increases, and as $V_0 \to \infty$ the cross section saturates to of order of the geometric cross section $\sigma_{\mathrm{geo}} = \pi R^2$.

This transition can easily be treated analytically for small momentum\footnote{An analytical treatment of the large-momentum limit ($R^{-1} \ll k$, corresponding to geometric optics) is more complicated. One can derive the $\sigma \propto V_0^2$ behavior to agree with \eqref{eq:born-landau}, but as $V_0$ grows the scattered wave naturally divides into two equal parts: a refracted component and a defracted component, each with total cross section $\pi R^2 = \sigma_{\mathrm{geo}}$.  (The latter destructively interferes with the incident wave to produce a shadow. It is fully concentrated in the forward direction and so is singular in the $ka \to \infty$ limit.  See, e.g., p.~1551 of Ref.~\cite{morse1953methods}.)  For large enough $V_0$, the potential becomes completely reflective, and the scattering is isotropic.}, $k \ll R^{-1}$ and $k \ll \sqrt{2m V_0} $.  In this case the scattering is isotropic and this crossover is well described\footnote{When $V_0 < 0$, we have $\tanh (R\sqrt{2m V_0}) /(R\sqrt{2m V_0})= \mathrm{tan} (R\sqrt{2m \vert V_0 \vert})/(R\sqrt{2m \vert V_0 \vert})$, which exhibits resonant behavior.} by (p.~549 of Ref.~\cite{landau1991quantum})
\be
\label{eq:s-wave-cross-section}
\sigma = 4\pi R^2 \left(1-\frac{\tanh (R\sqrt{2m V_0})}{R \sqrt{2m V_0}} \right),
\ee
which saturates at $4\pi R^2 = 4 \sigma_{\mathrm{geo}}$.

We will now confirm this qualitative behavior -- that the Born approximation breaks down when the total cross section approaches the geometric\footnote{Here, ``geometric'' includes the range of the Yukawa force, which may extend much more than an atomic spacing beyond the surface of the target.} cross section -- when scattering from a Yukawa potential.

The Born series for the differential scattering amplitude is \cite{Taylor:2006xx,Newton:1982xx}
\begin{align}\begin{split}
f(\pin,\pout) &= -(2\pi)^2 \mDM \sum_{s=1}^{\infty} \lambda^s T^{(s)}(\pin,\pout)
\end{split}\end{align}
where the expansion of the $T$ matrix is given by 
\begin{align}\begin{split}
T^{(1)}(\pin,\pout) &= \matrixelement{\pout}{\hat{V}}{\pin}  \\
T^{(2)}(\pin,\pout) &= \matrixelement{\pout}{\hat{V} \hat{G}^0 (k^2/2 \mDM + i \epsilon) \hat{V}}{\pin} \\
&\,\,\, \vdots\\
T^{(s)}(\pin,\pout) &= \matrixelement{\pout}{\hat{V} \left[\hat{G}^0 (k^2/2 \mDM + i \epsilon) \hat{V} \right]^{s-1}}{\pin}.
\end{split}\end{align}
Above, $\lambda$ is a dummy expansion parameter, $\epsilon$ is an infinitesimal guiding the contour integral, $\hat{V}$ is the scattering (operator) potential for the entire target, and $\hat{G}^0 (z) = (z - \hat{P}^2/2\mDM)^{-1}$ is the free-particle (operator) Green's function.  

Note that the convergence properties of the Born series depend on the momenta $(\pin,\pout)$ and, in particular, that the approximation can be valid for forward (small-angle) scattering long after it has broken down for hard (large-angle) collisions \cite{Everhart:1955xx, Mott:1965xx, Taylor:2006xx}.

The (leading-order) Born approximation for the differential scattering cross section \eqref{eqn:yukawa_xs} for a single Yukawa potential $V(r) = - (\gM \gDM/ 4\pi) e^{-r \mmed} /  r$ is obtained by expanding $T^{(1)}(\pin,\pout)$ in the position basis and performing the resulting Fourier transform of $V(r)$,
\begin{align}\begin{split}
\label{eq:fourier-v}
T^{(1)}(\pin,\pout) &= \matrixelement{\pout}{\hat{V}}{\pin} \\
&= \frac{1}{(2\pi)^3}\int \! d {\bf x} \, V(x) e^{i (\pout-\pin) \cdot {\bf x}} \\
&= \frac{4\pi}{(2\pi)^3} \int_0^\infty \! dr \, V(r) r \frac{\sin(\abs{\pout-\pin} r)}{\abs{\pout-\pin}}\\
&= -\frac{\gM \gDM}{(2\pi)^3} \frac{1}{\mmed^2 + (\pout-\pin)^2}
\end{split}\end{align}
and then setting $d\sigma/d\Omega = \abs{f(\pin,\pout)}^2$, $\lambda = 1$.  A rigid target of $N$ scattering centers is obtained with the replacement $V(r) \to \sum_{i=1}^{N} V(\abs{{\bf r} - {\bf x}_i})$, which attaches a net factor of $\sum_{i=1}^{N} e^{i (\pout-\pin) \cdot {\bf x}_i} \approx N \gtil(\pout-\pin)$ to the last line of \eqref{eq:fourier-v}.  This is the origin of the structure factor \eqref{eq:struct-fac-decomp} in the Born approximation.

We will now compute the second term in the series and find the regime when it is smaller than the first term.  It will further be clear that higher-order terms pick up increasing powers of small ratios, so that the first-order approximation provides a good estimate to the entire convergent series.

Inserting a complete set of momentum eigenstates $\pmid$, we get
\begin{align}\begin{split}
\label{eq:t2}
&T^{(2)}(\pin,\pout) = \matrixelement{\pout}{\hat{V} \hat{G}^0 (k^2/2 \mDM + i \epsilon) \hat{V}}{\pin}\\
&\qquad = 2\mDM \int \! d\pmid \, \frac{\matrixelement{\pout}{\hat{V}}{\pmid}\matrixelement{\pmid}{\hat{V}}{\pin}}{k^2-p^2 + i\epsilon}\\
&\qquad = 2\mDM N^2\frac{\gM^2 \gDM^2}{(2\pi)^6} \int \! d\pmid \, \frac{1}{k^2-p^2 + i\epsilon} \\
&\qquad\qquad\qquad \times \frac{\gtil(\pout-\pmid)}{\mmed^2 + (\pout-\pmid)^2}\frac{\gtil(\pmid-\pin)}{\mmed^2 + (\pmid-\pin)^2}
\end{split}\end{align}
The relevant scales are $R^{-1}$, $\mmed$, and $k$, yielding two dimensionless parameters. Now we consider how the ratio $\abs{T^{(2)}/T^{(1)}}$ behaves in the various limiting cases.

When $\mmed \ll R^{-1}$, the Yukawa potentials are strongly overlapping and extend far beyond the size of the object itself, so we can set $\gtil(q) = 1$. In other words, this is just the traditional single Yukawa potential with strength $V(r) \to NV(r)$. In this case the answer is (p.\ 292 of Ref.~\cite{Newton:1982xx})
\begin{align}\begin{split}
T^{(2)}(\pin,\pout) &= \frac{\mDM \gM^2 \gDM^2 N^2}{2^5\pi^4 q \xi} 
\\ &\quad \times  
\left[2 \tan^{-1} \left(\frac{q \mmed}{2 \xi}\right) + i \ln \left(\frac{1+k q/\xi}{1-k q/\xi}\right)\right]
\end{split}\end{align}
with $\xi^2 \equiv \mmed^4 + 4\mmed^2 k^2 +k^2 q^2$.  Then we have\footnote{In the case $\mmed \lesssim q \ll k$, there is an additional factor of $(4 \ln \frac{q}{\mmed})^2$.}
\begin{align}\begin{split}
\label{eq:ratio-smallm}
\abs{\frac{T^{(2)}(\pin,\pout)}{T^{(1)}(\pin,\pout)}} &= \frac{N \sqrt{\alphaDM\alphaM}}{\vDM} 
\begin{cases} 
k/\mmed, & \text{if}\ k \ll \mmed \\
1/2, & \text{if}\ q \ll \mmed \ll k \\
\left(2 \ln \frac{q}{\mmed}\right), & \text{if}\ \mmed \ll q \ll k\\
\end{cases}.
\end{split}\end{align}
Note that this ratio may be small even for large coupling so long as the range of the force $\lambdamed = \mmed^{-1}$  decreases fast enough.   This just confirms that the $N^2$ dependence, a characteristic of the coherence elastic scattering enhancement, persists for contact interactions so long as the total cross section is small compared to the geometric cross section of the charges.  This is ultimately because the Born approximation does not require the potential at any given location to be small or even finite, only that the potential's overall effect on the incident wavefunction is small.

Alternatively, when $\mmed \gg R^{-1}$ in Eq.~\eqref{eq:t2}, the potential of each nucleus will generally overlap with its neighbors (so long as $\mmed \lesssim \mathrm{\AA}^{-1}$) but will not extend much outside of the superposed object.  Intuitively we expect to roughly recover the behavior of the spherical square-well potential \eqref{eq:sphericalsquarewell}.  This is especially true when the DM cannot probe the interior of the object, so let us first assume $k \ll R^{-1}$. In this case, $\gtilsph({\bf p})$ is slowly varying over scales of order $k$, so we can replace $\gtilsph(\pout-\pmid)\gtilsph(\pmid-\pin) \to \vert\gtilsph(\pmid)\vert^2$.  Similarly, $\gtilsph(\pmid)$ ensures the integrand is suppressed for $p \gtrsim R^{-1}$, so we can ignore the terms $(\pmid-\pin)^2$ and $(\pout-\pmid)^2$ in the denominators.  This removes $q$ dependence.  Then
\begin{align}\begin{split}
&T^{(2)}(\pin,\pout) \approx \frac{2\mDM}{\mmed^4} \frac{\gM^2 \gDM^2}{(2\pi)^6} N^2 \int \! d\pmid \, \frac{\vert\gtil(\pmid)\vert^2}{k^2-p^2 + i\epsilon} \\
& = \frac{2\mDM}{\mmed^4} \frac{\gM^2 \gDM^2}{(2\pi)^6} N^2\frac{1}{R} \left(\frac{12 \pi^2}{5} + O(kR) \right)
\end{split}\end{align}
To get the second line we insert Eq.~\eqref{eq:sphere-dis}, compute the integral over $\pmid$, take the limit $\epsilon \to 0$, and expand in powers of $k R$.  This gives 
\begin{align}
\label{eq:ratio-largemlargek}
\abs{\frac{T^{(2)}(\pin,\pout)}{T^{(1)}(\pin,\pout)}} &\approx \frac{N \sqrt{\alphaDM\alphaM}}{\vDM} \left(\frac{12 k}{5 \mmed^2 R}\right).
\end{align}
We can check that this recovers our intuitive expectation that, if $k \ll R^{-1} \ll m$, the Born approximation breaks down when the total cross section becomes of order of the geometric cross section.  Indeed, $d\sigma_{\mathrm{Born}}/d\Omega =  \abs{f^{(1)}(\pin,\pout)} = 16 \pi^3 \mDM^2 \abs{T^{(1)}(\pin,\pout)}^2$ and
\begin{align}
\frac{\sigma_{\mathrm{Born}}}{\sigma_{\mathrm{geo}}} \approx \left( \frac{5}{6\pi}\right)^2 \abs{\frac{T^{(2)}(\pin,\pout)}{T^{(1)}(\pin,\pout)}}^2 
\end{align}

The last situation, when both $k \gg R^{-1}$ and $\mmed \gg R^{-1}$, is the most complicated because the validity of the Born approximation depends on $q = \abs{\pin - \pout}$.  First let us calculate for $q=0$, changing integration variables to $\pell \equiv \pin - \pmid$:
\begin{align}\begin{split}
&T^{(2)}(\pin,\pin) \\
&\qquad = \frac{2\mDM}{\mmed^4}\frac{\gM^2 \gDM^2}{(2\pi)^6} N^2 \int \! d\pell \, \frac{\vert\gtil(\pell)\vert^2}{2 \pin\cdot\pell - \ell^2 + i\epsilon}\\
&\qquad = \frac{2\mDM}{\mmed^4}\frac{\gM^2 \gDM^2}{(2\pi)^6} N^2 2\pi \int_0^\infty \! d\ell \, \vert\gtil(\ell)\vert^2 \frac{\ell}{2 k} \\
&\qquad\qquad\qquad\qquad\qquad\qquad \times \left[\ln \left(\frac{1-\ell/2k}{1+\ell/2k}\right) - i \pi\right]
\end{split}\end{align}
where to get the second equality we assume $\gtil(\pell) = \gtil(\ell)$ is spherically symmetric (as for our uniform spherical target) and then perform the integral over the solid angle using the residue theorem.  Specializing $\gtil \to \gtilsph$ and expanding in the small quantity $(kR)^{-1}$ we perform the integral and get
\begin{align}\begin{split}
&T^{(2)}(\pin,\pin) = \frac{3}{64 \pi^4}\frac{\mDM \gM^2 \gDM^2 N^2}{\mmed^4} \\
&\qquad\qquad\qquad\qquad \times\left[ -i \frac{3}{2 k R} -\frac{1}{(k R)^2} + O\left(k^{-3} R^{-3} \right)\right]
\end{split}\end{align}
which yields
\begin{align}
\label{eq:ratio-largemsmallk}
\abs{\frac{T^{(2)}(\pin,\pin)}{T^{(1)}(\pin,\pin)}} &\approx \frac{N \sqrt{\alphaDM\alphaM}}{\vDM} \left(\frac{3}{2 \mmed R}\right)^2.
\end{align}
Now, note that the quantities $T^{(1)}(\pin,\pout)$ and $T^{(2)}(\pin,\pout)$ should not vary too much for $q \lesssim R^{-1}$ because $\gtil(\ell) \approx 1$ for all $q \ll R^{-1}$. Since the structure factor cuts off the scattering for $q \gtrsim R^{-1}$, we expect the total decoherence (dominated by small momentum transfer) calculated in the Born approximation to be roughly correct so long as the above ratio is small, even in regimes where the Born approximation is invalid for the (very rare) hard scattering events.

For the purpose of calculating decoherence rates, we identify the regions of validity of the Born approximation by the condition that the appropriate ratio, Eq.~\eqref{eq:ratio-smallm}, \eqref{eq:ratio-largemlargek}, or \eqref{eq:ratio-largemsmallk}, is less than unity for the dominant momentum transfer \eqref{eq:momtransfer}.  Our sensitivity estimates in Figs.~\ref{fig:sensitivity-mdm-big} and \ref{fig:sensitivity-mdm-many} cut off when the required coupling $\alphaMsense$ violates these conditions.  With the notable exception of resonance behavior, we expect the cross section to quickly saturate at of order the geometric cross section, so the interferometers would be insensitive to DM in this regime regardless of strength of coupling.

\emph{Exact cross section.} The cross section for Yukawa scattering outside the Born regime cannot be computed in generality.  When the range of the potential is much smaller than the geometric size of the target, the total potential can be approximated by the spherical square-well potential \eqref{eq:sphericalsquarewell}. The exact scattering amplitude is
\be
\label{eq:scatt-amp}
f(\theta) = \frac{1}{k} \sum_{\ell = 0}^{\infty} (2 \ell + 1) f_\ell P_\ell(\cos \theta)
\ee
where $f_\ell = e^{i \delta_\ell} \sin \delta_\ell$ is the harmonic amplitude, $P_\ell$ is the Legendre polynomial, 
\be
\delta_\ell = \frac{j_\ell^\prime(k R) j_\ell(\kappa R) - (\kappa/k) j_\ell(k R) j_\ell^\prime(\kappa R)}{n_\ell^\prime(k R) j_\ell(\kappa R) - (\kappa/k) n_\ell(k R) j_\ell^\prime(\kappa R)},
\ee
is the phase shift, $j_\ell$ and $n_\ell$ are the spherical Bessel functions of the first and second kind\footnote{Recall that they are related to the ordinary Bessel functions of the first and second kind by $j_\ell(z) = \sqrt{\pi/(2z)}J_{\ell+1/2}(z)$ and $n_\ell(z) = \sqrt{\pi/(2z)}N_{\ell+1/2}(z)$.  Some authors use the alternative notation $Y_\ell$ and $y_\ell$ for the ordinary and spherical Bessel functions of the second kind.}, $\kappa = \sqrt{k^2 + 2 m V_0}$ is the momentum inside the potential, and primes denote derivatives.  The sum in Eq.~\eqref{eq:scatt-amp} is dominated by the first $O(kR)$ terms.  The efficiently computable spherical square well regime ($R \gg \mmed$, $k R \not\gg 1$) partially overlaps with the Born regime, and one can check that the answers agree in this region.

\section{Debye-Waller factor}
\label{sec:debye}

The Debye-Waller factor quantifies the fractional suppression of coherent elastic scattering due to the mean relative spatial uncertainty of the scattering centers \cite{squires1978introduction}.  For a given momentum transfer ${\bf q}$ it is given by
\begin{align}
\label{eq:debye}
e^{-2W} &= \abs{\left\langle e^{i {\bf q}\cdot {\bf y}} \right\rangle}^2 = e^{-q^2 \langle {\bf y} \rangle^2 /3}
\end{align}
where the second equality follows by assuming the nuclei positions ${\bf y}$ are normally distributed (i.e., in isotropic, harmonic potentials). When the target is perfectly rigid, $\langle {\bf y} \rangle = 0$, there is no suppression.  In the Debye model one can approximate the mean squared displacement at temperatures $T$ above about $100 \K$ by
\begin{align}
\langle {\bf y} \rangle^2 \approx \frac{4 k_{\mathrm{B}} T}{\pi c_s^2 a_0 \rho} = d_{\mathrm{300K}}^2 \left(\frac{T}{300\K}\right),
\end{align}
where the material properties are the speed of sound $c_s$, the atomic spacing $a_0$, and the density $\rho$. Here, $d_{\mathrm{300K}}$ is the rms displacement at room temperature, and one can check (e.g, \cite{singh1971debye-waller}) that $0.05 \AAA < d_{\mathrm{300K}} < 0.2 \AAA$ for all the targets we consider.  For cooler temperatures, ${\bf y}$ saturates at the finite zero-point motion $d_{\mathrm{0K}} < d_{\mathrm{300K}}$.  These are much smaller length scales associated with the very small momentum transfer (coherent over the entire target) which are the dominant source of decoherence.  Indeed, even for hard-sphere scattering with $\mDM = 10 \MeV$, we have $q^{-1} \le \lambdaDM /2 \approx 0.1 \AAA$.  This confirms the intuition that decoherence detection is most useful for when the interactions are too soft to excite detectable phonon modes.

\bibliographystyle{apsrev4-1}
\bibliography{deco_dm_bib}

\begin{thebibliography}{104}%
\makeatletter
\providecommand \@ifxundefined [1]{%
 \@ifx{#1\undefined}
}%
\providecommand \@ifnum [1]{%
 \ifnum #1\expandafter \@firstoftwo
 \else \expandafter \@secondoftwo
 \fi
}%
\providecommand \@ifx [1]{%
 \ifx #1\expandafter \@firstoftwo
 \else \expandafter \@secondoftwo
 \fi
}%
\providecommand \natexlab [1]{#1}%
\providecommand \enquote  [1]{``#1''}%
\providecommand \bibnamefont  [1]{#1}%
\providecommand \bibfnamefont [1]{#1}%
\providecommand \citenamefont [1]{#1}%
\providecommand \href@noop [0]{\@secondoftwo}%
\providecommand \href [0]{\begingroup \@sanitize@url \@href}%
\providecommand \@href[1]{\@@startlink{#1}\@@href}%
\providecommand \@@href[1]{\endgroup#1\@@endlink}%
\providecommand \@sanitize@url [0]{\catcode `\\12\catcode `\$12\catcode
  `\&12\catcode `\#12\catcode `\^12\catcode `\_12\catcode `\%12\relax}%
\providecommand \@@startlink[1]{}%
\providecommand \@@endlink[0]{}%
\providecommand \url  [0]{\begingroup\@sanitize@url \@url }%
\providecommand \@url [1]{\endgroup\@href {#1}{\urlprefix }}%
\providecommand \urlprefix  [0]{URL }%
\providecommand \Eprint [0]{\href }%
\providecommand \doibase [0]{http://dx.doi.org/}%
\providecommand \selectlanguage [0]{\@gobble}%
\providecommand \bibinfo  [0]{\@secondoftwo}%
\providecommand \bibfield  [0]{\@secondoftwo}%
\providecommand \translation [1]{[#1]}%
\providecommand \BibitemOpen [0]{}%
\providecommand \bibitemStop [0]{}%
\providecommand \bibitemNoStop [0]{.\EOS\space}%
\providecommand \EOS [0]{\spacefactor3000\relax}%
\providecommand \BibitemShut  [1]{\csname bibitem#1\endcsname}%
\let\auto@bib@innerbib\@empty
\bibitem [{\citenamefont {{\relax R.~Agnese et al. ({SuperCDMS}
  Collaboration)}}(2016)}]{supercdmscollaboration2016new}%
  \BibitemOpen
  \bibfield  {author} {\bibinfo {author} {\bibnamefont {{\relax R.~Agnese et
  al. ({SuperCDMS} Collaboration)}}},\ }\href {\doibase
  10.1103/PhysRevLett.116.071301} {\bibfield  {journal} {\bibinfo  {journal}
  {Physical Review Letters}\ }\textbf {\bibinfo {volume} {116}},\ \bibinfo
  {pages} {071301} (\bibinfo {year} {2016})}\BibitemShut {NoStop}%
\bibitem [{\citenamefont {{PandaX-II Collaboration}}\ \emph
  {et~al.}(2016)\citenamefont {{PandaX-II Collaboration}}, \citenamefont {Tan},
  \citenamefont {Xiao}, \citenamefont {Cui}, \citenamefont {Chen},
  \citenamefont {Chen}, \citenamefont {Fang}, \citenamefont {Fu}, \citenamefont
  {Giboni}, \citenamefont {Giuliani}, \citenamefont {Gong}, \citenamefont
  {Guo}, \citenamefont {Han}, \citenamefont {Hu}, \citenamefont {Huang},
  \citenamefont {Ji}, \citenamefont {Ju}, \citenamefont {Lei}, \citenamefont
  {Li}, \citenamefont {Li}, \citenamefont {Li}, \citenamefont {Liang},
  \citenamefont {Lin}, \citenamefont {Liu}, \citenamefont {Liu}, \citenamefont
  {Lorenzon}, \citenamefont {Ma}, \citenamefont {Mao}, \citenamefont {Ni},
  \citenamefont {Ren}, \citenamefont {Schubnell}, \citenamefont {Shen},
  \citenamefont {Shi}, \citenamefont {Wang}, \citenamefont {Wang},
  \citenamefont {Wang}, \citenamefont {Wang}, \citenamefont {Wang},
  \citenamefont {Wang}, \citenamefont {Wang}, \citenamefont {Wu}, \citenamefont
  {Xiao}, \citenamefont {Xie}, \citenamefont {Yan}, \citenamefont {Yang},
  \citenamefont {Yue}, \citenamefont {Zeng}, \citenamefont {Zhang},
  \citenamefont {Zhang}, \citenamefont {Zhang}, \citenamefont {Zhang},
  \citenamefont {Zhao}, \citenamefont {Zhou}, \citenamefont {Zhou},\ and\
  \citenamefont {Zhou}}]{pandax-iicollaboration2016dark}%
  \BibitemOpen
  \bibfield  {author} {\bibinfo {author} {\bibnamefont {{PandaX-II
  Collaboration}}}, \bibinfo {author} {\bibfnamefont {A.}~\bibnamefont {Tan}},
  \bibinfo {author} {\bibfnamefont {M.}~\bibnamefont {Xiao}}, \bibinfo {author}
  {\bibfnamefont {X.}~\bibnamefont {Cui}}, \bibinfo {author} {\bibfnamefont
  {X.}~\bibnamefont {Chen}}, \bibinfo {author} {\bibfnamefont {Y.}~\bibnamefont
  {Chen}}, \bibinfo {author} {\bibfnamefont {D.}~\bibnamefont {Fang}}, \bibinfo
  {author} {\bibfnamefont {C.}~\bibnamefont {Fu}}, \bibinfo {author}
  {\bibfnamefont {K.}~\bibnamefont {Giboni}}, \bibinfo {author} {\bibfnamefont
  {F.}~\bibnamefont {Giuliani}}, \bibinfo {author} {\bibfnamefont
  {H.}~\bibnamefont {Gong}}, \bibinfo {author} {\bibfnamefont {X.}~\bibnamefont
  {Guo}}, \bibinfo {author} {\bibfnamefont {K.}~\bibnamefont {Han}}, \bibinfo
  {author} {\bibfnamefont {S.}~\bibnamefont {Hu}}, \bibinfo {author}
  {\bibfnamefont {X.}~\bibnamefont {Huang}}, \bibinfo {author} {\bibfnamefont
  {X.}~\bibnamefont {Ji}}, \bibinfo {author} {\bibfnamefont {Y.}~\bibnamefont
  {Ju}}, \bibinfo {author} {\bibfnamefont {S.}~\bibnamefont {Lei}}, \bibinfo
  {author} {\bibfnamefont {S.}~\bibnamefont {Li}}, \bibinfo {author}
  {\bibfnamefont {X.}~\bibnamefont {Li}}, \bibinfo {author} {\bibfnamefont
  {X.}~\bibnamefont {Li}}, \bibinfo {author} {\bibfnamefont {H.}~\bibnamefont
  {Liang}}, \bibinfo {author} {\bibfnamefont {Q.}~\bibnamefont {Lin}}, \bibinfo
  {author} {\bibfnamefont {H.}~\bibnamefont {Liu}}, \bibinfo {author}
  {\bibfnamefont {J.}~\bibnamefont {Liu}}, \bibinfo {author} {\bibfnamefont
  {W.}~\bibnamefont {Lorenzon}}, \bibinfo {author} {\bibfnamefont
  {Y.}~\bibnamefont {Ma}}, \bibinfo {author} {\bibfnamefont {Y.}~\bibnamefont
  {Mao}}, \bibinfo {author} {\bibfnamefont {K.}~\bibnamefont {Ni}}, \bibinfo
  {author} {\bibfnamefont {X.}~\bibnamefont {Ren}}, \bibinfo {author}
  {\bibfnamefont {M.}~\bibnamefont {Schubnell}}, \bibinfo {author}
  {\bibfnamefont {M.}~\bibnamefont {Shen}}, \bibinfo {author} {\bibfnamefont
  {F.}~\bibnamefont {Shi}}, \bibinfo {author} {\bibfnamefont {H.}~\bibnamefont
  {Wang}}, \bibinfo {author} {\bibfnamefont {J.}~\bibnamefont {Wang}}, \bibinfo
  {author} {\bibfnamefont {M.}~\bibnamefont {Wang}}, \bibinfo {author}
  {\bibfnamefont {Q.}~\bibnamefont {Wang}}, \bibinfo {author} {\bibfnamefont
  {S.}~\bibnamefont {Wang}}, \bibinfo {author} {\bibfnamefont {X.}~\bibnamefont
  {Wang}}, \bibinfo {author} {\bibfnamefont {Z.}~\bibnamefont {Wang}}, \bibinfo
  {author} {\bibfnamefont {S.}~\bibnamefont {Wu}}, \bibinfo {author}
  {\bibfnamefont {X.}~\bibnamefont {Xiao}}, \bibinfo {author} {\bibfnamefont
  {P.}~\bibnamefont {Xie}}, \bibinfo {author} {\bibfnamefont {B.}~\bibnamefont
  {Yan}}, \bibinfo {author} {\bibfnamefont {Y.}~\bibnamefont {Yang}}, \bibinfo
  {author} {\bibfnamefont {J.}~\bibnamefont {Yue}}, \bibinfo {author}
  {\bibfnamefont {X.}~\bibnamefont {Zeng}}, \bibinfo {author} {\bibfnamefont
  {H.}~\bibnamefont {Zhang}}, \bibinfo {author} {\bibfnamefont
  {H.}~\bibnamefont {Zhang}}, \bibinfo {author} {\bibfnamefont
  {H.}~\bibnamefont {Zhang}}, \bibinfo {author} {\bibfnamefont
  {T.}~\bibnamefont {Zhang}}, \bibinfo {author} {\bibfnamefont
  {L.}~\bibnamefont {Zhao}}, \bibinfo {author} {\bibfnamefont {J.}~\bibnamefont
  {Zhou}}, \bibinfo {author} {\bibfnamefont {N.}~\bibnamefont {Zhou}}, \ and\
  \bibinfo {author} {\bibfnamefont {X.}~\bibnamefont {Zhou}},\ }\href {\doibase
  10.1103/PhysRevLett.117.121303} {\bibfield  {journal} {\bibinfo  {journal}
  {Physical Review Letters}\ }\textbf {\bibinfo {volume} {117}},\ \bibinfo
  {pages} {121303} (\bibinfo {year} {2016})}\BibitemShut {NoStop}%
\bibitem [{\citenamefont {Angloher}\ \emph {et~al.}(2016)\citenamefont
  {Angloher}, \citenamefont {Bento}, \citenamefont {Bucci}, \citenamefont
  {Canonica}, \citenamefont {Defay}, \citenamefont {Erb}, \citenamefont
  {Feilitzsch}, \citenamefont {Iachellini}, \citenamefont {Gorla},
  \citenamefont {G\"{u}tlein}, \citenamefont {Hauff}, \citenamefont {Jochum},
  \citenamefont {Kiefer}, \citenamefont {Kluck}, \citenamefont {Kraus},
  \citenamefont {Lanfranchi}, \citenamefont {Loebell}, \citenamefont
  {M\"{u}nster}, \citenamefont {Pagliarone}, \citenamefont {Petricca},
  \citenamefont {Potzel}, \citenamefont {Pr\"{o}bst}, \citenamefont {Reindl},
  \citenamefont {Sch\"{a}ffner}, \citenamefont {Schieck}, \citenamefont
  {Sch\"{o}nert}, \citenamefont {Seidel}, \citenamefont {Stodolsky},
  \citenamefont {Strandhagen}, \citenamefont {Strauss}, \citenamefont {Tanzke},
  \citenamefont {Thi}, \citenamefont {T\"{u}rko\u{g}lu}, \citenamefont
  {Uffinger}, \citenamefont {Ulrich}, \citenamefont {Usherov}, \citenamefont
  {Wawoczny}, \citenamefont {Willers}, \citenamefont {W\"{u}strich},\ and\
  \citenamefont {Z\"{o}ller}}]{angloher2016results}%
  \BibitemOpen
  \bibfield  {author} {\bibinfo {author} {\bibfnamefont {G.}~\bibnamefont
  {Angloher}}, \bibinfo {author} {\bibfnamefont {A.}~\bibnamefont {Bento}},
  \bibinfo {author} {\bibfnamefont {C.}~\bibnamefont {Bucci}}, \bibinfo
  {author} {\bibfnamefont {L.}~\bibnamefont {Canonica}}, \bibinfo {author}
  {\bibfnamefont {X.}~\bibnamefont {Defay}}, \bibinfo {author} {\bibfnamefont
  {A.}~\bibnamefont {Erb}}, \bibinfo {author} {\bibfnamefont {F.~v.}\
  \bibnamefont {Feilitzsch}}, \bibinfo {author} {\bibfnamefont {N.~F.}\
  \bibnamefont {Iachellini}}, \bibinfo {author} {\bibfnamefont
  {P.}~\bibnamefont {Gorla}}, \bibinfo {author} {\bibfnamefont
  {A.}~\bibnamefont {G\"{u}tlein}}, \bibinfo {author} {\bibfnamefont
  {D.}~\bibnamefont {Hauff}}, \bibinfo {author} {\bibfnamefont
  {J.}~\bibnamefont {Jochum}}, \bibinfo {author} {\bibfnamefont
  {M.}~\bibnamefont {Kiefer}}, \bibinfo {author} {\bibfnamefont
  {H.}~\bibnamefont {Kluck}}, \bibinfo {author} {\bibfnamefont
  {H.}~\bibnamefont {Kraus}}, \bibinfo {author} {\bibfnamefont {J.~C.}\
  \bibnamefont {Lanfranchi}}, \bibinfo {author} {\bibfnamefont
  {J.}~\bibnamefont {Loebell}}, \bibinfo {author} {\bibfnamefont
  {A.}~\bibnamefont {M\"{u}nster}}, \bibinfo {author} {\bibfnamefont
  {C.}~\bibnamefont {Pagliarone}}, \bibinfo {author} {\bibfnamefont
  {F.}~\bibnamefont {Petricca}}, \bibinfo {author} {\bibfnamefont
  {W.}~\bibnamefont {Potzel}}, \bibinfo {author} {\bibfnamefont
  {F.}~\bibnamefont {Pr\"{o}bst}}, \bibinfo {author} {\bibfnamefont
  {F.}~\bibnamefont {Reindl}}, \bibinfo {author} {\bibfnamefont
  {K.}~\bibnamefont {Sch\"{a}ffner}}, \bibinfo {author} {\bibfnamefont
  {J.}~\bibnamefont {Schieck}}, \bibinfo {author} {\bibfnamefont
  {S.}~\bibnamefont {Sch\"{o}nert}}, \bibinfo {author} {\bibfnamefont
  {W.}~\bibnamefont {Seidel}}, \bibinfo {author} {\bibfnamefont
  {L.}~\bibnamefont {Stodolsky}}, \bibinfo {author} {\bibfnamefont
  {C.}~\bibnamefont {Strandhagen}}, \bibinfo {author} {\bibfnamefont
  {R.}~\bibnamefont {Strauss}}, \bibinfo {author} {\bibfnamefont
  {A.}~\bibnamefont {Tanzke}}, \bibinfo {author} {\bibfnamefont {H.~H.~T.}\
  \bibnamefont {Thi}}, \bibinfo {author} {\bibfnamefont {C.}~\bibnamefont
  {T\"{u}rko\u{g}lu}}, \bibinfo {author} {\bibfnamefont {M.}~\bibnamefont
  {Uffinger}}, \bibinfo {author} {\bibfnamefont {A.}~\bibnamefont {Ulrich}},
  \bibinfo {author} {\bibfnamefont {I.}~\bibnamefont {Usherov}}, \bibinfo
  {author} {\bibfnamefont {S.}~\bibnamefont {Wawoczny}}, \bibinfo {author}
  {\bibfnamefont {M.}~\bibnamefont {Willers}}, \bibinfo {author} {\bibfnamefont
  {M.}~\bibnamefont {W\"{u}strich}}, \ and\ \bibinfo {author} {\bibfnamefont
  {A.}~\bibnamefont {Z\"{o}ller}},\ }\href {\doibase
  10.1140/epjc/s10052-016-3877-3} {\bibfield  {journal} {\bibinfo  {journal}
  {The European Physical Journal C}\ }\textbf {\bibinfo {volume} {76}},\
  \bibinfo {pages} {25} (\bibinfo {year} {2016})}\BibitemShut {NoStop}%
\bibitem [{\citenamefont {{LUX Collaboration}}\ \emph
  {et~al.}(2017)\citenamefont {{LUX Collaboration}}, \citenamefont {Akerib},
  \citenamefont {Alsum}, \citenamefont {Araújo}, \citenamefont {Bai},
  \citenamefont {Bailey}, \citenamefont {Balajthy}, \citenamefont {Beltrame},
  \citenamefont {Bernard}, \citenamefont {Bernstein}, \citenamefont
  {Biesiadzinski}, \citenamefont {Boulton}, \citenamefont {Bramante},
  \citenamefont {Brás}, \citenamefont {Byram}, \citenamefont {Cahn},
  \citenamefont {Carmona-Benitez}, \citenamefont {Chan}, \citenamefont
  {Chiller}, \citenamefont {Chiller}, \citenamefont {Currie}, \citenamefont
  {Cutter}, \citenamefont {Davison}, \citenamefont {Dobi}, \citenamefont
  {Dobson}, \citenamefont {Druszkiewicz}, \citenamefont {Edwards},
  \citenamefont {Faham}, \citenamefont {Fiorucci}, \citenamefont {Gaitskell},
  \citenamefont {Gehman}, \citenamefont {Ghag}, \citenamefont {Gibson},
  \citenamefont {Gilchriese}, \citenamefont {Hall}, \citenamefont {Hanhardt},
  \citenamefont {Haselschwardt}, \citenamefont {Hertel}, \citenamefont {Hogan},
  \citenamefont {Horn}, \citenamefont {Huang}, \citenamefont {Ignarra},
  \citenamefont {Ihm}, \citenamefont {Jacobsen}, \citenamefont {Ji},
  \citenamefont {Kamdin}, \citenamefont {Kazkaz}, \citenamefont {Khaitan},
  \citenamefont {Knoche}, \citenamefont {Larsen}, \citenamefont {Lee},
  \citenamefont {Lenardo}, \citenamefont {Lesko}, \citenamefont {Lindote},
  \citenamefont {Lopes}, \citenamefont {Manalaysay}, \citenamefont {Mannino},
  \citenamefont {Marzioni}, \citenamefont {McKinsey}, \citenamefont {Mei},
  \citenamefont {Mock}, \citenamefont {Moongweluwan}, \citenamefont {Morad},
  \citenamefont {Murphy}, \citenamefont {Nehrkorn}, \citenamefont {Nelson},
  \citenamefont {Neves}, \citenamefont {O’Sullivan}, \citenamefont
  {Oliver-Mallory}, \citenamefont {Palladino}, \citenamefont {Pease},
  \citenamefont {Phelps}, \citenamefont {Reichhart}, \citenamefont {Rhyne},
  \citenamefont {Shaw}, \citenamefont {Shutt}, \citenamefont {Silva},
  \citenamefont {Solmaz}, \citenamefont {Solovov}, \citenamefont {Sorensen},
  \citenamefont {Stephenson}, \citenamefont {Sumner}, \citenamefont {Szydagis},
  \citenamefont {Taylor}, \citenamefont {Taylor}, \citenamefont {Tennyson},
  \citenamefont {Terman}, \citenamefont {Tiedt}, \citenamefont {To},
  \citenamefont {Tripathi}, \citenamefont {Tvrznikova}, \citenamefont {Uvarov},
  \citenamefont {Verbus}, \citenamefont {Webb}, \citenamefont {White},
  \citenamefont {Whitis}, \citenamefont {Witherell}, \citenamefont {Wolfs},
  \citenamefont {Xu}, \citenamefont {Yazdani}, \citenamefont {Young},\ and\
  \citenamefont {Zhang}}]{luxcollaboration2017results}%
  \BibitemOpen
  \bibfield  {author} {\bibinfo {author} {\bibnamefont {{LUX Collaboration}}},
  \bibinfo {author} {\bibfnamefont {D.~S.}\ \bibnamefont {Akerib}}, \bibinfo
  {author} {\bibfnamefont {S.}~\bibnamefont {Alsum}}, \bibinfo {author}
  {\bibfnamefont {H.~M.}\ \bibnamefont {Araújo}}, \bibinfo {author}
  {\bibfnamefont {X.}~\bibnamefont {Bai}}, \bibinfo {author} {\bibfnamefont
  {A.~J.}\ \bibnamefont {Bailey}}, \bibinfo {author} {\bibfnamefont
  {J.}~\bibnamefont {Balajthy}}, \bibinfo {author} {\bibfnamefont
  {P.}~\bibnamefont {Beltrame}}, \bibinfo {author} {\bibfnamefont {E.~P.}\
  \bibnamefont {Bernard}}, \bibinfo {author} {\bibfnamefont {A.}~\bibnamefont
  {Bernstein}}, \bibinfo {author} {\bibfnamefont {T.~P.}\ \bibnamefont
  {Biesiadzinski}}, \bibinfo {author} {\bibfnamefont {E.~M.}\ \bibnamefont
  {Boulton}}, \bibinfo {author} {\bibfnamefont {R.}~\bibnamefont {Bramante}},
  \bibinfo {author} {\bibfnamefont {P.}~\bibnamefont {Brás}}, \bibinfo
  {author} {\bibfnamefont {D.}~\bibnamefont {Byram}}, \bibinfo {author}
  {\bibfnamefont {S.~B.}\ \bibnamefont {Cahn}}, \bibinfo {author}
  {\bibfnamefont {M.~C.}\ \bibnamefont {Carmona-Benitez}}, \bibinfo {author}
  {\bibfnamefont {C.}~\bibnamefont {Chan}}, \bibinfo {author} {\bibfnamefont
  {A.~A.}\ \bibnamefont {Chiller}}, \bibinfo {author} {\bibfnamefont
  {C.}~\bibnamefont {Chiller}}, \bibinfo {author} {\bibfnamefont
  {A.}~\bibnamefont {Currie}}, \bibinfo {author} {\bibfnamefont {J.~E.}\
  \bibnamefont {Cutter}}, \bibinfo {author} {\bibfnamefont {T.~J.~R.}\
  \bibnamefont {Davison}}, \bibinfo {author} {\bibfnamefont {A.}~\bibnamefont
  {Dobi}}, \bibinfo {author} {\bibfnamefont {J.~E.~Y.}\ \bibnamefont {Dobson}},
  \bibinfo {author} {\bibfnamefont {E.}~\bibnamefont {Druszkiewicz}}, \bibinfo
  {author} {\bibfnamefont {B.~N.}\ \bibnamefont {Edwards}}, \bibinfo {author}
  {\bibfnamefont {C.~H.}\ \bibnamefont {Faham}}, \bibinfo {author}
  {\bibfnamefont {S.}~\bibnamefont {Fiorucci}}, \bibinfo {author}
  {\bibfnamefont {R.~J.}\ \bibnamefont {Gaitskell}}, \bibinfo {author}
  {\bibfnamefont {V.~M.}\ \bibnamefont {Gehman}}, \bibinfo {author}
  {\bibfnamefont {C.}~\bibnamefont {Ghag}}, \bibinfo {author} {\bibfnamefont
  {K.~R.}\ \bibnamefont {Gibson}}, \bibinfo {author} {\bibfnamefont {M.~G.~D.}\
  \bibnamefont {Gilchriese}}, \bibinfo {author} {\bibfnamefont {C.~R.}\
  \bibnamefont {Hall}}, \bibinfo {author} {\bibfnamefont {M.}~\bibnamefont
  {Hanhardt}}, \bibinfo {author} {\bibfnamefont {S.~J.}\ \bibnamefont
  {Haselschwardt}}, \bibinfo {author} {\bibfnamefont {S.~A.}\ \bibnamefont
  {Hertel}}, \bibinfo {author} {\bibfnamefont {D.~P.}\ \bibnamefont {Hogan}},
  \bibinfo {author} {\bibfnamefont {M.}~\bibnamefont {Horn}}, \bibinfo {author}
  {\bibfnamefont {D.~Q.}\ \bibnamefont {Huang}}, \bibinfo {author}
  {\bibfnamefont {C.~M.}\ \bibnamefont {Ignarra}}, \bibinfo {author}
  {\bibfnamefont {M.}~\bibnamefont {Ihm}}, \bibinfo {author} {\bibfnamefont
  {R.~G.}\ \bibnamefont {Jacobsen}}, \bibinfo {author} {\bibfnamefont
  {W.}~\bibnamefont {Ji}}, \bibinfo {author} {\bibfnamefont {K.}~\bibnamefont
  {Kamdin}}, \bibinfo {author} {\bibfnamefont {K.}~\bibnamefont {Kazkaz}},
  \bibinfo {author} {\bibfnamefont {D.}~\bibnamefont {Khaitan}}, \bibinfo
  {author} {\bibfnamefont {R.}~\bibnamefont {Knoche}}, \bibinfo {author}
  {\bibfnamefont {N.~A.}\ \bibnamefont {Larsen}}, \bibinfo {author}
  {\bibfnamefont {C.}~\bibnamefont {Lee}}, \bibinfo {author} {\bibfnamefont
  {B.~G.}\ \bibnamefont {Lenardo}}, \bibinfo {author} {\bibfnamefont {K.~T.}\
  \bibnamefont {Lesko}}, \bibinfo {author} {\bibfnamefont {A.}~\bibnamefont
  {Lindote}}, \bibinfo {author} {\bibfnamefont {M.~I.}\ \bibnamefont {Lopes}},
  \bibinfo {author} {\bibfnamefont {A.}~\bibnamefont {Manalaysay}}, \bibinfo
  {author} {\bibfnamefont {R.~L.}\ \bibnamefont {Mannino}}, \bibinfo {author}
  {\bibfnamefont {M.~F.}\ \bibnamefont {Marzioni}}, \bibinfo {author}
  {\bibfnamefont {D.~N.}\ \bibnamefont {McKinsey}}, \bibinfo {author}
  {\bibfnamefont {D.-M.}\ \bibnamefont {Mei}}, \bibinfo {author} {\bibfnamefont
  {J.}~\bibnamefont {Mock}}, \bibinfo {author} {\bibfnamefont {M.}~\bibnamefont
  {Moongweluwan}}, \bibinfo {author} {\bibfnamefont {J.~A.}\ \bibnamefont
  {Morad}}, \bibinfo {author} {\bibfnamefont {A.~S.~J.}\ \bibnamefont
  {Murphy}}, \bibinfo {author} {\bibfnamefont {C.}~\bibnamefont {Nehrkorn}},
  \bibinfo {author} {\bibfnamefont {H.~N.}\ \bibnamefont {Nelson}}, \bibinfo
  {author} {\bibfnamefont {F.}~\bibnamefont {Neves}}, \bibinfo {author}
  {\bibfnamefont {K.}~\bibnamefont {O’Sullivan}}, \bibinfo {author}
  {\bibfnamefont {K.~C.}\ \bibnamefont {Oliver-Mallory}}, \bibinfo {author}
  {\bibfnamefont {K.~J.}\ \bibnamefont {Palladino}}, \bibinfo {author}
  {\bibfnamefont {E.~K.}\ \bibnamefont {Pease}}, \bibinfo {author}
  {\bibfnamefont {P.}~\bibnamefont {Phelps}}, \bibinfo {author} {\bibfnamefont
  {L.}~\bibnamefont {Reichhart}}, \bibinfo {author} {\bibfnamefont
  {C.}~\bibnamefont {Rhyne}}, \bibinfo {author} {\bibfnamefont
  {S.}~\bibnamefont {Shaw}}, \bibinfo {author} {\bibfnamefont {T.~A.}\
  \bibnamefont {Shutt}}, \bibinfo {author} {\bibfnamefont {C.}~\bibnamefont
  {Silva}}, \bibinfo {author} {\bibfnamefont {M.}~\bibnamefont {Solmaz}},
  \bibinfo {author} {\bibfnamefont {V.~N.}\ \bibnamefont {Solovov}}, \bibinfo
  {author} {\bibfnamefont {P.}~\bibnamefont {Sorensen}}, \bibinfo {author}
  {\bibfnamefont {S.}~\bibnamefont {Stephenson}}, \bibinfo {author}
  {\bibfnamefont {T.~J.}\ \bibnamefont {Sumner}}, \bibinfo {author}
  {\bibfnamefont {M.}~\bibnamefont {Szydagis}}, \bibinfo {author}
  {\bibfnamefont {D.~J.}\ \bibnamefont {Taylor}}, \bibinfo {author}
  {\bibfnamefont {W.~C.}\ \bibnamefont {Taylor}}, \bibinfo {author}
  {\bibfnamefont {B.~P.}\ \bibnamefont {Tennyson}}, \bibinfo {author}
  {\bibfnamefont {P.~A.}\ \bibnamefont {Terman}}, \bibinfo {author}
  {\bibfnamefont {D.~R.}\ \bibnamefont {Tiedt}}, \bibinfo {author}
  {\bibfnamefont {W.~H.}\ \bibnamefont {To}}, \bibinfo {author} {\bibfnamefont
  {M.}~\bibnamefont {Tripathi}}, \bibinfo {author} {\bibfnamefont
  {L.}~\bibnamefont {Tvrznikova}}, \bibinfo {author} {\bibfnamefont
  {S.}~\bibnamefont {Uvarov}}, \bibinfo {author} {\bibfnamefont {J.~R.}\
  \bibnamefont {Verbus}}, \bibinfo {author} {\bibfnamefont {R.~C.}\
  \bibnamefont {Webb}}, \bibinfo {author} {\bibfnamefont {J.~T.}\ \bibnamefont
  {White}}, \bibinfo {author} {\bibfnamefont {T.~J.}\ \bibnamefont {Whitis}},
  \bibinfo {author} {\bibfnamefont {M.~S.}\ \bibnamefont {Witherell}}, \bibinfo
  {author} {\bibfnamefont {F.~L.~H.}\ \bibnamefont {Wolfs}}, \bibinfo {author}
  {\bibfnamefont {J.}~\bibnamefont {Xu}}, \bibinfo {author} {\bibfnamefont
  {K.}~\bibnamefont {Yazdani}}, \bibinfo {author} {\bibfnamefont {S.~K.}\
  \bibnamefont {Young}}, \ and\ \bibinfo {author} {\bibfnamefont
  {C.}~\bibnamefont {Zhang}},\ }\href {\doibase 10.1103/PhysRevLett.118.021303}
  {\bibfield  {journal} {\bibinfo  {journal} {Physical Review Letters}\
  }\textbf {\bibinfo {volume} {118}},\ \bibinfo {pages} {021303} (\bibinfo
  {year} {2017})}\BibitemShut {NoStop}%
\bibitem [{\citenamefont {Essig}\ \emph
  {et~al.}(2012{\natexlab{a}})\citenamefont {Essig}, \citenamefont {Mardon},\
  and\ \citenamefont {Volansky}}]{Essig:2011nj}%
  \BibitemOpen
  \bibfield  {author} {\bibinfo {author} {\bibfnamefont {R.}~\bibnamefont
  {Essig}}, \bibinfo {author} {\bibfnamefont {J.}~\bibnamefont {Mardon}}, \
  and\ \bibinfo {author} {\bibfnamefont {T.}~\bibnamefont {Volansky}},\ }\href
  {\doibase 10.1103/PhysRevD.85.076007} {\bibfield  {journal} {\bibinfo
  {journal} {Phys.Rev.}\ }\textbf {\bibinfo {volume} {D85}},\ \bibinfo {pages}
  {076007} (\bibinfo {year} {2012}{\natexlab{a}})}\BibitemShut {NoStop}%
\bibitem [{\citenamefont {Graham}\ \emph {et~al.}(2012)\citenamefont {Graham},
  \citenamefont {Kaplan}, \citenamefont {Rajendran},\ and\ \citenamefont
  {Walters}}]{graham2012semiconductor}%
  \BibitemOpen
  \bibfield  {author} {\bibinfo {author} {\bibfnamefont {P.~W.}\ \bibnamefont
  {Graham}}, \bibinfo {author} {\bibfnamefont {D.~E.}\ \bibnamefont {Kaplan}},
  \bibinfo {author} {\bibfnamefont {S.}~\bibnamefont {Rajendran}}, \ and\
  \bibinfo {author} {\bibfnamefont {M.~T.}\ \bibnamefont {Walters}},\ }\href
  {\doibase 10.1016/j.dark.2012.09.001} {\bibfield  {journal} {\bibinfo
  {journal} {Physics of the Dark Universe}\ }\bibinfo {series} {Next Decade in
  Dark Matter and Dark Energy},\ \textbf {\bibinfo {volume} {1}},\ \bibinfo
  {pages} {32} (\bibinfo {year} {2012})}\BibitemShut {NoStop}%
\bibitem [{\citenamefont {{\relax R.~Essig et al. (Snowmass
  report)}}(2013)}]{essig2013dark}%
  \BibitemOpen
  \bibfield  {author} {\bibinfo {author} {\bibnamefont {{\relax R.~Essig et al.
  (Snowmass report)}}},\ }\href {http://arxiv.org/abs/1311.0029} {\bibfield
  {journal} {\bibinfo  {journal} {{arXiv:1311.0029}}\ } (\bibinfo {year}
  {2013})}\BibitemShut {NoStop}%
\bibitem [{\citenamefont {Lee}\ \emph {et~al.}(2015)\citenamefont {Lee},
  \citenamefont {Lisanti}, \citenamefont {Mishra-Sharma},\ and\ \citenamefont
  {Safdi}}]{lee2015modulation}%
  \BibitemOpen
  \bibfield  {author} {\bibinfo {author} {\bibfnamefont {S.~K.}\ \bibnamefont
  {Lee}}, \bibinfo {author} {\bibfnamefont {M.}~\bibnamefont {Lisanti}},
  \bibinfo {author} {\bibfnamefont {S.}~\bibnamefont {Mishra-Sharma}}, \ and\
  \bibinfo {author} {\bibfnamefont {B.~R.}\ \bibnamefont {Safdi}},\ }\href
  {\doibase 10.1103/PhysRevD.92.083517} {\bibfield  {journal} {\bibinfo
  {journal} {Physical Review D}\ }\textbf {\bibinfo {volume} {92}},\ \bibinfo
  {pages} {083517} (\bibinfo {year} {2015})}\BibitemShut {NoStop}%
\bibitem [{\citenamefont {Hochberg}\ \emph
  {et~al.}(2016{\natexlab{a}})\citenamefont {Hochberg}, \citenamefont {Pyle},
  \citenamefont {Zhao},\ and\ \citenamefont {Zurek}}]{hochberg2016detecting}%
  \BibitemOpen
  \bibfield  {author} {\bibinfo {author} {\bibfnamefont {Y.}~\bibnamefont
  {Hochberg}}, \bibinfo {author} {\bibfnamefont {M.}~\bibnamefont {Pyle}},
  \bibinfo {author} {\bibfnamefont {Y.}~\bibnamefont {Zhao}}, \ and\ \bibinfo
  {author} {\bibfnamefont {K.~M.}\ \bibnamefont {Zurek}},\ }\href {\doibase
  10.1007/JHEP08(2016)057} {\bibfield  {journal} {\bibinfo  {journal} {Journal
  of High Energy Physics}\ }\textbf {\bibinfo {volume} {2016}},\ \bibinfo
  {pages} {57} (\bibinfo {year} {2016}{\natexlab{a}})}\BibitemShut {NoStop}%
\bibitem [{\citenamefont {Hochberg}\ \emph
  {et~al.}(2016{\natexlab{b}})\citenamefont {Hochberg}, \citenamefont {Zhao},\
  and\ \citenamefont {Zurek}}]{hochberg2016superconducting}%
  \BibitemOpen
  \bibfield  {author} {\bibinfo {author} {\bibfnamefont {Y.}~\bibnamefont
  {Hochberg}}, \bibinfo {author} {\bibfnamefont {Y.}~\bibnamefont {Zhao}}, \
  and\ \bibinfo {author} {\bibfnamefont {K.~M.}\ \bibnamefont {Zurek}},\ }\href
  {\doibase 10.1103/PhysRevLett.116.011301} {\bibfield  {journal} {\bibinfo
  {journal} {Physical Review Letters}\ }\textbf {\bibinfo {volume} {116}},\
  \bibinfo {pages} {011301} (\bibinfo {year} {2016}{\natexlab{b}})}\BibitemShut
  {NoStop}%
\bibitem [{\citenamefont {Hochberg}\ \emph
  {et~al.}(2016{\natexlab{c}})\citenamefont {Hochberg}, \citenamefont {Kahn},
  \citenamefont {Lisanti}, \citenamefont {Tully},\ and\ \citenamefont
  {Zurek}}]{hochberg2016directional}%
  \BibitemOpen
  \bibfield  {author} {\bibinfo {author} {\bibfnamefont {Y.}~\bibnamefont
  {Hochberg}}, \bibinfo {author} {\bibfnamefont {Y.}~\bibnamefont {Kahn}},
  \bibinfo {author} {\bibfnamefont {M.}~\bibnamefont {Lisanti}}, \bibinfo
  {author} {\bibfnamefont {C.~G.}\ \bibnamefont {Tully}}, \ and\ \bibinfo
  {author} {\bibfnamefont {K.~M.}\ \bibnamefont {Zurek}},\ }\href
  {http://arxiv.org/abs/1606.08849} {\bibfield  {journal} {\bibinfo  {journal}
  {{arXiv:1606.08849}}\ } (\bibinfo {year} {2016}{\natexlab{c}})}\BibitemShut
  {NoStop}%
\bibitem [{\citenamefont {Essig}\ \emph
  {et~al.}(2012{\natexlab{b}})\citenamefont {Essig}, \citenamefont
  {Manalaysay}, \citenamefont {Mardon}, \citenamefont {Sorensen},\ and\
  \citenamefont {Volansky}}]{Essig:2012yx}%
  \BibitemOpen
  \bibfield  {author} {\bibinfo {author} {\bibfnamefont {R.}~\bibnamefont
  {Essig}}, \bibinfo {author} {\bibfnamefont {A.}~\bibnamefont {Manalaysay}},
  \bibinfo {author} {\bibfnamefont {J.}~\bibnamefont {Mardon}}, \bibinfo
  {author} {\bibfnamefont {P.}~\bibnamefont {Sorensen}}, \ and\ \bibinfo
  {author} {\bibfnamefont {T.}~\bibnamefont {Volansky}},\ }\href {\doibase
  10.1103/PhysRevLett.109.021301} {\bibfield  {journal} {\bibinfo  {journal}
  {Phys.Rev.Lett.}\ }\textbf {\bibinfo {volume} {109}},\ \bibinfo {pages}
  {021301} (\bibinfo {year} {2012}{\natexlab{b}})}\BibitemShut {NoStop}%
\bibitem [{\citenamefont {Essig}\ \emph {et~al.}(2013)\citenamefont {Essig},
  \citenamefont {Kuflik}, \citenamefont {{McDermott}}, \citenamefont
  {Volansky},\ and\ \citenamefont {Zurek}}]{essig2013constraining}%
  \BibitemOpen
  \bibfield  {author} {\bibinfo {author} {\bibfnamefont {R.}~\bibnamefont
  {Essig}}, \bibinfo {author} {\bibfnamefont {E.}~\bibnamefont {Kuflik}},
  \bibinfo {author} {\bibfnamefont {S.~D.}\ \bibnamefont {{McDermott}}},
  \bibinfo {author} {\bibfnamefont {T.}~\bibnamefont {Volansky}}, \ and\
  \bibinfo {author} {\bibfnamefont {K.~M.}\ \bibnamefont {Zurek}},\ }\href
  {\doibase 10.1007/JHEP11(2013)193} {\bibfield  {journal} {\bibinfo  {journal}
  {Journal of High Energy Physics}\ }\textbf {\bibinfo {volume} {2013}},\
  \bibinfo {pages} {1} (\bibinfo {year} {2013})}\BibitemShut {NoStop}%
\bibitem [{\citenamefont {Riedel}(2013)}]{Riedel:2012ur}%
  \BibitemOpen
  \bibfield  {author} {\bibinfo {author} {\bibfnamefont {C.~J.}\ \bibnamefont
  {Riedel}},\ }\href {\doibase 10.1103/PhysRevD.88.116005} {\bibfield
  {journal} {\bibinfo  {journal} {Physical Review D}\ }\textbf {\bibinfo
  {volume} {88}},\ \bibinfo {pages} {116005} (\bibinfo {year}
  {2013})}\BibitemShut {NoStop}%
\bibitem [{\citenamefont {Riedel}(2015)}]{Riedel:2015xx}%
  \BibitemOpen
  \bibfield  {author} {\bibinfo {author} {\bibfnamefont {C.~J.}\ \bibnamefont
  {Riedel}},\ }\href {\doibase 10.1103/PhysRevA.92.010101} {\bibfield
  {journal} {\bibinfo  {journal} {Physical Review A}\ }\textbf {\bibinfo
  {volume} {92}},\ \bibinfo {pages} {010101(R)} (\bibinfo {year}
  {2015})}\BibitemShut {NoStop}%
\bibitem [{\citenamefont {Peskin}\ and\ \citenamefont
  {Schroeder}(1995)}]{Peskin:1995ev}%
  \BibitemOpen
  \bibfield  {author} {\bibinfo {author} {\bibfnamefont {M.~E.}\ \bibnamefont
  {Peskin}}\ and\ \bibinfo {author} {\bibfnamefont {D.~V.}\ \bibnamefont
  {Schroeder}},\ }\href@noop {} {\emph {\bibinfo {title} {{An Introduction to
  quantum field theory}}}}\ (\bibinfo  {publisher} {Addison-Wesley},\ \bibinfo
  {year} {1995})\BibitemShut {NoStop}%
\bibitem [{\citenamefont {Chivukula}\ \emph {et~al.}(1990)\citenamefont
  {Chivukula}, \citenamefont {Cohen}, \citenamefont {Dimopoulos},\ and\
  \citenamefont {Walker}}]{Chivukula:1990xx}%
  \BibitemOpen
  \bibfield  {author} {\bibinfo {author} {\bibfnamefont {R.~S.}\ \bibnamefont
  {Chivukula}}, \bibinfo {author} {\bibfnamefont {A.~G.}\ \bibnamefont
  {Cohen}}, \bibinfo {author} {\bibfnamefont {S.}~\bibnamefont {Dimopoulos}}, \
  and\ \bibinfo {author} {\bibfnamefont {T.~P.}\ \bibnamefont {Walker}},\
  }\href {\doibase 10.1103/PhysRevLett.65.957} {\bibfield  {journal} {\bibinfo
  {journal} {Physical Review Letters}\ }\textbf {\bibinfo {volume} {65}},\
  \bibinfo {pages} {957} (\bibinfo {year} {1990})}\BibitemShut {NoStop}%
\bibitem [{\citenamefont {Hardy}\ and\ \citenamefont
  {Lasenby}(2016)}]{hardy2016stellar}%
  \BibitemOpen
  \bibfield  {author} {\bibinfo {author} {\bibfnamefont {E.}~\bibnamefont
  {Hardy}}\ and\ \bibinfo {author} {\bibfnamefont {R.}~\bibnamefont
  {Lasenby}},\ }\href {http://arxiv.org/abs/1611.05852} {\bibfield  {journal}
  {\bibinfo  {journal} {{arXiv:1611.05852}}\ } (\bibinfo {year}
  {2016})}\BibitemShut {NoStop}%
\bibitem [{\citenamefont {Leeb}\ and\ \citenamefont
  {Schmiedmayer}(1992)}]{Leeb:1992xx}%
  \BibitemOpen
  \bibfield  {author} {\bibinfo {author} {\bibfnamefont {H.}~\bibnamefont
  {Leeb}}\ and\ \bibinfo {author} {\bibfnamefont {J.}~\bibnamefont
  {Schmiedmayer}},\ }\href {\doibase 10.1103/PhysRevLett.68.1472} {\bibfield
  {journal} {\bibinfo  {journal} {Physical Review Letters}\ }\textbf {\bibinfo
  {volume} {68}},\ \bibinfo {pages} {1472} (\bibinfo {year}
  {1992})}\BibitemShut {NoStop}%
\bibitem [{\citenamefont {Nesvizhevsky}\ \emph {et~al.}(2008)\citenamefont
  {Nesvizhevsky}, \citenamefont {Pignol},\ and\ \citenamefont
  {Protasov}}]{Nesvizhevsky:2008xx}%
  \BibitemOpen
  \bibfield  {author} {\bibinfo {author} {\bibfnamefont {V.~V.}\ \bibnamefont
  {Nesvizhevsky}}, \bibinfo {author} {\bibfnamefont {G.}~\bibnamefont
  {Pignol}}, \ and\ \bibinfo {author} {\bibfnamefont {K.~V.}\ \bibnamefont
  {Protasov}},\ }\href {\doibase 10.1103/PhysRevD.77.034020} {\bibfield
  {journal} {\bibinfo  {journal} {Physical Review D}\ }\textbf {\bibinfo
  {volume} {77}},\ \bibinfo {pages} {034020} (\bibinfo {year}
  {2008})}\BibitemShut {NoStop}%
\bibitem [{\citenamefont {Adelberger}\ \emph {et~al.}(2003)\citenamefont
  {Adelberger}, \citenamefont {Heckel},\ and\ \citenamefont
  {Nelson}}]{Adelberger:2003xx}%
  \BibitemOpen
  \bibfield  {author} {\bibinfo {author} {\bibfnamefont {E.}~\bibnamefont
  {Adelberger}}, \bibinfo {author} {\bibfnamefont {B.}~\bibnamefont {Heckel}},
  \ and\ \bibinfo {author} {\bibfnamefont {A.}~\bibnamefont {Nelson}},\ }\href
  {\doibase 10.1146/annurev.nucl.53.041002.110503} {\bibfield  {journal}
  {\bibinfo  {journal} {Annual Review of Nuclear and Particle Science}\
  }\textbf {\bibinfo {volume} {53}},\ \bibinfo {pages} {77} (\bibinfo {year}
  {2003})}\BibitemShut {NoStop}%
\bibitem [{\citenamefont {Raffelt}(1996)}]{Raffelt:1996xx}%
  \BibitemOpen
  \bibfield  {author} {\bibinfo {author} {\bibfnamefont {G.~G.}\ \bibnamefont
  {Raffelt}},\ }\href@noop {} {\emph {\bibinfo {title} {Stars as Laboratories
  for Fundamental Physics: The Astrophysics of Neutrinos, Axions, and Other
  Weakly Interacting Particles}}}\ (\bibinfo  {publisher} {University of
  Chicago Press},\ \bibinfo {year} {1996})\BibitemShut {NoStop}%
\bibitem [{\citenamefont {Raffelt}(1999)}]{Raffelt:1999xx}%
  \BibitemOpen
  \bibfield  {author} {\bibinfo {author} {\bibfnamefont {G.~G.}\ \bibnamefont
  {Raffelt}},\ }\href {\doibase 10.1146/annurev.nucl.49.1.163} {\bibfield
  {journal} {\bibinfo  {journal} {Annual Review of Nuclear and Particle
  Science}\ }\textbf {\bibinfo {volume} {49}},\ \bibinfo {pages} {163}
  (\bibinfo {year} {1999})}\BibitemShut {NoStop}%
\bibitem [{\citenamefont {Rocha}\ \emph {et~al.}(2013)\citenamefont {Rocha},
  \citenamefont {Peter}, \citenamefont {Bullock}, \citenamefont {Kaplinghat},
  \citenamefont {Garrison-Kimmel}, \citenamefont {Onorbe},\ and\ \citenamefont
  {Moustakas}}]{Rocha:2012jg}%
  \BibitemOpen
  \bibfield  {author} {\bibinfo {author} {\bibfnamefont {M.}~\bibnamefont
  {Rocha}}, \bibinfo {author} {\bibfnamefont {A.~H.~G.}\ \bibnamefont {Peter}},
  \bibinfo {author} {\bibfnamefont {J.~S.}\ \bibnamefont {Bullock}}, \bibinfo
  {author} {\bibfnamefont {M.}~\bibnamefont {Kaplinghat}}, \bibinfo {author}
  {\bibfnamefont {S.}~\bibnamefont {Garrison-Kimmel}}, \bibinfo {author}
  {\bibfnamefont {J.}~\bibnamefont {Onorbe}}, \ and\ \bibinfo {author}
  {\bibfnamefont {L.~A.}\ \bibnamefont {Moustakas}},\ }\href {\doibase
  10.1093/mnras/sts514} {\bibfield  {journal} {\bibinfo  {journal} {Mon. Not.
  Roy. Astron. Soc.}\ }\textbf {\bibinfo {volume} {430}},\ \bibinfo {pages}
  {81} (\bibinfo {year} {2013})}\BibitemShut {NoStop}%
\bibitem [{\citenamefont {Fan}\ \emph {et~al.}(2013)\citenamefont {Fan},
  \citenamefont {Katz}, \citenamefont {Randall},\ and\ \citenamefont
  {Reece}}]{Fan:2013yva}%
  \BibitemOpen
  \bibfield  {author} {\bibinfo {author} {\bibfnamefont {J.}~\bibnamefont
  {Fan}}, \bibinfo {author} {\bibfnamefont {A.}~\bibnamefont {Katz}}, \bibinfo
  {author} {\bibfnamefont {L.}~\bibnamefont {Randall}}, \ and\ \bibinfo
  {author} {\bibfnamefont {M.}~\bibnamefont {Reece}},\ }\href {\doibase
  10.1016/j.dark.2013.07.001} {\bibfield  {journal} {\bibinfo  {journal}
  {Phys.Dark Univ.}\ }\textbf {\bibinfo {volume} {2}},\ \bibinfo {pages} {139}
  (\bibinfo {year} {2013})}\BibitemShut {NoStop}%
\bibitem [{\citenamefont {Kramer}\ and\ \citenamefont
  {Randall}(2016)}]{kramer2016updated}%
  \BibitemOpen
  \bibfield  {author} {\bibinfo {author} {\bibfnamefont {E.~D.}\ \bibnamefont
  {Kramer}}\ and\ \bibinfo {author} {\bibfnamefont {L.}~\bibnamefont
  {Randall}},\ }\href {\doibase 10.3847/0004-637X/824/2/116} {\bibfield
  {journal} {\bibinfo  {journal} {The Astrophysical Journal}\ }\textbf
  {\bibinfo {volume} {824}},\ \bibinfo {pages} {116} (\bibinfo {year}
  {2016})}\BibitemShut {NoStop}%
\bibitem [{\citenamefont {Shifman}\ \emph {et~al.}(1978)\citenamefont
  {Shifman}, \citenamefont {Vainshtein},\ and\ \citenamefont
  {Zakharov}}]{Shifman:1978zn}%
  \BibitemOpen
  \bibfield  {author} {\bibinfo {author} {\bibfnamefont {M.~A.}\ \bibnamefont
  {Shifman}}, \bibinfo {author} {\bibfnamefont {A.~I.}\ \bibnamefont
  {Vainshtein}}, \ and\ \bibinfo {author} {\bibfnamefont {V.~I.}\ \bibnamefont
  {Zakharov}},\ }\href {\doibase 10.1016/0370-2693(78)90481-1} {\bibfield
  {journal} {\bibinfo  {journal} {Phys. Lett.}\ }\textbf {\bibinfo {volume}
  {B78}},\ \bibinfo {pages} {443} (\bibinfo {year} {1978})}\BibitemShut
  {NoStop}%
\bibitem [{\citenamefont {Gerlich}\ \emph {et~al.}(2011)\citenamefont
  {Gerlich}, \citenamefont {Eibenberger}, \citenamefont {Tomandl},
  \citenamefont {Nimmrichter}, \citenamefont {Hornberger}, \citenamefont
  {Fagan}, \citenamefont {T\"{u}xen}, \citenamefont {Mayor},\ and\
  \citenamefont {Arndt}}]{Gerlich:2011xx}%
  \BibitemOpen
  \bibfield  {author} {\bibinfo {author} {\bibfnamefont {S.}~\bibnamefont
  {Gerlich}}, \bibinfo {author} {\bibfnamefont {S.}~\bibnamefont
  {Eibenberger}}, \bibinfo {author} {\bibfnamefont {M.}~\bibnamefont
  {Tomandl}}, \bibinfo {author} {\bibfnamefont {S.}~\bibnamefont
  {Nimmrichter}}, \bibinfo {author} {\bibfnamefont {K.}~\bibnamefont
  {Hornberger}}, \bibinfo {author} {\bibfnamefont {P.~J.}\ \bibnamefont
  {Fagan}}, \bibinfo {author} {\bibfnamefont {J.}~\bibnamefont {T\"{u}xen}},
  \bibinfo {author} {\bibfnamefont {M.}~\bibnamefont {Mayor}}, \ and\ \bibinfo
  {author} {\bibfnamefont {M.}~\bibnamefont {Arndt}},\ }\href
  {http://www.nature.com/ncomms/journal/v2/n4/abs/ncomms1263.html} {\bibfield
  {journal} {\bibinfo  {journal} {Nature communications}\ }\textbf {\bibinfo
  {volume} {2}},\ \bibinfo {pages} {263} (\bibinfo {year} {2011})}\BibitemShut
  {NoStop}%
\bibitem [{\citenamefont {Eibenberger}\ \emph {et~al.}(2013)\citenamefont
  {Eibenberger}, \citenamefont {Gerlich}, \citenamefont {Arndt}, \citenamefont
  {Mayor},\ and\ \citenamefont {T\"{u}xen}}]{Eibenberger:2013xx}%
  \BibitemOpen
  \bibfield  {author} {\bibinfo {author} {\bibfnamefont {S.}~\bibnamefont
  {Eibenberger}}, \bibinfo {author} {\bibfnamefont {S.}~\bibnamefont
  {Gerlich}}, \bibinfo {author} {\bibfnamefont {M.}~\bibnamefont {Arndt}},
  \bibinfo {author} {\bibfnamefont {M.}~\bibnamefont {Mayor}}, \ and\ \bibinfo
  {author} {\bibfnamefont {J.}~\bibnamefont {T\"{u}xen}},\ }\href
  {http://pubs.rsc.org/EN/content/articlehtml/2013/cp/c3cp51500a} {\bibfield
  {journal} {\bibinfo  {journal} {Physical Chemistry Chemical Physics}\
  }\textbf {\bibinfo {volume} {15}},\ \bibinfo {pages} {14696} (\bibinfo {year}
  {2013})}\BibitemShut {NoStop}%
\bibitem [{\citenamefont {Nimmrichter}\ \emph
  {et~al.}(2011{\natexlab{a}})\citenamefont {Nimmrichter}, \citenamefont
  {Haslinger}, \citenamefont {Hornberger},\ and\ \citenamefont
  {Arndt}}]{Nimmrichter:2011x1}%
  \BibitemOpen
  \bibfield  {author} {\bibinfo {author} {\bibfnamefont {S.}~\bibnamefont
  {Nimmrichter}}, \bibinfo {author} {\bibfnamefont {P.}~\bibnamefont
  {Haslinger}}, \bibinfo {author} {\bibfnamefont {K.}~\bibnamefont
  {Hornberger}}, \ and\ \bibinfo {author} {\bibfnamefont {M.}~\bibnamefont
  {Arndt}},\ }\href {\doibase 10.1088/1367-2630/13/7/075002} {\bibfield
  {journal} {\bibinfo  {journal} {New Journal of Physics}\ }\textbf {\bibinfo
  {volume} {13}},\ \bibinfo {pages} {075002} (\bibinfo {year}
  {2011}{\natexlab{a}})}\BibitemShut {NoStop}%
\bibitem [{\citenamefont {Nimmrichter}\ \emph
  {et~al.}(2011{\natexlab{b}})\citenamefont {Nimmrichter}, \citenamefont
  {Hornberger}, \citenamefont {Haslinger},\ and\ \citenamefont
  {Arndt}}]{Nimmrichter:2011x2}%
  \BibitemOpen
  \bibfield  {author} {\bibinfo {author} {\bibfnamefont {S.}~\bibnamefont
  {Nimmrichter}}, \bibinfo {author} {\bibfnamefont {K.}~\bibnamefont
  {Hornberger}}, \bibinfo {author} {\bibfnamefont {P.}~\bibnamefont
  {Haslinger}}, \ and\ \bibinfo {author} {\bibfnamefont {M.}~\bibnamefont
  {Arndt}},\ }\href {\doibase 10.1103/PhysRevA.83.043621} {\bibfield  {journal}
  {\bibinfo  {journal} {Physical Review A}\ }\textbf {\bibinfo {volume} {83}},\
  \bibinfo {pages} {043621} (\bibinfo {year} {2011}{\natexlab{b}})}\BibitemShut
  {NoStop}%
\bibitem [{\citenamefont {Arndt}\ and\ \citenamefont
  {Hornberger}(2014)}]{Arndt:2014xx}%
  \BibitemOpen
  \bibfield  {author} {\bibinfo {author} {\bibfnamefont {M.}~\bibnamefont
  {Arndt}}\ and\ \bibinfo {author} {\bibfnamefont {K.}~\bibnamefont
  {Hornberger}},\ }\href {\doibase 10.1038/nphys2863} {\bibfield  {journal}
  {\bibinfo  {journal} {Nature Physics}\ }\textbf {\bibinfo {volume} {10}},\
  \bibinfo {pages} {271} (\bibinfo {year} {2014})}\BibitemShut {NoStop}%
\bibitem [{\citenamefont {Bateman}\ \emph {et~al.}(2014)\citenamefont
  {Bateman}, \citenamefont {Nimmrichter}, \citenamefont {Hornberger},\ and\
  \citenamefont {Ulbricht}}]{Bateman:2014xx}%
  \BibitemOpen
  \bibfield  {author} {\bibinfo {author} {\bibfnamefont {J.}~\bibnamefont
  {Bateman}}, \bibinfo {author} {\bibfnamefont {S.}~\bibnamefont
  {Nimmrichter}}, \bibinfo {author} {\bibfnamefont {K.}~\bibnamefont
  {Hornberger}}, \ and\ \bibinfo {author} {\bibfnamefont {H.}~\bibnamefont
  {Ulbricht}},\ }\href {\doibase 10.1038/ncomms5788} {\bibfield  {journal}
  {\bibinfo  {journal} {Nature Communications}\ }\textbf {\bibinfo {volume}
  {5}},\ \bibinfo {pages} {4788} (\bibinfo {year} {2014})}\BibitemShut
  {NoStop}%
\bibitem [{\citenamefont {Geraci}\ and\ \citenamefont
  {Goldman}(2015)}]{geraci2015sensing}%
  \BibitemOpen
  \bibfield  {author} {\bibinfo {author} {\bibfnamefont {A.}~\bibnamefont
  {Geraci}}\ and\ \bibinfo {author} {\bibfnamefont {H.}~\bibnamefont
  {Goldman}},\ }\href {\doibase 10.1103/PhysRevD.92.062002} {\bibfield
  {journal} {\bibinfo  {journal} {Physical Review D}\ }\textbf {\bibinfo
  {volume} {92}},\ \bibinfo {pages} {062002} (\bibinfo {year}
  {2015})}\BibitemShut {NoStop}%
\bibitem [{\citenamefont {Wan}\ \emph {et~al.}(2016)\citenamefont {Wan},
  \citenamefont {Scala}, \citenamefont {Morley}, \citenamefont {Rahman},
  \citenamefont {Ulbricht}, \citenamefont {Bateman}, \citenamefont {Barker},
  \citenamefont {Bose},\ and\ \citenamefont {Kim}}]{wan2016free}%
  \BibitemOpen
  \bibfield  {author} {\bibinfo {author} {\bibfnamefont {C.}~\bibnamefont
  {Wan}}, \bibinfo {author} {\bibfnamefont {M.}~\bibnamefont {Scala}}, \bibinfo
  {author} {\bibfnamefont {G.~W.}\ \bibnamefont {Morley}}, \bibinfo {author}
  {\bibfnamefont {A.~A.}\ \bibnamefont {Rahman}}, \bibinfo {author}
  {\bibfnamefont {H.}~\bibnamefont {Ulbricht}}, \bibinfo {author}
  {\bibfnamefont {J.}~\bibnamefont {Bateman}}, \bibinfo {author} {\bibfnamefont
  {P.~F.}\ \bibnamefont {Barker}}, \bibinfo {author} {\bibfnamefont
  {S.}~\bibnamefont {Bose}}, \ and\ \bibinfo {author} {\bibfnamefont {M.~S.}\
  \bibnamefont {Kim}},\ }\href {\doibase 10.1103/PhysRevLett.117.143003}
  {\bibfield  {journal} {\bibinfo  {journal} {Physical Review Letters}\
  }\textbf {\bibinfo {volume} {117}},\ \bibinfo {pages} {143003} (\bibinfo
  {year} {2016})}\BibitemShut {NoStop}%
\bibitem [{\citenamefont {Kaltenbaek}()}]{Kaltenbaek:2012x1}%
  \BibitemOpen
  \bibfield  {author} {\bibinfo {author} {\bibfnamefont {R.}~\bibnamefont
  {Kaltenbaek}},\ }\href@noop {} {\enquote {\bibinfo {title} {Macroscopic
  quantum experiments in space using massive mechanical resonators {(MQES)} -
  technical note \#3},}\ }\bibinfo {note} {Tech. Rep., Study conducted under
  contract with the European Space Agency (2012)}\BibitemShut {NoStop}%
\bibitem [{\citenamefont {\relax R.~Kaltenbaek et~al.
  (MAQRO~Collaboration)}(2016)}]{Kaltenbaek:2015x1}%
  \BibitemOpen
  \bibfield  {author} {\bibinfo {author} {\bibnamefont {\relax R.~Kaltenbaek
  et~al. (MAQRO~Collaboration)}},\ }\href {\doibase
  10.1140/epjqt/s40507-016-0043-7} {\bibfield  {journal} {\bibinfo  {journal}
  {{EPJ} Quantum Technology}\ }\textbf {\bibinfo {volume} {3}},\ \bibinfo
  {pages} {5} (\bibinfo {year} {2016})}\BibitemShut {NoStop}%
\bibitem [{\citenamefont {Kaltenbaek}(2015)}]{Kaltenbaek:2015x2}%
  \BibitemOpen
  \bibfield  {author} {\bibinfo {author} {\bibfnamefont {R.}~\bibnamefont
  {Kaltenbaek}},\ }\href@noop {} {\enquote {\bibinfo {title} {Testing quantum
  physics in space using high-mass matter-wave interferometry},}\ } (\bibinfo
  {year} {2015}),\ \bibinfo {note} {{Proceedings of 50th Rencontres de Moriond;
  Gravitation: 100 Years after GR, pp. 141 [arXiv:1508.07796]}}\BibitemShut
  {NoStop}%
\bibitem [{\citenamefont {Pino}\ \emph {et~al.}(2016)\citenamefont {Pino},
  \citenamefont {{Prat-Camps}}, \citenamefont {Sinha}, \citenamefont
  {Venkatesh},\ and\ \citenamefont {{Romero-Isart}}}]{pino2016quantum}%
  \BibitemOpen
  \bibfield  {author} {\bibinfo {author} {\bibfnamefont {H.}~\bibnamefont
  {Pino}}, \bibinfo {author} {\bibfnamefont {J.}~\bibnamefont {{Prat-Camps}}},
  \bibinfo {author} {\bibfnamefont {K.}~\bibnamefont {Sinha}}, \bibinfo
  {author} {\bibfnamefont {B.~P.}\ \bibnamefont {Venkatesh}}, \ and\ \bibinfo
  {author} {\bibfnamefont {O.}~\bibnamefont {{Romero-Isart}}},\ }\href
  {http://arxiv.org/abs/1603.01553} {\bibfield  {journal} {\bibinfo  {journal}
  {{arXiv:1603.01553}}\ } (\bibinfo {year} {2016})}\BibitemShut {NoStop}%
\bibitem [{\citenamefont {Haslinger}\ \emph {et~al.}(2013)\citenamefont
  {Haslinger}, \citenamefont {D\"{o}rre}, \citenamefont {Geyer}, \citenamefont
  {Rodewald}, \citenamefont {Nimmrichter},\ and\ \citenamefont
  {Arndt}}]{Haslinger:2013xx}%
  \BibitemOpen
  \bibfield  {author} {\bibinfo {author} {\bibfnamefont {P.}~\bibnamefont
  {Haslinger}}, \bibinfo {author} {\bibfnamefont {N.}~\bibnamefont
  {D\"{o}rre}}, \bibinfo {author} {\bibfnamefont {P.}~\bibnamefont {Geyer}},
  \bibinfo {author} {\bibfnamefont {J.}~\bibnamefont {Rodewald}}, \bibinfo
  {author} {\bibfnamefont {S.}~\bibnamefont {Nimmrichter}}, \ and\ \bibinfo
  {author} {\bibfnamefont {M.}~\bibnamefont {Arndt}},\ }\href {\doibase
  10.1038/nphys2542} {\bibfield  {journal} {\bibinfo  {journal} {Nature
  Physics}\ }\textbf {\bibinfo {volume} {9}},\ \bibinfo {pages} {144} (\bibinfo
  {year} {2013})}\BibitemShut {NoStop}%
\bibitem [{\citenamefont {Romero-Isart}(2016)}]{romero-isart2016coherent}%
  \BibitemOpen
  \bibfield  {author} {\bibinfo {author} {\bibfnamefont {O.}~\bibnamefont
  {Romero-Isart}},\ }\href {http://arxiv.org/abs/1612.04290} {\bibfield
  {journal} {\bibinfo  {journal} {{arXiv:1612.04290}}\ } (\bibinfo {year}
  {2016})}\BibitemShut {NoStop}%
\bibitem [{\citenamefont {Gieseler}\ \emph {et~al.}(2012)\citenamefont
  {Gieseler}, \citenamefont {Deutsch}, \citenamefont {Quidant},\ and\
  \citenamefont {Novotny}}]{Gieseler:2012xx}%
  \BibitemOpen
  \bibfield  {author} {\bibinfo {author} {\bibfnamefont {J.}~\bibnamefont
  {Gieseler}}, \bibinfo {author} {\bibfnamefont {B.}~\bibnamefont {Deutsch}},
  \bibinfo {author} {\bibfnamefont {R.}~\bibnamefont {Quidant}}, \ and\
  \bibinfo {author} {\bibfnamefont {L.}~\bibnamefont {Novotny}},\ }\href
  {\doibase 10.1103/PhysRevLett.109.103603} {\bibfield  {journal} {\bibinfo
  {journal} {Physical Review Letters}\ }\textbf {\bibinfo {volume} {109}},\
  \bibinfo {pages} {103603} (\bibinfo {year} {2012})}\BibitemShut {NoStop}%
\bibitem [{\citenamefont {Chang}\ \emph {et~al.}(2010)\citenamefont {Chang},
  \citenamefont {Regal}, \citenamefont {Papp}, \citenamefont {Wilson},
  \citenamefont {Ye}, \citenamefont {Painter}, \citenamefont {Kimble},\ and\
  \citenamefont {Zoller}}]{chang2010cavity}%
  \BibitemOpen
  \bibfield  {author} {\bibinfo {author} {\bibfnamefont {D.~E.}\ \bibnamefont
  {Chang}}, \bibinfo {author} {\bibfnamefont {C.~A.}\ \bibnamefont {Regal}},
  \bibinfo {author} {\bibfnamefont {S.~B.}\ \bibnamefont {Papp}}, \bibinfo
  {author} {\bibfnamefont {D.~J.}\ \bibnamefont {Wilson}}, \bibinfo {author}
  {\bibfnamefont {J.}~\bibnamefont {Ye}}, \bibinfo {author} {\bibfnamefont
  {O.}~\bibnamefont {Painter}}, \bibinfo {author} {\bibfnamefont {H.~J.}\
  \bibnamefont {Kimble}}, \ and\ \bibinfo {author} {\bibfnamefont
  {P.}~\bibnamefont {Zoller}},\ }\href {\doibase 10.1073/pnas.0912969107}
  {\bibfield  {journal} {\bibinfo  {journal} {Proceedings of the National
  Academy of Sciences}\ }\textbf {\bibinfo {volume} {107}},\ \bibinfo {pages}
  {1005} (\bibinfo {year} {2010})}\BibitemShut {NoStop}%
\bibitem [{\citenamefont {Li}\ \emph {et~al.}(2011)\citenamefont {Li},
  \citenamefont {Kheifets},\ and\ \citenamefont {Raizen}}]{Li:2011xx}%
  \BibitemOpen
  \bibfield  {author} {\bibinfo {author} {\bibfnamefont {T.}~\bibnamefont
  {Li}}, \bibinfo {author} {\bibfnamefont {S.}~\bibnamefont {Kheifets}}, \ and\
  \bibinfo {author} {\bibfnamefont {M.~G.}\ \bibnamefont {Raizen}},\ }\href
  {\doibase 10.1038/nphys1952} {\bibfield  {journal} {\bibinfo  {journal}
  {Nature Physics}\ }\textbf {\bibinfo {volume} {7}},\ \bibinfo {pages} {527}
  (\bibinfo {year} {2011})}\BibitemShut {NoStop}%
\bibitem [{\citenamefont {Kiesel}\ \emph {et~al.}(2013)\citenamefont {Kiesel},
  \citenamefont {Blaser}, \citenamefont {Deli\'{c}}, \citenamefont {Grass},
  \citenamefont {Kaltenbaek},\ and\ \citenamefont
  {Aspelmeyer}}]{kiesel2013cavity}%
  \BibitemOpen
  \bibfield  {author} {\bibinfo {author} {\bibfnamefont {N.}~\bibnamefont
  {Kiesel}}, \bibinfo {author} {\bibfnamefont {F.}~\bibnamefont {Blaser}},
  \bibinfo {author} {\bibfnamefont {U.}~\bibnamefont {Deli\'{c}}}, \bibinfo
  {author} {\bibfnamefont {D.}~\bibnamefont {Grass}}, \bibinfo {author}
  {\bibfnamefont {R.}~\bibnamefont {Kaltenbaek}}, \ and\ \bibinfo {author}
  {\bibfnamefont {M.}~\bibnamefont {Aspelmeyer}},\ }\href {\doibase
  10.1073/pnas.1309167110} {\bibfield  {journal} {\bibinfo  {journal}
  {Proceedings of the National Academy of Sciences}\ }\textbf {\bibinfo
  {volume} {110}},\ \bibinfo {pages} {14180} (\bibinfo {year}
  {2013})}\BibitemShut {NoStop}%
\bibitem [{\citenamefont {Fonseca}\ \emph {et~al.}(2016)\citenamefont
  {Fonseca}, \citenamefont {Aranas}, \citenamefont {Millen}, \citenamefont
  {Monteiro},\ and\ \citenamefont {Barker}}]{fonseca2016nonlinear}%
  \BibitemOpen
  \bibfield  {author} {\bibinfo {author} {\bibfnamefont {P.~Z.~G.}\
  \bibnamefont {Fonseca}}, \bibinfo {author} {\bibfnamefont {E.~B.}\
  \bibnamefont {Aranas}}, \bibinfo {author} {\bibfnamefont {J.}~\bibnamefont
  {Millen}}, \bibinfo {author} {\bibfnamefont {T.~S.}\ \bibnamefont
  {Monteiro}}, \ and\ \bibinfo {author} {\bibfnamefont {P.~F.}\ \bibnamefont
  {Barker}},\ }\href {\doibase 10.1103/PhysRevLett.117.173602} {\bibfield
  {journal} {\bibinfo  {journal} {Physical Review Letters}\ }\textbf {\bibinfo
  {volume} {117}},\ \bibinfo {pages} {173602} (\bibinfo {year}
  {2016})}\BibitemShut {NoStop}%
\bibitem [{\citenamefont {Millen}\ \emph {et~al.}(2015)\citenamefont {Millen},
  \citenamefont {Fonseca}, \citenamefont {Mavrogordatos}, \citenamefont
  {Monteiro},\ and\ \citenamefont {Barker}}]{millen2015cavity}%
  \BibitemOpen
  \bibfield  {author} {\bibinfo {author} {\bibfnamefont {J.}~\bibnamefont
  {Millen}}, \bibinfo {author} {\bibfnamefont {P.~Z.~G.}\ \bibnamefont
  {Fonseca}}, \bibinfo {author} {\bibfnamefont {T.}~\bibnamefont
  {Mavrogordatos}}, \bibinfo {author} {\bibfnamefont {T.~S.}\ \bibnamefont
  {Monteiro}}, \ and\ \bibinfo {author} {\bibfnamefont {P.~F.}\ \bibnamefont
  {Barker}},\ }\href {\doibase 10.1103/PhysRevLett.114.123602} {\bibfield
  {journal} {\bibinfo  {journal} {Physical Review Letters}\ }\textbf {\bibinfo
  {volume} {114}},\ \bibinfo {pages} {123602} (\bibinfo {year}
  {2015})}\BibitemShut {NoStop}%
\bibitem [{\citenamefont {Vovrosh}\ \emph {et~al.}(2017)\citenamefont
  {Vovrosh}, \citenamefont {Rashid}, \citenamefont {Hempston}, \citenamefont
  {Bateman}, \citenamefont {Paternostro},\ and\ \citenamefont
  {Ulbricht}}]{vovrosh2016controlling}%
  \BibitemOpen
  \bibfield  {author} {\bibinfo {author} {\bibfnamefont {J.}~\bibnamefont
  {Vovrosh}}, \bibinfo {author} {\bibfnamefont {M.}~\bibnamefont {Rashid}},
  \bibinfo {author} {\bibfnamefont {D.}~\bibnamefont {Hempston}}, \bibinfo
  {author} {\bibfnamefont {J.}~\bibnamefont {Bateman}}, \bibinfo {author}
  {\bibfnamefont {M.}~\bibnamefont {Paternostro}}, \ and\ \bibinfo {author}
  {\bibfnamefont {H.}~\bibnamefont {Ulbricht}},\ }\href {\doibase
  10.1364/JOSAB.34.001421} {\bibfield  {journal} {\bibinfo  {journal} {Journal
  of the Optical Society of America B}\ }\textbf {\bibinfo {volume} {34}},\
  \bibinfo {pages} {1421–1428} (\bibinfo {year} {2017})}\BibitemShut
  {NoStop}%
\bibitem [{\citenamefont {O'Connell}\ \emph {et~al.}(2010)\citenamefont
  {O'Connell}, \citenamefont {Hofheinz}, \citenamefont {Ansmann}, \citenamefont
  {Bialczak}, \citenamefont {Lenander}, \citenamefont {Lucero}, \citenamefont
  {Neeley}, \citenamefont {Sank}, \citenamefont {Wang}, \citenamefont {Weides},
  \citenamefont {Wenner}, \citenamefont {Martinis},\ and\ \citenamefont
  {Cleland}}]{oconnell2010quantum}%
  \BibitemOpen
  \bibfield  {author} {\bibinfo {author} {\bibfnamefont {A.}~\bibnamefont
  {O'Connell}}, \bibinfo {author} {\bibfnamefont {M.}~\bibnamefont {Hofheinz}},
  \bibinfo {author} {\bibfnamefont {M.}~\bibnamefont {Ansmann}}, \bibinfo
  {author} {\bibfnamefont {R.}~\bibnamefont {Bialczak}}, \bibinfo {author}
  {\bibfnamefont {M.}~\bibnamefont {Lenander}}, \bibinfo {author}
  {\bibfnamefont {E.}~\bibnamefont {Lucero}}, \bibinfo {author} {\bibfnamefont
  {M.}~\bibnamefont {Neeley}}, \bibinfo {author} {\bibfnamefont
  {D.}~\bibnamefont {Sank}}, \bibinfo {author} {\bibfnamefont {H.}~\bibnamefont
  {Wang}}, \bibinfo {author} {\bibfnamefont {M.}~\bibnamefont {Weides}},
  \bibinfo {author} {\bibfnamefont {J.}~\bibnamefont {Wenner}}, \bibinfo
  {author} {\bibfnamefont {J.~M.}\ \bibnamefont {Martinis}}, \ and\ \bibinfo
  {author} {\bibfnamefont {A.~N.}\ \bibnamefont {Cleland}},\ }\href@noop {}
  {\bibfield  {journal} {\bibinfo  {journal} {Nature}\ }\textbf {\bibinfo
  {volume} {464}},\ \bibinfo {pages} {697} (\bibinfo {year}
  {2010})}\BibitemShut {NoStop}%
\bibitem [{\citenamefont {Chan}\ \emph {et~al.}(2011)\citenamefont {Chan},
  \citenamefont {Alegre}, \citenamefont {Safavi-Naeini}, \citenamefont {Hill},
  \citenamefont {Krause}, \citenamefont {Gröblacher}, \citenamefont
  {Aspelmeyer},\ and\ \citenamefont {Painter}}]{chan2011laser}%
  \BibitemOpen
  \bibfield  {author} {\bibinfo {author} {\bibfnamefont {J.}~\bibnamefont
  {Chan}}, \bibinfo {author} {\bibfnamefont {T.~P.}\ \bibnamefont {Alegre}},
  \bibinfo {author} {\bibfnamefont {A.~H.}\ \bibnamefont {Safavi-Naeini}},
  \bibinfo {author} {\bibfnamefont {J.~T.}\ \bibnamefont {Hill}}, \bibinfo
  {author} {\bibfnamefont {A.}~\bibnamefont {Krause}}, \bibinfo {author}
  {\bibfnamefont {S.}~\bibnamefont {Gröblacher}}, \bibinfo {author}
  {\bibfnamefont {M.}~\bibnamefont {Aspelmeyer}}, \ and\ \bibinfo {author}
  {\bibfnamefont {O.}~\bibnamefont {Painter}},\ }\href@noop {} {\bibfield
  {journal} {\bibinfo  {journal} {Nature}\ }\textbf {\bibinfo {volume} {478}},\
  \bibinfo {pages} {89} (\bibinfo {year} {2011})}\BibitemShut {NoStop}%
\bibitem [{\citenamefont {Teufel}\ \emph {et~al.}(2011)\citenamefont {Teufel},
  \citenamefont {Donner}, \citenamefont {Li}, \citenamefont {Harlow},
  \citenamefont {Allman}, \citenamefont {Cicak}, \citenamefont {Sirois},
  \citenamefont {Whittaker}, \citenamefont {Lehnert},\ and\ \citenamefont
  {Simmonds}}]{teufel2011sideband}%
  \BibitemOpen
  \bibfield  {author} {\bibinfo {author} {\bibfnamefont {J.~D.}\ \bibnamefont
  {Teufel}}, \bibinfo {author} {\bibfnamefont {T.}~\bibnamefont {Donner}},
  \bibinfo {author} {\bibfnamefont {D.}~\bibnamefont {Li}}, \bibinfo {author}
  {\bibfnamefont {J.~W.}\ \bibnamefont {Harlow}}, \bibinfo {author}
  {\bibfnamefont {M.~S.}\ \bibnamefont {Allman}}, \bibinfo {author}
  {\bibfnamefont {K.}~\bibnamefont {Cicak}}, \bibinfo {author} {\bibfnamefont
  {A.~J.}\ \bibnamefont {Sirois}}, \bibinfo {author} {\bibfnamefont {J.~D.}\
  \bibnamefont {Whittaker}}, \bibinfo {author} {\bibfnamefont {K.~W.}\
  \bibnamefont {Lehnert}}, \ and\ \bibinfo {author} {\bibfnamefont {R.~W.}\
  \bibnamefont {Simmonds}},\ }\href {\doibase 10.1038/nature10261} {\bibfield
  {journal} {\bibinfo  {journal} {Nature}\ }\textbf {\bibinfo {volume} {475}},\
  \bibinfo {pages} {359} (\bibinfo {year} {2011})}\BibitemShut {NoStop}%
\bibitem [{\citenamefont {Palomaki}\ \emph {et~al.}(2013)\citenamefont
  {Palomaki}, \citenamefont {Teufel}, \citenamefont {Simmonds},\ and\
  \citenamefont {Lehnert}}]{palomaki2013entangling}%
  \BibitemOpen
  \bibfield  {author} {\bibinfo {author} {\bibfnamefont {T.~A.}\ \bibnamefont
  {Palomaki}}, \bibinfo {author} {\bibfnamefont {J.~D.}\ \bibnamefont
  {Teufel}}, \bibinfo {author} {\bibfnamefont {R.~W.}\ \bibnamefont
  {Simmonds}}, \ and\ \bibinfo {author} {\bibfnamefont {K.~W.}\ \bibnamefont
  {Lehnert}},\ }\href {\doibase 10.1126/science.1244563} {\bibfield  {journal}
  {\bibinfo  {journal} {Science}\ }\textbf {\bibinfo {volume} {342}},\ \bibinfo
  {pages} {710} (\bibinfo {year} {2013})}\BibitemShut {NoStop}%
\bibitem [{\citenamefont {Clark}\ \emph {et~al.}(2017)\citenamefont {Clark},
  \citenamefont {Lecocq}, \citenamefont {Simmonds}, \citenamefont {Aumentado},\
  and\ \citenamefont {Teufel}}]{clark2017sideband}%
  \BibitemOpen
  \bibfield  {author} {\bibinfo {author} {\bibfnamefont {J.~B.}\ \bibnamefont
  {Clark}}, \bibinfo {author} {\bibfnamefont {F.}~\bibnamefont {Lecocq}},
  \bibinfo {author} {\bibfnamefont {R.~W.}\ \bibnamefont {Simmonds}}, \bibinfo
  {author} {\bibfnamefont {J.}~\bibnamefont {Aumentado}}, \ and\ \bibinfo
  {author} {\bibfnamefont {J.~D.}\ \bibnamefont {Teufel}},\ }\href {\doibase
  10.1038/nature20604} {\bibfield  {journal} {\bibinfo  {journal} {Nature}\
  }\textbf {\bibinfo {volume} {541}},\ \bibinfo {pages} {191} (\bibinfo {year}
  {2017})}\BibitemShut {NoStop}%
\bibitem [{\citenamefont {Aspelmeyer}\ \emph
  {et~al.}(2014{\natexlab{a}})\citenamefont {Aspelmeyer}, \citenamefont
  {Kippenberg},\ and\ \citenamefont {Marquardt}}]{aspelmeyer2014cavity-a}%
  \BibitemOpen
  \bibfield  {author} {\bibinfo {author} {\bibfnamefont {M.}~\bibnamefont
  {Aspelmeyer}}, \bibinfo {author} {\bibfnamefont {T.~J.}\ \bibnamefont
  {Kippenberg}}, \ and\ \bibinfo {author} {\bibfnamefont {F.}~\bibnamefont
  {Marquardt}},\ }\href {\doibase 10.1103/RevModPhys.86.1391} {\bibfield
  {journal} {\bibinfo  {journal} {Reviews of Modern Physics}\ }\textbf
  {\bibinfo {volume} {86}},\ \bibinfo {pages} {1391} (\bibinfo {year}
  {2014}{\natexlab{a}})}\BibitemShut {NoStop}%
\bibitem [{\citenamefont {Aspelmeyer}\ \emph
  {et~al.}(2014{\natexlab{b}})\citenamefont {Aspelmeyer}, \citenamefont
  {Kippenberg},\ and\ \citenamefont {Marquardt}}]{aspelmeyer2014cavity-b}%
  \BibitemOpen
  \bibinfo {editor} {\bibfnamefont {M.}~\bibnamefont {Aspelmeyer}}, \bibinfo
  {editor} {\bibfnamefont {T.~J.}\ \bibnamefont {Kippenberg}}, \ and\ \bibinfo
  {editor} {\bibfnamefont {F.}~\bibnamefont {Marquardt}},\ eds.,\ \href@noop {}
  {\emph {\bibinfo {title} {Cavity Optomechanics - Nano- and Micromechanical
  Resonators Interacting with Light}}}\ (\bibinfo  {publisher} {Springer Berlin
  Heidelberg},\ \bibinfo {address} {Berlin, Heidelberg},\ \bibinfo {year}
  {2014})\BibitemShut {NoStop}%
\bibitem [{\citenamefont {Corbitt}\ \emph {et~al.}(2007)\citenamefont
  {Corbitt}, \citenamefont {Chen}, \citenamefont {Innerhofer}, \citenamefont
  {{M\"{u}ller-Ebhardt}}, \citenamefont {Ottaway}, \citenamefont {Rehbein},
  \citenamefont {Sigg}, \citenamefont {Whitcomb}, \citenamefont {Wipf},\ and\
  \citenamefont {Mavalvala}}]{corbitt2007all-optical}%
  \BibitemOpen
  \bibfield  {author} {\bibinfo {author} {\bibfnamefont {T.}~\bibnamefont
  {Corbitt}}, \bibinfo {author} {\bibfnamefont {Y.}~\bibnamefont {Chen}},
  \bibinfo {author} {\bibfnamefont {E.}~\bibnamefont {Innerhofer}}, \bibinfo
  {author} {\bibfnamefont {H.}~\bibnamefont {{M\"{u}ller-Ebhardt}}}, \bibinfo
  {author} {\bibfnamefont {D.}~\bibnamefont {Ottaway}}, \bibinfo {author}
  {\bibfnamefont {H.}~\bibnamefont {Rehbein}}, \bibinfo {author} {\bibfnamefont
  {D.}~\bibnamefont {Sigg}}, \bibinfo {author} {\bibfnamefont {S.}~\bibnamefont
  {Whitcomb}}, \bibinfo {author} {\bibfnamefont {C.}~\bibnamefont {Wipf}}, \
  and\ \bibinfo {author} {\bibfnamefont {N.}~\bibnamefont {Mavalvala}},\ }\href
  {\doibase 10.1103/PhysRevLett.98.150802} {\bibfield  {journal} {\bibinfo
  {journal} {Physical Review Letters}\ }\textbf {\bibinfo {volume} {98}},\
  \bibinfo {pages} {150802} (\bibinfo {year} {2007})}\BibitemShut {NoStop}%
\bibitem [{\citenamefont {{\relax B.~Abbott et al. ({LIGO} {Scientific}
  Collaboration)}}(2009)}]{abbott2009observation}%
  \BibitemOpen
  \bibfield  {author} {\bibinfo {author} {\bibnamefont {{\relax B.~Abbott et
  al. ({LIGO} {Scientific} Collaboration)}}},\ }\href {\doibase
  10.1088/1367-2630/11/7/073032} {\bibfield  {journal} {\bibinfo  {journal}
  {New Journal of Physics}\ }\textbf {\bibinfo {volume} {11}},\ \bibinfo
  {pages} {073032} (\bibinfo {year} {2009})}\BibitemShut {NoStop}%
\bibitem [{\citenamefont {Cohadon}\ \emph {et~al.}(2014)\citenamefont
  {Cohadon}, \citenamefont {Schnabel},\ and\ \citenamefont
  {Aspelmeyer}}]{cohadon2014suspended}%
  \BibitemOpen
  \bibfield  {author} {\bibinfo {author} {\bibfnamefont {P.}~\bibnamefont
  {Cohadon}}, \bibinfo {author} {\bibfnamefont {R.}~\bibnamefont {Schnabel}}, \
  and\ \bibinfo {author} {\bibfnamefont {M.}~\bibnamefont {Aspelmeyer}},\ }in\
  \href@noop {} {\emph {\bibinfo {booktitle} {Cavity Optomechanics}}},\
  \bibinfo {series and number} {Quantum Science and Technology},\ \bibinfo
  {editor} {edited by\ \bibinfo {editor} {\bibfnamefont {M.}~\bibnamefont
  {Aspelmeyer}}, \bibinfo {editor} {\bibfnamefont {T.~J.}\ \bibnamefont
  {Kippenberg}}, \ and\ \bibinfo {editor} {\bibfnamefont {F.}~\bibnamefont
  {Marquardt}}}\ (\bibinfo  {publisher} {Springer Berlin Heidelberg},\ \bibinfo
  {year} {2014})\ p.~\bibinfo {pages} {57}\BibitemShut {NoStop}%
\bibitem [{\citenamefont {Weaver}\ \emph {et~al.}(2016)\citenamefont {Weaver},
  \citenamefont {Pepper}, \citenamefont {Luna}, \citenamefont {Buters},
  \citenamefont {Eerkens}, \citenamefont {Welker}, \citenamefont {Perock},
  \citenamefont {Heeck}, \citenamefont {Man},\ and\ \citenamefont
  {Bouwmeester}}]{weaver2016nested}%
  \BibitemOpen
  \bibfield  {author} {\bibinfo {author} {\bibfnamefont {M.~J.}\ \bibnamefont
  {Weaver}}, \bibinfo {author} {\bibfnamefont {B.}~\bibnamefont {Pepper}},
  \bibinfo {author} {\bibfnamefont {F.}~\bibnamefont {Luna}}, \bibinfo {author}
  {\bibfnamefont {F.~M.}\ \bibnamefont {Buters}}, \bibinfo {author}
  {\bibfnamefont {H.~J.}\ \bibnamefont {Eerkens}}, \bibinfo {author}
  {\bibfnamefont {G.}~\bibnamefont {Welker}}, \bibinfo {author} {\bibfnamefont
  {B.}~\bibnamefont {Perock}}, \bibinfo {author} {\bibfnamefont
  {K.}~\bibnamefont {Heeck}}, \bibinfo {author} {\bibfnamefont {S.~d.}\
  \bibnamefont {Man}}, \ and\ \bibinfo {author} {\bibfnamefont
  {D.}~\bibnamefont {Bouwmeester}},\ }\href {\doibase 10.1063/1.4939828}
  {\bibfield  {journal} {\bibinfo  {journal} {Applied Physics Letters}\
  }\textbf {\bibinfo {volume} {108}},\ \bibinfo {pages} {033501} (\bibinfo
  {year} {2016})}\BibitemShut {NoStop}%
\bibitem [{\citenamefont {M\"{u}ller}\ \emph {et~al.}(2015)\citenamefont
  {M\"{u}ller}, \citenamefont {Reinhardt},\ and\ \citenamefont
  {Sankey}}]{mueller2015enhanced}%
  \BibitemOpen
  \bibfield  {author} {\bibinfo {author} {\bibfnamefont {T.}~\bibnamefont
  {M\"{u}ller}}, \bibinfo {author} {\bibfnamefont {C.}~\bibnamefont
  {Reinhardt}}, \ and\ \bibinfo {author} {\bibfnamefont {J.~C.}\ \bibnamefont
  {Sankey}},\ }\href {\doibase 10.1103/PhysRevA.91.053849} {\bibfield
  {journal} {\bibinfo  {journal} {Physical Review A}\ }\textbf {\bibinfo
  {volume} {91}},\ \bibinfo {pages} {053849} (\bibinfo {year}
  {2015})}\BibitemShut {NoStop}%
\bibitem [{\citenamefont {Page}\ \emph {et~al.}(2016)\citenamefont {Page},
  \citenamefont {Zhao}, \citenamefont {Blair}, \citenamefont {Ju},
  \citenamefont {Ma}, \citenamefont {Pan}, \citenamefont {Chao}, \citenamefont
  {Mitrofanov},\ and\ \citenamefont {Sadeghian}}]{page2016thermal}%
  \BibitemOpen
  \bibfield  {author} {\bibinfo {author} {\bibfnamefont {M.~A.}\ \bibnamefont
  {Page}}, \bibinfo {author} {\bibfnamefont {C.}~\bibnamefont {Zhao}}, \bibinfo
  {author} {\bibfnamefont {D.~G.}\ \bibnamefont {Blair}}, \bibinfo {author}
  {\bibfnamefont {L.}~\bibnamefont {Ju}}, \bibinfo {author} {\bibfnamefont
  {Y.}~\bibnamefont {Ma}}, \bibinfo {author} {\bibfnamefont {H.-W.}\
  \bibnamefont {Pan}}, \bibinfo {author} {\bibfnamefont {S.}~\bibnamefont
  {Chao}}, \bibinfo {author} {\bibfnamefont {V.~P.}\ \bibnamefont
  {Mitrofanov}}, \ and\ \bibinfo {author} {\bibfnamefont {H.}~\bibnamefont
  {Sadeghian}},\ }\href {\doibase 10.1088/0022-3727/49/45/455104} {\bibfield
  {journal} {\bibinfo  {journal} {Journal of Physics D: Applied Physics}\
  }\textbf {\bibinfo {volume} {49}},\ \bibinfo {pages} {455104} (\bibinfo
  {year} {2016})}\BibitemShut {NoStop}%
\bibitem [{\citenamefont {Singh}\ \emph {et~al.}(2010)\citenamefont {Singh},
  \citenamefont {Phelps}, \citenamefont {Goldbaum}, \citenamefont {Wright},\
  and\ \citenamefont {Meystre}}]{singh2010all-optical}%
  \BibitemOpen
  \bibfield  {author} {\bibinfo {author} {\bibfnamefont {S.}~\bibnamefont
  {Singh}}, \bibinfo {author} {\bibfnamefont {G.~A.}\ \bibnamefont {Phelps}},
  \bibinfo {author} {\bibfnamefont {D.~S.}\ \bibnamefont {Goldbaum}}, \bibinfo
  {author} {\bibfnamefont {E.~M.}\ \bibnamefont {Wright}}, \ and\ \bibinfo
  {author} {\bibfnamefont {P.}~\bibnamefont {Meystre}},\ }\href {\doibase
  10.1103/PhysRevLett.105.213602} {\bibfield  {journal} {\bibinfo  {journal}
  {Physical Review Letters}\ }\textbf {\bibinfo {volume} {105}},\ \bibinfo
  {pages} {213602} (\bibinfo {year} {2010})}\BibitemShut {NoStop}%
\bibitem [{\citenamefont {Guccione}\ \emph {et~al.}(2013)\citenamefont
  {Guccione}, \citenamefont {Hosseini}, \citenamefont {Adlong}, \citenamefont
  {Johnsson}, \citenamefont {Hope}, \citenamefont {Buchler},\ and\
  \citenamefont {Lam}}]{guccione2013scattering-free}%
  \BibitemOpen
  \bibfield  {author} {\bibinfo {author} {\bibfnamefont {G.}~\bibnamefont
  {Guccione}}, \bibinfo {author} {\bibfnamefont {M.}~\bibnamefont {Hosseini}},
  \bibinfo {author} {\bibfnamefont {S.}~\bibnamefont {Adlong}}, \bibinfo
  {author} {\bibfnamefont {M.~T.}\ \bibnamefont {Johnsson}}, \bibinfo {author}
  {\bibfnamefont {J.}~\bibnamefont {Hope}}, \bibinfo {author} {\bibfnamefont
  {B.~C.}\ \bibnamefont {Buchler}}, \ and\ \bibinfo {author} {\bibfnamefont
  {P.~K.}\ \bibnamefont {Lam}},\ }\href {\doibase
  10.1103/PhysRevLett.111.183001} {\bibfield  {journal} {\bibinfo  {journal}
  {Physical Review Letters}\ }\textbf {\bibinfo {volume} {111}},\ \bibinfo
  {pages} {183001} (\bibinfo {year} {2013})}\BibitemShut {NoStop}%
\bibitem [{\citenamefont {Catena}\ and\ \citenamefont
  {Ullio}(2010)}]{catena2010novel}%
  \BibitemOpen
  \bibfield  {author} {\bibinfo {author} {\bibfnamefont {R.}~\bibnamefont
  {Catena}}\ and\ \bibinfo {author} {\bibfnamefont {P.}~\bibnamefont {Ullio}},\
  }\href {\doibase 10.1088/1475-7516/2010/08/004} {\bibfield  {journal}
  {\bibinfo  {journal} {Journal of Cosmology and Astroparticle Physics}\
  }\textbf {\bibinfo {volume} {2010}},\ \bibinfo {pages} {004} (\bibinfo {year}
  {2010})}\BibitemShut {NoStop}%
\bibitem [{\citenamefont {Viel}\ \emph {et~al.}(2013)\citenamefont {Viel},
  \citenamefont {Becker}, \citenamefont {Bolton},\ and\ \citenamefont
  {Haehnelt}}]{viel:2013xx}%
  \BibitemOpen
  \bibfield  {author} {\bibinfo {author} {\bibfnamefont {M.}~\bibnamefont
  {Viel}}, \bibinfo {author} {\bibfnamefont {G.~D.}\ \bibnamefont {Becker}},
  \bibinfo {author} {\bibfnamefont {J.~S.}\ \bibnamefont {Bolton}}, \ and\
  \bibinfo {author} {\bibfnamefont {M.~G.}\ \bibnamefont {Haehnelt}},\ }\href
  {\doibase 10.1103/PhysRevD.88.043502} {\bibfield  {journal} {\bibinfo
  {journal} {Physical Review D}\ }\textbf {\bibinfo {volume} {88}},\ \bibinfo
  {pages} {043502} (\bibinfo {year} {2013})}\BibitemShut {NoStop}%
\bibitem [{\citenamefont {Smith}\ \emph {et~al.}(2007)\citenamefont {Smith},
  \citenamefont {Ruchti}, \citenamefont {Helmi}, \citenamefont {Wyse},
  \citenamefont {Fulbright}, \citenamefont {Freeman}, \citenamefont {Navarro},
  \citenamefont {Seabroke}, \citenamefont {Steinmetz}, \citenamefont
  {Williams}, \citenamefont {Bienaym\'{e}}, \citenamefont {Binney},
  \citenamefont {{Bland-Hawthorn}}, \citenamefont {Dehnen}, \citenamefont
  {Gibson}, \citenamefont {Gilmore}, \citenamefont {Grebel}, \citenamefont
  {Munari}, \citenamefont {Parker}, \citenamefont {Scholz}, \citenamefont
  {Siebert}, \citenamefont {Watson},\ and\ \citenamefont
  {Zwitter}}]{smith2007rave}%
  \BibitemOpen
  \bibfield  {author} {\bibinfo {author} {\bibfnamefont {M.~C.}\ \bibnamefont
  {Smith}}, \bibinfo {author} {\bibfnamefont {G.~R.}\ \bibnamefont {Ruchti}},
  \bibinfo {author} {\bibfnamefont {A.}~\bibnamefont {Helmi}}, \bibinfo
  {author} {\bibfnamefont {R.~F.~G.}\ \bibnamefont {Wyse}}, \bibinfo {author}
  {\bibfnamefont {J.~P.}\ \bibnamefont {Fulbright}}, \bibinfo {author}
  {\bibfnamefont {K.~C.}\ \bibnamefont {Freeman}}, \bibinfo {author}
  {\bibfnamefont {J.~F.}\ \bibnamefont {Navarro}}, \bibinfo {author}
  {\bibfnamefont {G.~M.}\ \bibnamefont {Seabroke}}, \bibinfo {author}
  {\bibfnamefont {M.}~\bibnamefont {Steinmetz}}, \bibinfo {author}
  {\bibfnamefont {M.}~\bibnamefont {Williams}}, \bibinfo {author}
  {\bibfnamefont {O.}~\bibnamefont {Bienaym\'{e}}}, \bibinfo {author}
  {\bibfnamefont {J.}~\bibnamefont {Binney}}, \bibinfo {author} {\bibfnamefont
  {J.}~\bibnamefont {{Bland-Hawthorn}}}, \bibinfo {author} {\bibfnamefont
  {W.}~\bibnamefont {Dehnen}}, \bibinfo {author} {\bibfnamefont {B.~K.}\
  \bibnamefont {Gibson}}, \bibinfo {author} {\bibfnamefont {G.}~\bibnamefont
  {Gilmore}}, \bibinfo {author} {\bibfnamefont {E.~K.}\ \bibnamefont {Grebel}},
  \bibinfo {author} {\bibfnamefont {U.}~\bibnamefont {Munari}}, \bibinfo
  {author} {\bibfnamefont {Q.~A.}\ \bibnamefont {Parker}}, \bibinfo {author}
  {\bibfnamefont {R.}~\bibnamefont {Scholz}}, \bibinfo {author} {\bibfnamefont
  {A.}~\bibnamefont {Siebert}}, \bibinfo {author} {\bibfnamefont {F.~G.}\
  \bibnamefont {Watson}}, \ and\ \bibinfo {author} {\bibfnamefont
  {T.}~\bibnamefont {Zwitter}},\ }\href {\doibase
  10.1111/j.1365-2966.2007.11964.x} {\bibfield  {journal} {\bibinfo  {journal}
  {Monthly Notices of the Royal Astronomical Society}\ }\textbf {\bibinfo
  {volume} {379}},\ \bibinfo {pages} {755} (\bibinfo {year}
  {2007})}\BibitemShut {NoStop}%
\bibitem [{\citenamefont {Joos}\ and\ \citenamefont {Zeh}(1985)}]{Joos:1985xx}%
  \BibitemOpen
  \bibfield  {author} {\bibinfo {author} {\bibfnamefont {E.}~\bibnamefont
  {Joos}}\ and\ \bibinfo {author} {\bibfnamefont {H.~D.}\ \bibnamefont {Zeh}},\
  }\href {\doibase 10.1007/BF01725541} {\bibfield  {journal} {\bibinfo
  {journal} {Zeitschrift f\"{u}r Physik B Condensed Matter}\ }\textbf {\bibinfo
  {volume} {59}},\ \bibinfo {pages} {223{\textendash}243} (\bibinfo {year}
  {1985})}\BibitemShut {NoStop}%
\bibitem [{\citenamefont {Gallis}\ and\ \citenamefont
  {Fleming}(1990)}]{Gallis:1990xx}%
  \BibitemOpen
  \bibfield  {author} {\bibinfo {author} {\bibfnamefont {M.~R.}\ \bibnamefont
  {Gallis}}\ and\ \bibinfo {author} {\bibfnamefont {G.~N.}\ \bibnamefont
  {Fleming}},\ }\href {\doibase 10.1103/PhysRevA.42.38} {\bibfield  {journal}
  {\bibinfo  {journal} {Physical Review A}\ }\textbf {\bibinfo {volume} {42}},\
  \bibinfo {pages} {38} (\bibinfo {year} {1990})}\BibitemShut {NoStop}%
\bibitem [{\citenamefont {Schlosshauer}(2008)}]{Schlosshauer:2008xx}%
  \BibitemOpen
  \bibfield  {author} {\bibinfo {author} {\bibfnamefont {M.}~\bibnamefont
  {Schlosshauer}},\ }\href@noop {} {\emph {\bibinfo {title} {Decoherence and
  the {Quantum-to-Classical} Transition}}}\ (\bibinfo  {publisher}
  {{Springer-Verlag}},\ \bibinfo {address} {Berlin},\ \bibinfo {year}
  {2008})\BibitemShut {NoStop}%
\bibitem [{\citenamefont {Bennett}(1981)}]{bennett1981upper}%
  \BibitemOpen
  \bibfield  {author} {\bibinfo {author} {\bibfnamefont {H.~S.}\ \bibnamefont
  {Bennett}},\ }\href@noop {} {\bibfield  {journal} {\bibinfo  {journal}
  {Journal of Research of the National Bureau of Standards}\ }\textbf {\bibinfo
  {volume} {86}},\ \bibinfo {pages} {503} (\bibinfo {year} {1981})}\BibitemShut
  {NoStop}%
\bibitem [{\citenamefont {Petraki}\ \emph {et~al.}(2016)\citenamefont
  {Petraki}, \citenamefont {Postma},\ and\ \citenamefont {{de
  Vries}}}]{petraki2016radiative}%
  \BibitemOpen
  \bibfield  {author} {\bibinfo {author} {\bibfnamefont {K.}~\bibnamefont
  {Petraki}}, \bibinfo {author} {\bibfnamefont {M.}~\bibnamefont {Postma}}, \
  and\ \bibinfo {author} {\bibfnamefont {J.}~\bibnamefont {{de Vries}}},\
  }\href@noop {} {\bibfield  {journal} {\bibinfo  {journal} {arXiv preprint
  arXiv:1611.01394}\ } (\bibinfo {year} {2016})}\BibitemShut {NoStop}%
\bibitem [{\citenamefont {Taylor}(2006)}]{Taylor:2006xx}%
  \BibitemOpen
  \bibfield  {author} {\bibinfo {author} {\bibfnamefont {J.~R.}\ \bibnamefont
  {Taylor}},\ }\href@noop {} {\emph {\bibinfo {title} {Scattering theory: the
  quantum theory of nonrelativistic collisions}}}\ (\bibinfo  {publisher}
  {Dover Publications},\ \bibinfo {address} {Mineola, {NY}},\ \bibinfo {year}
  {2006})\BibitemShut {NoStop}%
\bibitem [{\citenamefont {Squires}(1978{\natexlab{a}})}]{Squires:1978xx}%
  \BibitemOpen
  \bibfield  {author} {\bibinfo {author} {\bibfnamefont {G.~L.}\ \bibnamefont
  {Squires}},\ }\href@noop {} {\emph {\bibinfo {title} {Introduction to the
  Theory of Thermal Neutron Scattering}}}\ (\bibinfo  {publisher} {Cambridge
  University Press},\ \bibinfo {address} {New York},\ \bibinfo {year}
  {1978})\BibitemShut {NoStop}%
\bibitem [{\citenamefont {Rembold}\ \emph {et~al.}(2014)\citenamefont
  {Rembold}, \citenamefont {Sch\"{u}tz}, \citenamefont {Chang}, \citenamefont
  {Stefanov}, \citenamefont {Pooch}, \citenamefont {Hwang}, \citenamefont
  {G\"{u}nther},\ and\ \citenamefont {Stibor}}]{rembold2014correction}%
  \BibitemOpen
  \bibfield  {author} {\bibinfo {author} {\bibfnamefont {A.}~\bibnamefont
  {Rembold}}, \bibinfo {author} {\bibfnamefont {G.}~\bibnamefont {Sch\"{u}tz}},
  \bibinfo {author} {\bibfnamefont {W.~T.}\ \bibnamefont {Chang}}, \bibinfo
  {author} {\bibfnamefont {A.}~\bibnamefont {Stefanov}}, \bibinfo {author}
  {\bibfnamefont {A.}~\bibnamefont {Pooch}}, \bibinfo {author} {\bibfnamefont
  {I.~S.}\ \bibnamefont {Hwang}}, \bibinfo {author} {\bibfnamefont
  {A.}~\bibnamefont {G\"{u}nther}}, \ and\ \bibinfo {author} {\bibfnamefont
  {A.}~\bibnamefont {Stibor}},\ }\href {\doibase 10.1103/PhysRevA.89.033635}
  {\bibfield  {journal} {\bibinfo  {journal} {Physical Review A}\ }\textbf
  {\bibinfo {volume} {89}},\ \bibinfo {pages} {033635} (\bibinfo {year}
  {2014})}\BibitemShut {NoStop}%
\bibitem [{\citenamefont {Günther}\ \emph {et~al.}(2015)\citenamefont
  {Günther}, \citenamefont {Rembold}, \citenamefont {Schütz},\ and\
  \citenamefont {Stibor}}]{gunther2015multifrequency}%
  \BibitemOpen
  \bibfield  {author} {\bibinfo {author} {\bibfnamefont {A.}~\bibnamefont
  {Günther}}, \bibinfo {author} {\bibfnamefont {A.}~\bibnamefont {Rembold}},
  \bibinfo {author} {\bibfnamefont {G.}~\bibnamefont {Schütz}}, \ and\
  \bibinfo {author} {\bibfnamefont {A.}~\bibnamefont {Stibor}},\ }\href
  {\doibase 10.1103/PhysRevA.92.053607} {\bibfield  {journal} {\bibinfo
  {journal} {Physical Review A}\ }\textbf {\bibinfo {volume} {92}},\ \bibinfo
  {pages} {053607} (\bibinfo {year} {2015})}\BibitemShut {NoStop}%
\bibitem [{\citenamefont {Rembold}\ \emph {et~al.}(2017)\citenamefont
  {Rembold}, \citenamefont {Röpke}, \citenamefont {Schütz}, \citenamefont
  {Fortágh}, \citenamefont {Stibor},\ and\ \citenamefont
  {Günther}}]{rembold2017secondorder}%
  \BibitemOpen
  \bibfield  {author} {\bibinfo {author} {\bibfnamefont {A.}~\bibnamefont
  {Rembold}}, \bibinfo {author} {\bibfnamefont {R.}~\bibnamefont {Röpke}},
  \bibinfo {author} {\bibfnamefont {G.}~\bibnamefont {Schütz}}, \bibinfo
  {author} {\bibfnamefont {J.}~\bibnamefont {Fortágh}}, \bibinfo {author}
  {\bibfnamefont {A.}~\bibnamefont {Stibor}}, \ and\ \bibinfo {author}
  {\bibfnamefont {A.}~\bibnamefont {Günther}},\ }\href@noop {} {\bibfield
  {journal} {\bibinfo  {journal} {arXiv:1703.07819}\ } (\bibinfo {year}
  {2017})}\BibitemShut {NoStop}%
\bibitem [{\citenamefont {Cherwinka}\ \emph {et~al.}(2012)\citenamefont
  {Cherwinka}, \citenamefont {Co}, \citenamefont {Cowen}, \citenamefont
  {Grant}, \citenamefont {Halzen}, \citenamefont {Heeger}, \citenamefont {Hsu},
  \citenamefont {Karle}, \citenamefont {Kudryavtsev}, \citenamefont {Maruyama},
  \citenamefont {Pettus}, \citenamefont {Robinson},\ and\ \citenamefont
  {Spooner}}]{cherwinka2011search}%
  \BibitemOpen
  \bibfield  {author} {\bibinfo {author} {\bibfnamefont {J.}~\bibnamefont
  {Cherwinka}}, \bibinfo {author} {\bibfnamefont {R.}~\bibnamefont {Co}},
  \bibinfo {author} {\bibfnamefont {D.~F.}\ \bibnamefont {Cowen}}, \bibinfo
  {author} {\bibfnamefont {D.}~\bibnamefont {Grant}}, \bibinfo {author}
  {\bibfnamefont {F.}~\bibnamefont {Halzen}}, \bibinfo {author} {\bibfnamefont
  {K.~M.}\ \bibnamefont {Heeger}}, \bibinfo {author} {\bibfnamefont
  {L.}~\bibnamefont {Hsu}}, \bibinfo {author} {\bibfnamefont {A.}~\bibnamefont
  {Karle}}, \bibinfo {author} {\bibfnamefont {V.~A.}\ \bibnamefont
  {Kudryavtsev}}, \bibinfo {author} {\bibfnamefont {R.}~\bibnamefont
  {Maruyama}}, \bibinfo {author} {\bibfnamefont {W.}~\bibnamefont {Pettus}},
  \bibinfo {author} {\bibfnamefont {M.}~\bibnamefont {Robinson}}, \ and\
  \bibinfo {author} {\bibfnamefont {N.~J.~C.}\ \bibnamefont {Spooner}},\ }\href
  {http://www.sciencedirect.com/science/article/pii/S0927650512000564}
  {\bibfield  {journal} {\bibinfo  {journal} {Astroparticle Physics}\ }\textbf
  {\bibinfo {volume} {35}},\ \bibinfo {pages} {749} (\bibinfo {year}
  {2012})}\BibitemShut {NoStop}%
\bibitem [{\citenamefont {Aalseth}\ \emph {et~al.}(2011)\citenamefont
  {Aalseth}, \citenamefont {Barbeau}, \citenamefont {Colaresi}, \citenamefont
  {Collar}, \citenamefont {Diaz~Leon}, \citenamefont {Fast}, \citenamefont
  {Fields}, \citenamefont {Hossbach}, \citenamefont {Keillor}, \citenamefont
  {Kephart}, \citenamefont {Knecht}, \citenamefont {Marino}, \citenamefont
  {Miley}, \citenamefont {Miller}, \citenamefont {Orrell}, \citenamefont
  {Radford}, \citenamefont {Wilkerson},\ and\ \citenamefont
  {Yocum}}]{aalseth2011search}%
  \BibitemOpen
  \bibfield  {author} {\bibinfo {author} {\bibfnamefont {C.~E.}\ \bibnamefont
  {Aalseth}}, \bibinfo {author} {\bibfnamefont {P.~S.}\ \bibnamefont
  {Barbeau}}, \bibinfo {author} {\bibfnamefont {J.}~\bibnamefont {Colaresi}},
  \bibinfo {author} {\bibfnamefont {J.~I.}\ \bibnamefont {Collar}}, \bibinfo
  {author} {\bibfnamefont {J.}~\bibnamefont {Diaz~Leon}}, \bibinfo {author}
  {\bibfnamefont {J.~E.}\ \bibnamefont {Fast}}, \bibinfo {author}
  {\bibfnamefont {N.}~\bibnamefont {Fields}}, \bibinfo {author} {\bibfnamefont
  {T.~W.}\ \bibnamefont {Hossbach}}, \bibinfo {author} {\bibfnamefont {M.~E.}\
  \bibnamefont {Keillor}}, \bibinfo {author} {\bibfnamefont {J.~D.}\
  \bibnamefont {Kephart}}, \bibinfo {author} {\bibfnamefont {A.}~\bibnamefont
  {Knecht}}, \bibinfo {author} {\bibfnamefont {M.~G.}\ \bibnamefont {Marino}},
  \bibinfo {author} {\bibfnamefont {H.~S.}\ \bibnamefont {Miley}}, \bibinfo
  {author} {\bibfnamefont {M.~L.}\ \bibnamefont {Miller}}, \bibinfo {author}
  {\bibfnamefont {J.~L.}\ \bibnamefont {Orrell}}, \bibinfo {author}
  {\bibfnamefont {D.~C.}\ \bibnamefont {Radford}}, \bibinfo {author}
  {\bibfnamefont {J.~F.}\ \bibnamefont {Wilkerson}}, \ and\ \bibinfo {author}
  {\bibfnamefont {K.~M.}\ \bibnamefont {Yocum}},\ }\href {\doibase
  10.1103/PhysRevLett.107.141301} {\bibfield  {journal} {\bibinfo  {journal}
  {Physical Review Letters}\ }\textbf {\bibinfo {volume} {107}},\ \bibinfo
  {pages} {141301} (\bibinfo {year} {2011})}\BibitemShut {NoStop}%
\bibitem [{\citenamefont {{Herrero-Garcia}}\ \emph {et~al.}(2012)\citenamefont
  {{Herrero-Garcia}}, \citenamefont {Schwetz},\ and\ \citenamefont
  {Zupan}}]{herrero-garcia2012annual}%
  \BibitemOpen
  \bibfield  {author} {\bibinfo {author} {\bibfnamefont {J.}~\bibnamefont
  {{Herrero-Garcia}}}, \bibinfo {author} {\bibfnamefont {T.}~\bibnamefont
  {Schwetz}}, \ and\ \bibinfo {author} {\bibfnamefont {J.}~\bibnamefont
  {Zupan}},\ }\href {\doibase 10.1088/1475-7516/2012/03/005} {\bibfield
  {journal} {\bibinfo  {journal} {Journal of Cosmology and Astroparticle
  Physics}\ }\textbf {\bibinfo {volume} {2012}},\ \bibinfo {pages} {005}
  (\bibinfo {year} {2012})}\BibitemShut {NoStop}%
\bibitem [{\citenamefont {Lee}\ \emph {et~al.}(2014)\citenamefont {Lee},
  \citenamefont {Lisanti}, \citenamefont {Peter},\ and\ \citenamefont
  {Safdi}}]{lee2014effect}%
  \BibitemOpen
  \bibfield  {author} {\bibinfo {author} {\bibfnamefont {S.~K.}\ \bibnamefont
  {Lee}}, \bibinfo {author} {\bibfnamefont {M.}~\bibnamefont {Lisanti}},
  \bibinfo {author} {\bibfnamefont {A.~H.~G.}\ \bibnamefont {Peter}}, \ and\
  \bibinfo {author} {\bibfnamefont {B.~R.}\ \bibnamefont {Safdi}},\ }\href
  {\doibase 10.1103/PhysRevLett.112.011301} {\bibfield  {journal} {\bibinfo
  {journal} {Physical Review Letters}\ }\textbf {\bibinfo {volume} {112}},\
  \bibinfo {pages} {011301} (\bibinfo {year} {2014})}\BibitemShut {NoStop}%
\bibitem [{\citenamefont {Britto}\ and\ \citenamefont
  {Meyers}(2015)}]{britto2015monthly}%
  \BibitemOpen
  \bibfield  {author} {\bibinfo {author} {\bibfnamefont {V.}~\bibnamefont
  {Britto}}\ and\ \bibinfo {author} {\bibfnamefont {J.}~\bibnamefont
  {Meyers}},\ }\href {\doibase 10.1088/1475-7516/2015/11/006} {\bibfield
  {journal} {\bibinfo  {journal} {Journal of Cosmology and Astroparticle
  Physics}\ }\textbf {\bibinfo {volume} {2015}},\ \bibinfo {pages} {006}
  (\bibinfo {year} {2015})}\BibitemShut {NoStop}%
\bibitem [{\citenamefont {Damour}\ and\ \citenamefont
  {Krauss}(1999)}]{damour1999new}%
  \BibitemOpen
  \bibfield  {author} {\bibinfo {author} {\bibfnamefont {T.}~\bibnamefont
  {Damour}}\ and\ \bibinfo {author} {\bibfnamefont {L.~M.}\ \bibnamefont
  {Krauss}},\ }\href {\doibase 10.1103/PhysRevD.59.063509} {\bibfield
  {journal} {\bibinfo  {journal} {Physical Review D}\ }\textbf {\bibinfo
  {volume} {59}},\ \bibinfo {pages} {063509} (\bibinfo {year}
  {1999})}\BibitemShut {NoStop}%
\bibitem [{\citenamefont {Peter}(2008)}]{peter2008particle}%
  \BibitemOpen
  \bibfield  {author} {\bibinfo {author} {\bibfnamefont {A.~H.~G.}\
  \bibnamefont {Peter}},\ }\emph {\bibinfo {title} {Particle Dark Matter in the
  Solar System}},\ \href@noop {} {Ph.D. thesis},\ \bibinfo  {school}
  {California Intitute of Technology} (\bibinfo {year} {2008})\BibitemShut
  {NoStop}%
\bibitem [{\citenamefont {Adler}(2009{\natexlab{a}})}]{adler2009flyby}%
  \BibitemOpen
  \bibfield  {author} {\bibinfo {author} {\bibfnamefont {S.~L.}\ \bibnamefont
  {Adler}},\ }\href {\doibase 10.1103/PhysRevD.79.023505} {\bibfield  {journal}
  {\bibinfo  {journal} {Physical Review D}\ }\textbf {\bibinfo {volume} {79}},\
  \bibinfo {pages} {023505} (\bibinfo {year} {2009}{\natexlab{a}})}\BibitemShut
  {NoStop}%
\bibitem [{\citenamefont {Edsjo}\ and\ \citenamefont
  {Peter}(2010)}]{edsjo2010comments}%
  \BibitemOpen
  \bibfield  {author} {\bibinfo {author} {\bibfnamefont {J.}~\bibnamefont
  {Edsjo}}\ and\ \bibinfo {author} {\bibfnamefont {A.~H.~G.}\ \bibnamefont
  {Peter}},\ }\href {http://arxiv.org/abs/1004.5258} {\bibfield  {journal}
  {\bibinfo  {journal} {{arXiv:1004.5258}}\ } (\bibinfo {year}
  {2010})}\BibitemShut {NoStop}%
\bibitem [{\citenamefont {Iorio}(2006)}]{iorio2006solar}%
  \BibitemOpen
  \bibfield  {author} {\bibinfo {author} {\bibfnamefont {L.}~\bibnamefont
  {Iorio}},\ }\href {\doibase 10.1088/1475-7516/2006/05/002} {\bibfield
  {journal} {\bibinfo  {journal} {Journal of Cosmology and Astroparticle
  Physics}\ }\textbf {\bibinfo {volume} {2006}},\ \bibinfo {pages} {002}
  (\bibinfo {year} {2006})}\BibitemShut {NoStop}%
\bibitem [{\citenamefont {Khriplovich}(2007)}]{khriplovich2007density}%
  \BibitemOpen
  \bibfield  {author} {\bibinfo {author} {\bibfnamefont {I.~B.}\ \bibnamefont
  {Khriplovich}},\ }\href {\doibase 10.1142/S0218271807010869} {\bibfield
  {journal} {\bibinfo  {journal} {International Journal of Modern Physics D}\
  }\textbf {\bibinfo {volume} {16}},\ \bibinfo {pages} {1475} (\bibinfo {year}
  {2007})}\BibitemShut {NoStop}%
\bibitem [{\citenamefont {Fr\`{e}re}\ \emph {et~al.}(2008)\citenamefont
  {Fr\`{e}re}, \citenamefont {Ling},\ and\ \citenamefont
  {Vertongen}}]{frere2008bound}%
  \BibitemOpen
  \bibfield  {author} {\bibinfo {author} {\bibfnamefont {J.}~\bibnamefont
  {Fr\`{e}re}}, \bibinfo {author} {\bibfnamefont {F.}~\bibnamefont {Ling}}, \
  and\ \bibinfo {author} {\bibfnamefont {G.}~\bibnamefont {Vertongen}},\ }\href
  {\doibase 10.1103/PhysRevD.77.083005} {\bibfield  {journal} {\bibinfo
  {journal} {Physical Review D}\ }\textbf {\bibinfo {volume} {77}},\ \bibinfo
  {pages} {083005} (\bibinfo {year} {2008})}\BibitemShut {NoStop}%
\bibitem [{\citenamefont {Pitjev}\ and\ \citenamefont
  {Pitjeva}(2013)}]{pitjev2013constraints}%
  \BibitemOpen
  \bibfield  {author} {\bibinfo {author} {\bibfnamefont {N.~P.}\ \bibnamefont
  {Pitjev}}\ and\ \bibinfo {author} {\bibfnamefont {E.~V.}\ \bibnamefont
  {Pitjeva}},\ }\href {\doibase 10.1134/S1063773713020060} {\bibfield
  {journal} {\bibinfo  {journal} {Astronomy Letters}\ }\textbf {\bibinfo
  {volume} {39}},\ \bibinfo {pages} {141} (\bibinfo {year} {2013})}\BibitemShut
  {NoStop}%
\bibitem [{\citenamefont {Adler}(2008)}]{adler2008placing}%
  \BibitemOpen
  \bibfield  {author} {\bibinfo {author} {\bibfnamefont {S.~L.}\ \bibnamefont
  {Adler}},\ }\href {\doibase 10.1088/1751-8113/41/41/412002} {\bibfield
  {journal} {\bibinfo  {journal} {Journal of Physics A: Mathematical and
  Theoretical}\ }\textbf {\bibinfo {volume} {41}},\ \bibinfo {pages} {412002}
  (\bibinfo {year} {2008})}\BibitemShut {NoStop}%
\bibitem [{\citenamefont {Bateman}\ \emph {et~al.}(2015)\citenamefont
  {Bateman}, \citenamefont {{McHardy}}, \citenamefont {Merle}, \citenamefont
  {Morris},\ and\ \citenamefont {Ulbricht}}]{bateman2015existence}%
  \BibitemOpen
  \bibfield  {author} {\bibinfo {author} {\bibfnamefont {J.}~\bibnamefont
  {Bateman}}, \bibinfo {author} {\bibfnamefont {I.}~\bibnamefont {{McHardy}}},
  \bibinfo {author} {\bibfnamefont {A.}~\bibnamefont {Merle}}, \bibinfo
  {author} {\bibfnamefont {T.~R.}\ \bibnamefont {Morris}}, \ and\ \bibinfo
  {author} {\bibfnamefont {H.}~\bibnamefont {Ulbricht}},\ }\href
  {http://www.nature.com/srep/2015/150127/srep08058/full/srep08058.html}
  {\bibfield  {journal} {\bibinfo  {journal} {Scientific Reports}\ }\textbf
  {\bibinfo {volume} {5}},\ \bibinfo {pages} {8058} (\bibinfo {year}
  {2015})}\BibitemShut {NoStop}%
\bibitem [{\citenamefont {Lin}\ \emph {et~al.}(2012)\citenamefont {Lin},
  \citenamefont {Yu},\ and\ \citenamefont {Zurek}}]{Lin:2011gj}%
  \BibitemOpen
  \bibfield  {author} {\bibinfo {author} {\bibfnamefont {T.}~\bibnamefont
  {Lin}}, \bibinfo {author} {\bibfnamefont {H.-B.}\ \bibnamefont {Yu}}, \ and\
  \bibinfo {author} {\bibfnamefont {K.~M.}\ \bibnamefont {Zurek}},\ }\href
  {\doibase 10.1103/PhysRevD.85.063503} {\bibfield  {journal} {\bibinfo
  {journal} {Phys.Rev.}\ }\textbf {\bibinfo {volume} {D85}},\ \bibinfo {pages}
  {063503} (\bibinfo {year} {2012})}\BibitemShut {NoStop}%
\bibitem [{\citenamefont {Madhavacheril}\ \emph {et~al.}(2014)\citenamefont
  {Madhavacheril}, \citenamefont {Sehgal},\ and\ \citenamefont
  {Slatyer}}]{madhavacheril2014current}%
  \BibitemOpen
  \bibfield  {author} {\bibinfo {author} {\bibfnamefont {M.~S.}\ \bibnamefont
  {Madhavacheril}}, \bibinfo {author} {\bibfnamefont {N.}~\bibnamefont
  {Sehgal}}, \ and\ \bibinfo {author} {\bibfnamefont {T.~R.}\ \bibnamefont
  {Slatyer}},\ }\href {\doibase 10.1103/PhysRevD.89.103508} {\bibfield
  {journal} {\bibinfo  {journal} {Physical Review D}\ }\textbf {\bibinfo
  {volume} {89}},\ \bibinfo {pages} {103508} (\bibinfo {year}
  {2014})}\BibitemShut {NoStop}%
\bibitem [{\citenamefont {Roberts}\ and\ \citenamefont
  {Ursell}(1960)}]{Roberts:1960xx}%
  \BibitemOpen
  \bibfield  {author} {\bibinfo {author} {\bibfnamefont {P.~H.}\ \bibnamefont
  {Roberts}}\ and\ \bibinfo {author} {\bibfnamefont {H.~D.}\ \bibnamefont
  {Ursell}},\ }\href {\doibase 10.1098/rsta.1960.0008} {\bibfield  {journal}
  {\bibinfo  {journal} {Philosophical Transactions of the Royal Society of
  London A: Mathematical, Physical and Engineering Sciences}\ }\textbf
  {\bibinfo {volume} {252}},\ \bibinfo {pages} {317} (\bibinfo {year}
  {1960})}\BibitemShut {NoStop}%
\bibitem [{\citenamefont {Adler}(2009{\natexlab{b}})}]{Adler:2009xx}%
  \BibitemOpen
  \bibfield  {author} {\bibinfo {author} {\bibfnamefont {S.~L.}\ \bibnamefont
  {Adler}},\ }\href {\doibase 10.1103/PhysRevD.79.023505} {\bibfield  {journal}
  {\bibinfo  {journal} {Physical Review D}\ }\textbf {\bibinfo {volume} {79}},\
  \bibinfo {pages} {023505} (\bibinfo {year} {2009}{\natexlab{b}})}\BibitemShut
  {NoStop}%
\bibitem [{\citenamefont {Hornberger}\ \emph {et~al.}(2009)\citenamefont
  {Hornberger}, \citenamefont {Gerlich}, \citenamefont {Ulbricht},
  \citenamefont {Hackerm\"{u}ller}, \citenamefont {Nimmrichter}, \citenamefont
  {Goldt}, \citenamefont {Boltalina},\ and\ \citenamefont
  {Arndt}}]{Hornberger:2009xx}%
  \BibitemOpen
  \bibfield  {author} {\bibinfo {author} {\bibfnamefont {K.}~\bibnamefont
  {Hornberger}}, \bibinfo {author} {\bibfnamefont {S.}~\bibnamefont {Gerlich}},
  \bibinfo {author} {\bibfnamefont {H.}~\bibnamefont {Ulbricht}}, \bibinfo
  {author} {\bibfnamefont {L.}~\bibnamefont {Hackerm\"{u}ller}}, \bibinfo
  {author} {\bibfnamefont {S.}~\bibnamefont {Nimmrichter}}, \bibinfo {author}
  {\bibfnamefont {I.~V.}\ \bibnamefont {Goldt}}, \bibinfo {author}
  {\bibfnamefont {O.}~\bibnamefont {Boltalina}}, \ and\ \bibinfo {author}
  {\bibfnamefont {M.}~\bibnamefont {Arndt}},\ }\href {\doibase
  10.1088/1367-2630/11/4/043032} {\bibfield  {journal} {\bibinfo  {journal}
  {New Journal of Physics}\ }\textbf {\bibinfo {volume} {11}},\ \bibinfo
  {pages} {043032} (\bibinfo {year} {2009})}\BibitemShut {NoStop}%
\bibitem [{\citenamefont {Arndt}\ \emph {et~al.}(1999)\citenamefont {Arndt},
  \citenamefont {Nairz}, \citenamefont {{Vos-Andreae}}, \citenamefont {Keller},
  \citenamefont {van~der Zouw},\ and\ \citenamefont
  {Zeilinger}}]{Arndt:1999xx}%
  \BibitemOpen
  \bibfield  {author} {\bibinfo {author} {\bibfnamefont {M.}~\bibnamefont
  {Arndt}}, \bibinfo {author} {\bibfnamefont {O.}~\bibnamefont {Nairz}},
  \bibinfo {author} {\bibfnamefont {J.}~\bibnamefont {{Vos-Andreae}}}, \bibinfo
  {author} {\bibfnamefont {C.}~\bibnamefont {Keller}}, \bibinfo {author}
  {\bibfnamefont {G.}~\bibnamefont {van~der Zouw}}, \ and\ \bibinfo {author}
  {\bibfnamefont {A.}~\bibnamefont {Zeilinger}},\ }\href
  {http://www.nature.com/nature/journal/v401/n6754/abs/401680a0.html}
  {\bibfield  {journal} {\bibinfo  {journal} {Nature}\ }\textbf {\bibinfo
  {volume} {401}},\ \bibinfo {pages} {680} (\bibinfo {year}
  {1999})}\BibitemShut {NoStop}%
\bibitem [{\citenamefont {Landau}\ and\ \citenamefont
  {Lifshitz}(1977)}]{landau1991quantum}%
  \BibitemOpen
  \bibfield  {author} {\bibinfo {author} {\bibfnamefont {L.}~\bibnamefont
  {Landau}}\ and\ \bibinfo {author} {\bibfnamefont {E.}~\bibnamefont
  {Lifshitz}},\ }\href@noop {} {\emph {\bibinfo {title} {Course of Theoretical
  Physics, Vol. 3 - Quantum Mechanics: Non-relativistic Theory}}},\ \bibinfo
  {edition} {3rd}\ ed.\ (\bibinfo  {publisher} {Pergamon Press},\ \bibinfo
  {year} {1977})\BibitemShut {NoStop}%
\bibitem [{\citenamefont {Morse}\ and\ \citenamefont
  {Feshbach}(1953)}]{morse1953methods}%
  \BibitemOpen
  \bibfield  {author} {\bibinfo {author} {\bibfnamefont {P.~M.}\ \bibnamefont
  {Morse}}\ and\ \bibinfo {author} {\bibfnamefont {H.}~\bibnamefont
  {Feshbach}},\ }\href@noop {} {\emph {\bibinfo {title} {Methods of Theoretical
  Physics}}}\ (\bibinfo  {publisher} {{McGraw-Hill}},\ \bibinfo {year}
  {1953})\BibitemShut {NoStop}%
\bibitem [{\citenamefont {Newton}(1982)}]{Newton:1982xx}%
  \BibitemOpen
  \bibfield  {author} {\bibinfo {author} {\bibfnamefont {R.~G.}\ \bibnamefont
  {Newton}},\ }\href@noop {} {\emph {\bibinfo {title} {Scattering Theory of
  Waves and Particles}}}\ (\bibinfo  {publisher} {Courier Dover Publications},\
  \bibinfo {year} {1982})\BibitemShut {NoStop}%
\bibitem [{\citenamefont {Everhart}\ \emph {et~al.}(1955)\citenamefont
  {Everhart}, \citenamefont {Stone},\ and\ \citenamefont
  {Carbone}}]{Everhart:1955xx}%
  \BibitemOpen
  \bibfield  {author} {\bibinfo {author} {\bibfnamefont {E.}~\bibnamefont
  {Everhart}}, \bibinfo {author} {\bibfnamefont {G.}~\bibnamefont {Stone}}, \
  and\ \bibinfo {author} {\bibfnamefont {R.~J.}\ \bibnamefont {Carbone}},\
  }\href {\doibase 10.1103/PhysRev.99.1287} {\bibfield  {journal} {\bibinfo
  {journal} {Physical Review}\ }\textbf {\bibinfo {volume} {99}},\ \bibinfo
  {pages} {1287} (\bibinfo {year} {1955})}\BibitemShut {NoStop}%
\bibitem [{\citenamefont {Mott}\ and\ \citenamefont
  {Massey}(1965)}]{Mott:1965xx}%
  \BibitemOpen
  \bibfield  {author} {\bibinfo {author} {\bibfnamefont {N.~F.}\ \bibnamefont
  {Mott}}\ and\ \bibinfo {author} {\bibfnamefont {H.~S.~W.}\ \bibnamefont
  {Massey}},\ }\href@noop {} {\emph {\bibinfo {title} {Theory of Atomic
  Collisions}}},\ \bibinfo {edition} {3rd}\ ed.\ (\bibinfo  {publisher}
  {Oxford},\ \bibinfo {address} {Oxford},\ \bibinfo {year} {1965})\BibitemShut
  {NoStop}%
\bibitem [{\citenamefont
  {Squires}(1978{\natexlab{b}})}]{squires1978introduction}%
  \BibitemOpen
  \bibfield  {author} {\bibinfo {author} {\bibfnamefont {G.~L.}\ \bibnamefont
  {Squires}},\ }\href@noop {} {\emph {\bibinfo {title} {Introduction to the
  Theory of Thermal Neutron Scattering}}}\ (\bibinfo  {publisher} {Cambridge
  University Press},\ \bibinfo {address} {New York},\ \bibinfo {year}
  {1978})\BibitemShut {NoStop}%
\bibitem [{\citenamefont {Singh}\ and\ \citenamefont
  {Sharma}(1971)}]{singh1971debye-waller}%
  \BibitemOpen
  \bibfield  {author} {\bibinfo {author} {\bibfnamefont {N.}~\bibnamefont
  {Singh}}\ and\ \bibinfo {author} {\bibfnamefont {P.~K.}\ \bibnamefont
  {Sharma}},\ }\href {\doibase 10.1103/PhysRevB.3.1141} {\bibfield  {journal}
  {\bibinfo  {journal} {Physical Review B}\ }\textbf {\bibinfo {volume} {3}},\
  \bibinfo {pages} {1141} (\bibinfo {year} {1971})}\BibitemShut {NoStop}%
\end{thebibliography}%

\end{document}